\documentclass[11pt]{article}
\usepackage[usenames,dvipsnames,svgnames,table]{xcolor} 
\usepackage[obeyspaces,hyphens,spaces]{url}
\usepackage{jcapmod}
\usepackage[english]{babel}
\usepackage{amsmath, amssymb, amsbsy, amstext, amsthm,mathtools}
\usepackage{hyperref}
\usepackage{graphicx}
\usepackage{amsfonts}
\usepackage{fouridx}
\usepackage{cases}
\usepackage{bm}
\usepackage{bbm}
\usepackage{cleveref}
\usepackage{slashed}
\usepackage{mdframed}
\usepackage{tensor}
\usepackage{simpler-wick}
\usepackage{setspace}

\usepackage{braket}
\usepackage{subcaption}

\usepackage{tikz}
\usetikzlibrary{calc,shadings,patterns,tikzmark,fadings,decorations.markings, decorations.pathmorphing}

\definecolor{Blue}{rgb}{0.25, 0.41, 0.88}
\definecolor{Red}{rgb}{0.92,0.,0.}
\definecolor{darkorange}{rgb}{1.0,0.549,0.}
\definecolor{cobalt}{RGB}{44, 98, 120}
\definecolor{Mathematica1}{rgb}{0.368417, 0.506779, 0.709798}
\definecolor{Mathematica2}{rgb}{0.880722, 0.611041, 0.142051}
\definecolor{Mathematica3}{rgb}{0.560181, 0.691569, 0.194885}
\definecolor{Mathematica4}{rgb}{0.922526, 0.385626, 0.209179}
\definecolor{Mathematica5}{rgb}{0.528488, 0.470624, 0.701351}
\definecolor{Mathematica6}{rgb}{0.772079, 0.431554, 0.102387}
\definecolor{Mathematica7}{rgb}{0.363898, 0.618501, 0.782349}
\definecolor{Mathematica8}{rgb}{1, 0.75, 0}
\definecolor{Mathematica9}{rgb}{0.647624, 0.37816, 0.614037}
\definecolor{plotBlue}{RGB}{94, 130, 181}
\definecolor{plotRed}{RGB}{233, 85, 54}
\definecolor{plotGreen}{RGB}{142, 176, 50}
\definecolor{plotPurple}{RGB}{135, 120, 178}
\definecolor{cornellRed}{HTML}{B31B1B}
\definecolor{cornellBlue}{HTML}{0068AC}
\definecolor{cornellGreen}{HTML}{6EB43F}

\newcolumntype{C}[1]{>{\centering\let\newline\\\arraybackslash\hspace{0pt}}m{#1}}

\def\e{{\lab{e}}}

\newcommand{\SHc}[1]{{Y^{\raisebox{1pt}{$\scriptstyle*$}}}_{\hspace{-7pt}#1}}

\newcommand{\tFo}[4]{{}_2 F_1\!\left[\genfrac..{-1pt}{0}{\raisebox{-1pt}{$#1, \,\, #2$}}{\raisebox{1pt}{$#3$}} \, \bigg| \, #4\, \right]}
\newcommand{\pFq}[5]{{}_{#1} F_{#2}\!\left[\genfrac..{-1pt}{0}{\raisebox{-1pt}{$#3$}}{\raisebox{1pt}{$#4$}} \, \bigg| \, #5\, \right]}

\newcommand{\G}[2]{\Gamma\!\left[\genfrac..{-1pt}{0}{\raisebox{-1pt}{$#1$}}{\raisebox{1pt}{$#2$}}\right]}

\makeatletter
\newlength{\apb@width}
\newcommand{\autoparbox}[2][c]{\settowidth{\apb@width}{#2}\parbox[#1]{\apb@width}{#2}}

\makeatother

\makeatletter
\newsavebox\myboxA
\newsavebox\myboxB
\newlength\mylenA

\newcommand*\xoverline[2][0.75]{
    \sbox{\myboxA}{$\m@th#2$}%
    \setbox\myboxB\null
    \ht\myboxB=\ht\myboxA%
    \dp\myboxB=\dp\myboxA%
    \wd\myboxB=#1\wd\myboxA
    \sbox\myboxB{$\m@th\overline{\copy\myboxB}$}
    \setlength\mylenA{\the\wd\myboxA}
    \addtolength\mylenA{-\the\wd\myboxB}%
    \ifdim\wd\myboxB<\wd\myboxA%
       \rlap{\hskip 0.5\mylenA\usebox\myboxB}{\usebox\myboxA}%
    \else
        \hskip -0.5\mylenA\rlap{\usebox\myboxA}{\hskip 0.5\mylenA\usebox\myboxB}%
    \fi}
\makeatother

\numberwithin{equation}{section}

\newcommand{\ud}{\mathrm{d}}
\newcommand{\lab}[1]{{\mathrm{#1}}}
\newcommand{\mb}[1]{{\mathbf{#1}}}
\newcommand{\minus}{{\scalebox {0.8}[1.0]{$-$}}}
\newcommand{\sminus}{{\scalebox {0.6}[0.85]{$-$}}}

\newcommand{\nord}[1]{{:\mathrel{#1}:}}

\theoremstyle{definition}

\DeclareRobustCommand{\SkipTocEntry}[4]{}

\newcommand{\es}{\hspace{0.5pt}}

\definecolor{pyBlue}{RGB}{31, 119, 180}
\definecolor{pyRed}{RGB}{214, 39, 40}
\definecolor{pyGreen}{RGB}{44, 160, 44}
\definecolor{pyBlue2}{RGB}{0, 111, 237}
\definecolor{pyRed2}{RGB}{224, 52, 36}

\newcommand{\slab}[1]{{\textsc{#1}}}

\newcommand{\subp}{{\scriptscriptstyle +}}
\newcommand{\subm}{{\scriptscriptstyle -}}

\tikzstyle{intSty}=[draw=white, thick, line width=0.24mm]
\tikzstyle{inflSty}=[cornellRed]
\tikzstyle{vertexProp}=[densely dashed, line width=0.3mm, dash phase=1pt]
\tikzstyle{sigmaProp}=[line width=0.3mm]
\tikzstyle{intVertSty}=[draw=white, line width=0.2mm]
\def\extVert{0.06}
\def\intVert{0.08}

\begin{document}

\pagenumbering{roman}
\begin{titlepage}
\baselineskip=15.5pt \thispagestyle{empty}

\bigskip\

\vspace{1cm}
\begin{center}
{\fontsize{18}{24}\selectfont  {\bfseries Compact Scalars at the Cosmological Collider}}
\end{center}
\vspace{0.1cm}
\begin{center}
{\fontsize{12}{18}\selectfont Priyesh Chakraborty and John Stout} 
\end{center}

\begin{center}
\vskip8pt
\textit{Department of Physics, Harvard University, Cambridge, MA 02138, USA}

\end{center}

\vspace{1.2cm}
\hrule \vspace{0.3cm}
\noindent {\bf Abstract}\\[0.1cm]
	We study the dynamics of scalar fields with compact field spaces, or axions, in de Sitter space. We argue that the field space topology can qualitatively affect the physics of these fields beyond just which terms are allowed in their actions. We argue that the sharpest difference is for massless fields---the free massless noncompact scalar field does not admit a two-point function that is both de Sitter-invariant and well-behaved at long distances, while the massless compact scalar does. As proof that this difference can be observable, we show that the long-distance behavior of a heavy scalar field, and thus its cosmological collider signal, can qualitatively change depending on whether it interacts with a light compact or noncompact scalar field. We find an interesting interplay between the circumference of the field space and the Hubble scale. When the field space is much larger than Hubble, the compact field behaves similarly to a light noncompact field and forces the heavy field to dilute much faster than any free field can. However, depending on how much smaller the field space is compared to Hubble, the compact field can cause the heavy scalar to decay either faster or slower than any free field and so we conclude that there can be qualitative and observable consequences of the field space's topology in inflationary correlation~functions.

\vskip10pt
\hrule
\vskip10pt

\end{titlepage}

\newcommand\emd{\xi}
\newcommand\imd{\zeta}

\thispagestyle{empty}
\setcounter{page}{2}
\begin{spacing}{1.03}
\tableofcontents
\end{spacing}

\clearpage
\pagenumbering{arabic}
\setcounter{page}{1}

\newpage

\newpage

\section{Introduction}

	It is hard to work in high energy physics these days without running into a compact scalar field. Also commonly called an axion or an axion-like particle, such fields were originally proposed to solve the strong CP problem~\cite{Peccei:1977ur,Peccei:1977hh,Weinberg:1977ma,Wilczek:1977pj} but have since seen use in a wide range of applications in beyond the Standard Model physics~\cite{Hook:2018dlk,Reece:2023czb}. They could be a component of dark matter~\cite{Preskill:1982cy,Dine:1982ah,Abbott:1982af}, the inflaton~\cite{Freese:1990rb}, or generate matter/antimatter asymmetry~\cite{Alexander:2004us}, just to name a few examples. Their core utility arises from their compact field space, i.e. that it is a circle rather than the entire real line, which allows them to enjoy a robust mechanism---quantum tunneling, essentially---that protects the qualitative form of their potential from currently incalculable quantum corrections. Beyond their utility for phenomenology, they also seem to be a generic feature of quantum gravitational theories at low energies, arising in great numbers and across wide ranges of masses in particularly well-studied lampposts of the string landscape~\cite{Svrcek:2006yi, Arvanitaki:2009fg,Demirtas:2018akl,Mehta:2021pwf,Demirtas:2021gsq,Gendler:2023kjt}. 

	Given the fundamental role it plays in the structure of the theory, are there any observational consequences of an axion's compact field space? Said differently, a compact scalar field $\varphi$ enjoys the \emph{gauge symmetry} $\varphi(x) \sim \varphi(x) + 2 \pi$, which is \emph{not} the same as a discrete global shift symmetry~$\varphi \to \varphi + 2 \pi$, even if it is often treated as such. Can we detect the difference? We know the answer for electromagnetism, where the $\lab{U}(1)$ gauge symmetry~$A_\mu(x) \sim A_\mu(x) + \partial_\mu \Lambda(x)$ removes the photon's longitudinal mode and ensures that it has long-range correlations---there is no chance of confusing it for a $\lab{U}(1)$ global symmetry. What about the axion?
	The goal of this paper is to explain how compact and noncompact scalar fields can have \emph{qualitatively} different behavior in de Sitter space, and thus the gauge symmetry $\varphi(x) \sim \varphi(x) + 2 \pi$ can have distinct observational consequences, even if these fields are described by exactly the same action.

	The standard answer to the above question is a bit boring and not particularly dynamical: to distinguish between compact and noncompact scalars, we simply catalog all states that couple to the scalar and determine whether or not there exist axion strings. These are strings in which the scalar field winds around its field space an integer number of times as we move around the string once, and thus only exist if the scalar is compact. Here, however, the word ``simply'' is doing a lot of work. The tension of these strings may be extremely high if the axion is \emph{fundamental}~\cite{Dolan:2017vmn,Reece:2018zvv,Heidenreich:2021yda}, arising from the dimensional reduction of a higher-dimensional gauge field, and so we may have no hope of ever creating or observing such an object. It seems similarly hopeless to detect whether the axion's potential is periodic or not if we are constrained to only lie in a single minimum and perform scattering experiments. We need \emph{global} information about the field space, not just local.

	Fortunately, things are different in de Sitter. As we explain in 
	Section~\ref{sec:compactMink}, compact and noncompact scalar fields differ in how we perform the path integral. That is, given a scalar field $\varphi$, do we impose the gauge constraint $\varphi(x) \sim \varphi(x) + 2 \pi$ and sum over all paths that respect this identification or not? This can often be turned into a decision about how we integrate over the spatially isotropic \emph{zero mode} of the field, which we denote $\varphi_0$, which becomes the only degree of freedom that knows about the field space's compactness. It is very difficult to distinguish between a compact and noncompact scalar field in Minkowski space because, there, this zero mode has infinite action and thus freezes out. The field's wavefunction cannot spread around its field space and figure out that it is a circle. However, since the Hubble volume is finite, the analogous mode in de Sitter is dynamically active and it can even dominate correlation functions~\cite{Allen:1985ux,Starobinsky:1986fx,Starobinsky:1994bd,Gorbenko:2019rza,Rajaraman:2010xd,Hollands:2011we,Beneke:2012kn,Burgess:2009bs,Burgess:2010dd,Chen:2016nrs,LopezNacir:2016gzi,Mirbabayi:2019qtx,Mirbabayi:2020vyt,Cohen:2020php,Baumgart:2020oby,Cohen:2021fzf}. For instance, loop corrections~\cite{Marolf:2010zp} induced by a light scalar field~$\varphi$ in de Sitter, with $m_\varphi < \frac{3}{2} H$, are enhanced~\cite{Lu:2021wxu,Chakraborty:2023qbp} by powers of the inverse mass $m_\varphi^{\sminus 1}$ in the light limit, due to the violent fluctuations of $\varphi$'s zero mode.

	\begin{figure}
		\centering
		\subfloat[\label{fig:varCompNon}]{\includegraphics[scale=1.05]{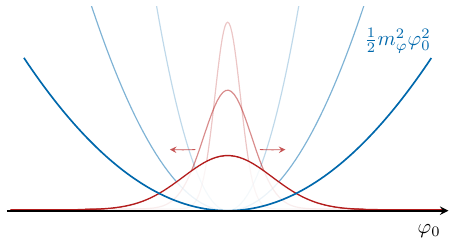}}\qquad
		\subfloat[\label{fig:varCompComp}]{\includegraphics[trim={-0.9cm 0 0 0}, scale=1.05]{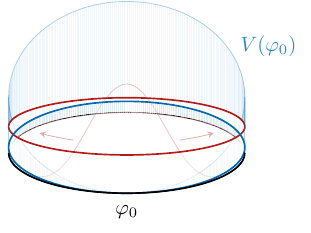}}
		\caption{The central idea of this paper. As the mass or potential of a scalar field $\varphi$ goes to zero, its dynamics in de Sitter cause it to explore its entire field space. For noncompact fields~[\textbf{left}], this allows the field to have arbitrarily large fluctuations and the variance of its spatially isotropic mode $\varphi_0$ diverges, $\langle \varphi_0^2 \rangle \propto (3 H^4)/(8 \pi^2 m_\varphi^2)$ for a quadratic potential. A compact field, however, only has so much field space to explore~[\textbf{right}] and its fluctuations remain under control even as its potential vanishes. We thus expect qualitative differences between the two in de Sitter.  \label{fig:varianceComparison}}
	\end{figure}

	The difference between compact and noncompact scalars is readily apparent in Starobinsky's stochastic picture~\cite{Starobinsky:1986fx,Starobinsky:1994bd,Gorbenko:2019rza}. In this framework, the scalar field is split into sub- and super-horizon modes and an effective description for the super-horizon modes is constructed. This effective description is in terms of the classical probability distribution $p(\varphi_0)$ for the light field's one-point function~$\langle \varphi(x) \rangle = \varphi_0$, which describes the distribution of the spatially isotropic zero mode $\varphi_0$ when averaged over Hubble patches. This probability distribution evolves according to the Fokker-Planck equation
	\begin{equation}
		\frac{\partial}{\partial t} \es p(\varphi_0, t) = \frac{H^3}{8\pi^2} \frac{\partial^2}{\partial \varphi_0^2}\es p(\varphi_0, t) +  \frac{1}{3 H} \frac{\partial}{\partial \varphi_0}\!\left(\! p(\varphi_0,t) \, \frac{\partial V(\varphi_0)}{\partial \varphi_0}\!\right)\,, \label{eq:fokker}
	\end{equation}
	with $V(\varphi)$ the field's potential, and if the field is compact $\varphi \sim \varphi + 2 \pi$ we must impose periodic boundary conditions~$p(\varphi + 2\pi, t) = p(\varphi, t)$~\cite{Graham:2018jyp}. As illustrated in Figure~\ref{fig:varCompNon}, in the limit that the noncompact field's potential vanishes and the scalar becomes massless, the equilibrium distribution $p(\varphi_0) \propto \exp\!\big(\minus \frac{8}{3} \pi^2  V(\varphi_0)/H^4\big)$ spreads out over the entire field space and the variance~$\langle \varphi_0^2\rangle$ diverges. This is the origin of the large loop corrections mentioned above, as the enormous fluctuations of the nearly massless noncompact field drive the theory to strong coupling.

	The compact field is clearly different. In the massless limit, its equilibrium distribution will also spread out over its field space, as shown in Figure~\ref{fig:varCompComp}. However, the variance (suitably defined) of the equilibrium distribution no longer diverges as the field is constrained to the circle. The compact field fluctuations at equilibrium are thus always under control, even when the field is massless. We should also expect qualitative differences between the dynamics of the two, even if they have the same potential. After all (\ref{eq:fokker}) is a Schr\"{o}dinger-like equation, and the quantum mechanical analog of this is the comparing a particle that lives on a circle and to one that lives in a one-dimensional crystal. Even though these may be described by exactly the same Hamiltonian, they have distinct physics: the former has a discrete set of energy eigenstates while the latter has a continuum of Bloch waves with no energy gap. 

	As one might expect from the discussion above, we will show that perhaps the sharpest difference between compact and noncompact scalars in de Sitter is for free massless fields. Said most simply, the massless compact scalar exists, admitting a well-behaved de Sitter-invariant propagator, while its noncompact counterpart does not~\cite{Allen:1985ux}. The compact scalar thus provides a realization of a massless scalar field in de Sitter space that is more natural than other attempts, which either explicitly project out the field's ill-behaved zero mode or restrict to certain shift-symmetric observables~\cite{Kirsten:1993ug,Tolley:2001gg,Page:2012fn}. Both elements are present here but are natural consequences of the gauge symmetry $\varphi \sim \varphi + 2 \pi$, which there is strong theoretical pressure towards, rather than ad hoc constraints placed on the field or its observables. 

	We also identify a potential inflationary observation channel that can distinguish between light compact and noncompact scalars. We will focus on the so-called ``cosmological collider'' signal~\cite{Chen:2009we,Chen:2009zp,Chen:2010xka, Baumann:2011nk,Chen:2012ge,Arkani-Hamed:2015bza,Assassi:2012zq,Noumi:2012vr,Lee:2016vti,Flauger:2016idt,Wang:2022eop} that a heavy scalar field $\sigma$, with mass $m_\sigma > \frac{3}{2} H$, can impart onto the primordial three-point function or bispectrum of the co-moving curvature perturbation $\zeta_{\mb{k}}$, which we parameterize as 
	\begin{equation}
		\langle \zeta_{\mb{k}_1} \zeta_{\mb{k}_2} \zeta_{\mb{k}_3} \rangle \equiv \frac{(2 \pi)^4 P^2_\zeta}{(k_1 k_2 k_3)^2}  (2 \pi)^3 \delta^{(3)}\big(\mb{k}_1 + \mb{k}_2 + \mb{k}_3) S(k_1, k_2, k_3)\,.
	\end{equation}
	Here, $P_\zeta \simeq 2 \times 10^{\sminus 9}$ is the amplitude of the scalar power spectrum, while $S(k_1, k_2, k_3)$ is the dimensionless ``shape'' function that only depends on the magnitudes of the momenta~$k_i = |\mb{k}_i|$. 

	If the heavy field $\sigma$ couples directly to $\zeta$, it will be spontaneously created from the vacuum, evolving freely and accumulating a phase at a rate set by its rest mass frequency, before eventually decaying and correlating the fluctuations of $\zeta$ at three distinct points. Diagrammatically, this process can be represented as 
	\begin{equation}
			\def\circSize{0.6}
			\begin{tikzpicture}[thick, baseline=-28pt]
				\coordinate (c1) at (-1.0, -1.75);
				\coordinate (c2) at (1.0, -1.75);
				\coordinate (c3) at (-1.75, 0);
				\coordinate (c4) at (-0.25, 0);
				\coordinate (c5) at (1.75, 0);

				\begin{scope}[shift={(0, -1.75)}]
					
					\draw (c1) -- (c2) node[midway, below] {$\sigma$};
				\end{scope}

				\draw[inflSty] (c1) -- (c3) node[midway, shift={(-0.3, 0)}] {$\zeta_{\mb{k}_1}$};
				\draw[inflSty] (c4) -- (c1) node[midway, shift={(0.4, 0)}] {$\zeta_{\mb{k}_3}$};
				\draw[inflSty] (c2) -- (c5) node[midway, shift={(0.4, 0)}] {$\zeta_{\mb{k}_2}$};
				\draw[line width=0.6mm, gray] (-2.75, 0) -- (2.75, 0);
				\fill[cornellRed, intSty] (c1) circle (0.07);
				\fill[cornellRed, intSty] (c2) circle (0.07);
				\fill[cornellRed, intSty] (c3) circle (0.07);
				\fill[cornellRed, intSty] (c4) circle (0.07);
				\fill[cornellRed, intSty] (c5) circle (0.07);

				\end{tikzpicture}\,, \label{eq:treeLevelSig}
		\end{equation}
		where the top-most gray line denotes the time at which inflation ends.
		In the squeezed limit, $k_1 \approx k_2 \gg k_3$, this free propagation imparts a characteristic oscillatory signature onto the bispectrum's shape of the form
		\begin{equation}\label{eq:cosmo_coll}
			S(k_1 \approx k_2 \gg k_3) \sim \mathcal{A} \left(\frac{k_3}{k_1}\right)^{\!\gamma}\!\!\es\es \sin \!\left[\omega \log\left(\frac{k_3}{k_1}\right) + \delta\right]\,,
		\end{equation}
		where the parameters $\mathcal{A}$, $\gamma$, $\omega$ and $\delta$ correspond to the amplitude, rate of decay, frequency, and phase of these oscillations and are all calculable for a given model. For the tree-level exchange of a heavy scalar $\sigma$ shown above, the decay rate $\gamma$ and frequency $\omega$ are determined entirely by $\sigma$'s free propagation in de Sitter space. The decay rate $\gamma = \frac{1}{2}$ is determined by how quickly~$\sigma$ fluctuations dilute due to Hubble expansion and is the same for all free scalar fields, while $\omega = \sqrt{m_\sigma^2/H^2 - 9/4}$ is its frequency at ``rest'' in de Sitter space.

		Interactions with a light scalar $\varphi$ can alter how $\sigma$ propagates over large distances~\cite{Lu:2021wxu}, generating a non-trivial self-energy and changing the exponent $\gamma$ from its free field value. Diagrammatically, such interactions change the process in (\ref{eq:treeLevelSig}) to
			\begin{equation}
				\def\circSize{0.6}
				\begin{tikzpicture}[thick, baseline=-28pt]
					\coordinate (c1) at (-1.5, -1.75);
					\coordinate (c2) at (1.5, -1.75);
					\coordinate (c3) at (-2.25, 0);
					\coordinate (c4) at (-0.75, 0);
					\coordinate (c5) at (2.25, 0);

					\begin{scope}[shift={(0, -1.75)}]
						
						\draw (c1) -- (-\circSize, 0) node[midway, below] {$\sigma$};
						\draw (c2) -- (\circSize, 0) node[midway, below] {$\sigma$};
						\begin{scope}
							\fill[white] (0, 0) circle (\circSize);
							\clip[draw] (0, 0) circle (\circSize);
							\foreach \x in {-0.575, -0.5, ..., 0.575} 
							{	
								\draw[rotate=45, thin] (-\circSize, \x) -- (\circSize, \x);
							}
						\end{scope}
						\fill[black, intSty] (-\circSize, 0) circle (0.07);
						\fill[black, intSty] (\circSize, 0) circle (0.07);
					\end{scope}

					\draw[inflSty] (c1) -- (c3) node[midway, shift={(-0.3, 0)}] {$\zeta_{\mb{k}_1}$};
					\draw[inflSty] (c4) -- (c1) node[midway, shift={(0.4, 0)}] {$\zeta_{\mb{k}_3}$};
					\draw[inflSty] (c2) -- (c5) node[midway, shift={(0.4, 0)}] {$\zeta_{\mb{k}_2}$};
					\draw[line width=0.6mm, gray] (-2.75, 0) -- (2.75, 0);
					\fill[cornellRed, intSty] (c1) circle (0.07);
					\fill[cornellRed, intSty] (c2) circle (0.07);
					\fill[cornellRed, intSty] (c3) circle (0.07);
					\fill[cornellRed, intSty] (c4) circle (0.07);
					\fill[cornellRed, intSty] (c5) circle (0.07);

				\end{tikzpicture}\,,
			\end{equation}
			where we use a hatched blob to denote the exact $\sigma$ propagator. Specifically, it has been shown that interactions with light weakly coupled noncompact scalars always cause $\sigma$ to decay \emph{faster} than any free field can~\cite{Marolf:2010zp,Lu:2021wxu,Chakraborty:2023qbp}, with $\gamma > \frac{1}{2}$. That is, the light field provides an additional channel into which $\sigma$ can decay and it can decay faster and faster the lighter $\varphi$ is, though the perturbative expansion is quickly invalidated as $m_\varphi \to 0$. This also prompts a small issue: if our universe is full of hundreds of light weakly coupled scalars, would the interactions with this horde not force~$\gamma \gg 1$ and cause the cosmological collider signal (\ref{eq:cosmo_coll}) to decay too quickly to be observed?

			We explain how to compute the corrections to $\sigma$'s long-distance behavior, and thus its observable cosmological collider signal, from light and massless compact scalar fields. By careful analytic continuation from Euclidean to Lorentzian signature, we find that there are \emph{qualitative} differences between the signals generated by compact and noncompact scalars, and so the field space's topology is a critical ingredient in determining the physics of light fields in de Sitter. 

			Specifically, we show that a light compact field $\varphi$'s effect on $\sigma$ depends sensitively on the ratio of the physical circumference of $\varphi$'s field space $2 \pi f$ to the Hubble scale~$H$, where~$f$ is often called the axion decay constant. In the noncompact limit, $2 \pi f \gg H$, we show that $\varphi$ generates corrections somewhat similar to those of a light noncompact scalar, allowing $\sigma$ to decay faster than any free field. However, unlike the noncompact scalar, these corrections are well-behaved and finite in the massless limit, as the compact field space regulates any large fluctuations.

			The largest qualitative differences between compact and noncompact fields arise when the circumference $2 \pi f$ is smaller than $H$. Since the Hubble scale during inflation could have been as high as $H \sim \mathcal{O}\big(10^{14} \, \lab{GeV}\big)$, this corresponds to a wide range of theoretically reasonable~\cite{Mehta:2021pwf} decay constants.  As we shrink the field space to be smaller than Hubble, $2 \pi f/H \to 0$, there is an interesting sort of resonant effect and we find that the correction to $\sigma$'s decay rate begins to oscillate in sign, enabling $\sigma$ to decay either \emph{faster} or \emph{slower} than any free field. Furthermore, we find that the correction to $\gamma$ is exponentially suppressed as $H/(2\pi f) \to \infty$, and so $\sigma$ can interact with a throng of light compact fields and still be relatively long-lived as long as those fields have small enough field spaces. Furthermore, this \emph{signal} from the compact scalar is largest when~$2 \pi f \sim H$. Interestingly, axions with decay constants $f \lesssim \mathcal{O}(10^{14} \, \lab{GeV})$ are generic in stringy examples~\cite{Mehta:2021pwf}, and so this regime is plausibly phenomenologically relevant.

	\noindent \textbf{Outline} In Section~\ref{sec:review}, we review the basic properties of free quantum field theory in de Sitter space, describing how to analytically continue any de Sitter-invariant two-point function from Euclidean to Lorentzian signature via the Watson-Sommerfeld transform and its inverse, the Lorentzian inversion formula. We then review how free massive noncompact fields behave in de Sitter space and describe their propagators. 

	In Section~\ref{sec:compactMink}, we review how the gauge symmetry $\varphi \sim \varphi + 2 \pi$ is implemented at the level of the path integral measure. We do so first in the simplest system where this arises: the particle on the circle. We explain how this gauge symmetry is implemented by summing over ``winding'' trajectories and restricting the zero mode integral to its fundamental domain, $\varphi_0 \in [0, 2 \pi)$. We then extend this to compact scalar fields in Minkowski space, and explain why the field space topology is so difficult to detect there.

	In Section~\ref{sec:compactDS}, we describe massless compact scalar fields in de Sitter space. While correlation functions like $\langle \varphi(x) \varphi(y)\rangle$ are ill-behaved and diverge at long distances, we show that gauge-invariant correlation functions of the vertex operators $\mathcal{V}(x) \propto \e^{i \varphi(x)}$ are both well-defined and well-behaved. In particular, we study the Bunch-Davies vertex propagator $\langle \mathcal{V}(x) \mathcal{V}^\dagger(y) \rangle$ in both position and momentum space, and show that is appropriately well-behaved at long distances.

	In Section~\ref{sec:compact_int}, we compute the correction to the long-distance behavior of a heavy field $\sigma$ in the presence of a light compact field $\varphi$, in a simple toy model that mimics an axion interacting with a saxion. The compactness of $\varphi$ requires that we perturb in the aforementioned vertex operators, and there we address many of the subtleties that arise.

	Finally, we present our conclusions in Section~\ref{sec:conclusions}. In Appendix~\ref{app:formulary} we provide definitions of and relations between the various special functions used throughout the main text. Finally, in Appendix~\ref{app:vertexTwo} we compute the corrected long-distance behavior of the vertex two-point function in the model of Section~\ref{sec:compact_int}.

\newpage
\section{de Sitter Quantum Field Theory} \label{sec:review}
	
	In this section, we review the basic aspects of free quantum field theory in de Sitter space that are relevant for the later sections. Specifically, in \S\ref{sec:dsPos} we describe the geometry of de Sitter space in both Euclidean and Lorentzian signatures. We then describe how to analytically continue de Sitter-invariant two-point functions from one signature to the other. In \S\ref{sec:dsMom}, we describe the momentum space representation of these two-point functions, reviewing the inversion formula that determines the representation appropriate for analytic continuation to Lorentzian signature. Finally, in \S\ref{sec:dsFree}, we review the basic properties of free massive scalar fields in de Sitter space.

		\subsection{de Sitter in Position Space} \label{sec:dsPos}

			de Sitter space is the unique maximally symmetric spacetime with constant positive curvature. In $D = d+1$ spacetime dimensions, its metric is given by~\cite{Spradlin:2001pw,Anninos:2012qw}
			\begin{equation}
				\ud s^2 = \ell^2 \left[\minus \ud t^2 + \cosh^2 t \, \ud \Omega_d\right]
			\end{equation}
			in so-called \emph{global coordinates}, where $\ud \Omega_d^2 = \ud \theta_1^2 + \sin^2 \theta_1 \, \ud \Omega_{d-1}^2$ is the standard round metric on the $d$-dimensional sphere, with $\theta_1, \ldots, \theta_{d-2}\in [0, \pi]$ and $\theta_{d-1} \in [0, 2 \pi)$, while the global time~$t \in \mathbb{R}$. The radius of curvature $\ell$ is related to the Hubble constant $H = \ell^{\sminus 1}$. Throughout, we will find it extremely convenient to work in Hubble units $H = 1$, such that all quantities are dimensionless and measured in units of Hubble. Dimensions can be restored in any expression by restoring appropriate powers of $\ell$ or $H$. Furthermore, we will work in arbitrary spacetime dimension~$D=2\alpha +1$, only taking $\alpha \to \frac{3}{2}$ at the very end of any calculation.

			We want to study interacting quantum field theories of compact scalars in Lorentzian de Sitter space. However, it will be helpful both conceptually and technically to \emph{define} these theories in Lorentzian signature via analytic continuation from Euclidean de Sitter space. Euclidean de Sitter space is also known as the sphere $\lab{S}^D$, whose metric is the standard round metric 
			\begin{equation}
				\ud s^2 = g_{\mu \nu} \ud x^\mu \, \ud x^\nu=  \ud \Omega_D^2 = \ud \tau^2 + \sin^2 \tau\, \ud \Omega_d^2\,, \label{eq:euclDSMet}
			\end{equation}
			with $\tau \in [0, \pi]$. We may parameterize an arbitrary point $x$ on this $(d+1)$-dimensional sphere via its $\tau$ coordinate and a unit vector $\mb{x}$ on the $d$-dimensional sub-sphere, $x = (\tau, \mb{x})$, where $|\mb{x}|^2=1$. 

			Perturbation theory in Euclidean signature converges to define an interacting $\lab{SO}(D+1)$-invariant state as long as the linearized field theory admits an $\lab{SO}(D+1)$-invariant propagator, and so interacting Euclidean correlators will be both $\lab{SO}(D+1)$-invariant (or covariant) and satisfy the Euclidean Schwinger-Dyson equations. We may then analytically continue these correlation functions to Lorentzian signature by taking $\tau \to i t + \frac{\pi}{2}$, combined with an appropriate~$i \epsilon$-prescription to select a particular operator ordering. These Lorentzian correlation functions will automatically be invariant under the de Sitter isometry group and satisfy the Lorentzian Schwinger-Dyson equations, thus defining a consistent de Sitter-invariant state. Technically, this way of defining the theory is useful because Euclidean de Sitter space has finite volume and thus any potential IR divergences are automatically regulated. Conceptually, it is useful because we will find that it is easiest to distinguish between the path integral measures of compact and noncompact scalars in Euclidean de Sitter space.

			We will study the two-point functions of local operators in de Sitter space. Any function of two points on the sphere that is invariant under its isometry group, and thus continues to a Lorentzian de Sitter-invariant two-point function, may be parameterized via the \emph{embedding distance},
			\begin{equation}
				\xi_{12} \equiv \xi(x_1, x_2) = \cos \tau_1 \, \cos \tau_2 + \sin \tau_1 \, \sin \tau_2\, (\mb{x}_1 \cdot \mb{x}_2)\,, \label{eq:edDef}
			\end{equation}
			where $\mb{x}_1 \cdot \mb{x}_2$ is the standard dot product in $\mathbb{R}^d$. We will often drop the subscript on $\xi_{12} \to \xi$ when there is no risk of ambiguity.
			Geometrically, $\xi$ can be understood as the cosine of the angle subtended by the great arc connecting the two points if we embed the $\lab{S}^D$ into $\mathbb{R}^{D+1}$. Clearly, in Euclidean signature it must lie in the interval $\xi \in [\minus 1, 1]$. Upon analytic continuation to Lorentzian signature, $\tau_i \to i t_i + \frac{\pi}{2}$, it may take values on the entire real line $\xi \in \mathbb{R}$. For instance, if the two points are connected via a spacelike geodesic, $\xi$ will still lie on the interval $\xi \in [\minus 1, 1]$, while $\xi > 1$ if they are connected by a timelike geodesic. If the two points are either coincident~$x_1 \to x_2$ or separated by a null geodesic, $\xi =1$. Finally, $\xi < \minus 1$ if the two points cannot be connected by any geodesic. This last case is particularly important for cosmological observations, as taking two points to future infinity with fixed spatial separation sends $\xi \to \minus \infty$.

			Many of our expressions, e.g. (\ref{eq:lorentzianInversion}), will be more naturally written in terms of the variable
			\begin{equation}
				\zeta = \frac{2}{\xi - 1}\,,\label{eq:imdDef}
			\end{equation}
			in which the limits $\xi \to \pm \infty$ correspond to $\zeta \to 0^\pm$. Because it is related to the long-time or long-distance limit, we will refer to the region $|\zeta| \in [0, 1)$ as the ``infrared,'' while we will call the region $|\zeta| \in [1, \infty)$ the ``ultraviolet.''

		\subsection{de Sitter in Momentum Space} \label{sec:dsMom}

			A very useful feature of Euclidean de Sitter space is that there exists a useful momentum space representation tied to the de Sitter isometry group. We may decompose any scalar field into the hyperspherical harmonics $Y_{\mb{J}}(x)$,
			\begin{equation}
				\sigma(x) = \sum_{\mb{J}} \sigma_{\mb{J}} Y_{\mb{J}}(x)\,.
			\end{equation}
			These harmonics are the $D$-dimensional analogs of the familiar spherical harmonics, which provide a complete and orthogonal basis of scalar functions on the sphere $\lab{S}^D$,
			\begin{equation}
				\int_{\lab{S}^D} \!\ud \Omega_D \, Y_{\mb{J}}(x) \es \SHc{\mb{K}}(x) = \delta_{\mb{J}\mb{K}}\quad \text{and} \quad \sum_{\mb{J}} Y_{\mb{J}}(x) \SHc{\mb{J}}(y) = \delta^{(D)}(x - y)/\sqrt{g}\,, \label{eq:shOrthog}
			\end{equation}
			so that $\sigma_{\mb{J}} = \int\!\ud \Omega_D\, \sigma(x) \SHc{\mb{J}}(x)$. These harmonics are labeled by a vector $\mb{J} = (J, m_1, \ldots, m_d)$ with $J \in \mathbb{N}$ a non-negative integer which we will call the \emph{total angular momentum} and $\mb{m} = (m_1, m_2, \ldots, m_d) \in \mathbb{Z}^d$ are a set of integers such that $J \geq m_1 \geq m_2 \geq \cdots \geq |m_d|$ which call the \emph{magnetic quantum numbers}. Most importantly, these harmonics are eigenfunctions of the Laplacian on $\lab{S}^D$,
			\begin{equation}
				\nabla^2 Y_{\mb{J}}(x) = -J(J + 2\alpha) Y_{\mb{J}}(x)\,, \label{eq:shCasimir}
			\end{equation}
			whose eigenvalue only depends on total angular momentum $J$. We will not need explicit forms for these harmonics, though a special role will be played by the \emph{zero mode} with $\mb{J} = 0$,
			\begin{equation}
				Y_0(x) = \sqrt{\frac{\Gamma(\alpha + 1)}{2 \pi^{\alpha+1}}} = \frac{1}{\sqrt{\lab{vol}\, \lab{S}^D}}\,,
			\end{equation} 
			which is just the constant profile on the sphere.

			Any scalar function $H(x, y)$ of two points on the sphere also admits a decomposition in terms of these hyperspherical harmonics, which greatly simplifies whenever $H(x, y) = H(\xi)$ is invariant under the de Sitter isometry group. In this case, the \emph{momentum space representation} of $H(x, y)$, which we denote $[H]_{\mb{J}}$, only depends on the total angular momentum $[H]_\mb{J} = [H]_J$. Using
			\begin{equation}
				\sum_{\mb{m}} Y_{J \mb{m}}(x) \SHc{J\mb{m}}(y) = \frac{\Gamma(\alpha)}{2 \pi^{\alpha + 1}} (J+\alpha) C_J^{\alpha}(\xi)\,,
			\end{equation} 
			we see that any de Sitter-invariant two-point function can be expanded in terms of the Gegenbauer polynomials~$C_{J}^{\alpha}(\xi)$,
			\begin{equation}
				H(\xi) = \sum_{\mb{J}} [H]_J Y_{\mb{J}}(x) \SHc{\mb{J}}(y) = \frac{\Gamma(\alpha)}{2 \pi^{\alpha+1}} \sum_{J = 0}^{\infty} (J+\alpha)[H]_J C_{J}^{\alpha}(\xi)\,, \label{eq:hJDef}
			\end{equation}
			which are a complete set of orthogonal polynomials defined by (\ref{eq:gegC}) on the interval $\xi \in [\minus 1, 1]$.

			Given a function $H(\xi)$, we may extract its momentum space representation $[H]_J$ for nonnegative integer $J$ via the Euclidean inversion formula,
			\begin{equation}
				[H]_J =  \frac{(4 \pi)^\alpha \Gamma(\alpha) \Gamma(J+1)}{\Gamma(J + 2 \alpha)} \int_{\sminus 1}^{1}\!\ud \emd \, \big(1 - \emd^2\big)^{\alpha - \frac{1}{2}} C_{J}^{\alpha}(\emd) H(\emd)\,. \label{eq:euclideanInv}
			\end{equation}
			Strictly speaking, the expansion (\ref{eq:hJDef}) only converges in $\xi \in [\minus 1, 1]$ when the inversion formula (\ref{eq:euclideanInv}) does, and vice versa. The Gegenbauer polynomials can faithfully represent any smooth function on the interval as long as it is not too singular at the endpoints. Likewise, since~$C_{J}^{\alpha}(\minus 1) \sim J^{2 \alpha - 1}$ as $J \to \infty$, cf. (\ref{eq:gegEndpoints}),  the expansion (\ref{eq:hJDef}) only converges if $[H]_J$ decays faster than~$J^{\sminus(2 \alpha + 1)}$ as $J \to \infty$. This will not be true for the functions we work with when $\alpha = \frac{3}{2}$.  We will instead keep $\alpha = \frac{1}{2}(3 - \epsilon)$ arbitrary throughout and use dimensional regularization to remove any $\epsilon$-divergences, \emph{defining} $H(\xi)$ and $[H]_J$ via analytic continuation in~$\alpha$. 

			Unfortunately, the expansion (\ref{eq:hJDef}) is rather useless if we want to evaluate the function $H(\xi)$ at Lorentzian separations $|\xi| \gg 1$, as Gegenbauer polynomial expansions are only guaranteed to converge along the interval $\xi \in [\minus 1, 1]$. We instead define the analytic continuation of (\ref{eq:hJDef}) to arbitrary $\xi$ via the Watson-Sommerfeld transform~\cite{Newton:2002stw,Hogervorst:2021uvp,Correia:2020xtr,DiPietro:2021sjt,Chakraborty:2023qbp}
			\begin{equation}
				H(\xi) = \frac{\Gamma(\alpha)}{2 \pi^{\alpha}} \! \int_{\gamma} \frac{\ud J}{2 \pi i} \frac{(J+\alpha) [H]_J}{\sin \pi J} C_{J}^{\alpha}(\minus \xi)\,, \label{eq:watsonSom}
			\end{equation}
			which expresses the sum as an integral over a contour $\gamma$ which lies slightly to the left and runs parallel to the $\lab{Im}\, J$-axis. Here, $[H]_J$ is a function on the complex $J$-plane which agrees with the coefficients appearing in (\ref{eq:hJDef}) at the nonnegative integers. More precisely, the $[H]_J$ appearing in (\ref{eq:watsonSom}) is an \emph{interpolation} of the coefficients appearing in (\ref{eq:hJDef}), though we will not distinguish between the two notationally.  Crucially, for (\ref{eq:watsonSom}) to both converge for arbitrary $\xi$ and to agree with the original sum (\ref{eq:hJDef}) on the interval $\xi \in [\minus 1, 1]$, the interpolation $[H]_J$ must be well-behaved and decay as $|J| \to \infty$ in the right-half $J$-plane.

			Given a function $H(\xi)$ defined for all $\xi \in \mathbb{C}$, how then do we invert (\ref{eq:watsonSom}) and find its well-behaved momentum space representation $[H]_J$? Since the Gegenbauer polynomials grow exponentially away from the $\lab{Re}\, J$-axis, $C_{J}^{\alpha}(\minus 1) \propto \exp(\pi |\es \lab{Im}\, J\es |)$, the representation provided by~(\ref{eq:euclideanInv}) does not work. Instead, we must rely on the Froissart-Gribov formula~\cite{Newton:2002stw,Hogervorst:2021uvp,Correia:2020xtr,DiPietro:2021sjt,Loparco:2023rug,Chakraborty:2023qbp}
			\begin{equation}
				[H]_J = \frac{(4 \pi)^\alpha \Gamma(\alpha) \Gamma(J+1)}{\Gamma(J+2\alpha)} \oint_{\mathcal{C}} \frac{\ud \emd}{2 \pi i}\, \big(\emd^2 - 1\big)^{\alpha - \frac{1}{2}} Q_J^\alpha (\emd) H(\emd)\,. \label{eq:lorentzianInversion}
			\end{equation}
			where $\mathcal{C}$ is a contour that wraps the interval $\xi \in [\minus 1, 1]$ counterclockwise, while the $Q_J^\alpha(\xi)$ are the Gegenbauer $Q$-functions defined in (\ref{eq:gegQ}). This provides the unique extension of the coefficients appearing in (\ref{eq:hJDef}) from nonnegative integers to all $J \in \mathbb{C}$ (save for potential singularities) that does not grow exponentially as $|J| \to \infty$.

			The two-point functions we work with will all have a discontinuity along $\xi \in [1, \infty)$, or analogously $\zeta \in [0, \infty)$, reflecting the fact that the ordering of time-like separated operators matters. We may thus deform the contour\footnote{Technically, (\ref{eq:lorentzianInversion}) only applies to functions which are analytic in a region surrounding the interval $\emd \in [\minus 1, 1]$. This is not true for the functions we consider which all have a discontinuity along $\emd \in [1,\infty)$. This is related to the fact that products of free propagators diverge more and more strongly in the coincident limit $\emd \to 1$. This singular behavior strongly depends on $\alpha$ and so, in keeping with our general strategy, we will work at small enough $\alpha$ so that (\ref{eq:lInvForm}) applies and then \emph{define} the $[H]_J$ at $\alpha =\frac{3}{2}$ by analytic continuation. Of course, $[H]_J$ may then diverge as $\alpha \to \frac{3}{2}$, but these are the UV divergences one typically encounters in any loop calculation and may be absorbed by local counterterms.} in (\ref{eq:lorentzianInversion}) to run along this discontinuity,
			\begin{equation}
				[H]_J = - \frac{2 \pi^{\alpha+1}\Gamma(J+1)}{4^J\es \Gamma(J+\alpha+1)} \int_{0}^{\infty}\!\frac{\ud \imd}{2 \pi i} \, \imd^{J-1}\,  \tFo{J+\alpha + \frac{1}{2}}{J+1}{2 J + 2 \alpha + 1}{\minus \imd} \, \lab{disc}\, H(\imd)\,, \label{eq:lInvForm}
			\end{equation}
			with the discontinuity defined as
			\begin{equation}
				\lab{disc} \, H(\imd) = \lim_{\epsilon \to 0^+} H(\imd + i \epsilon) - H(\imd - i \epsilon)\,.
			\end{equation}
			This \emph{Lorentzian inversion formula} (\ref{eq:lInvForm}) is the main tool we will use in this work---it defines the momentum space representation of $H(\emd)$ as long as $\lab{Re} \, J$ is large enough and $\alpha$ is small enough so that the integral converges at its endpoints $\imd \to 0$ and $\imd \to \infty$, respectively. Indeed, all of our loop calculations will eventually reduce to studying integrals of the form (\ref{eq:lInvForm}) with different choices of $H(\zeta)$ depending on which fields run in the loop.

	\subsection{Free Scalar Fields in de Sitter} \label{sec:dsFree}
		Finally, we now review the properties of free fields in de Sitter space. A free, massive, minimally-coupled scalar field $\sigma$ in Euclidean de Sitter space, with Euclidean action
		\begin{equation}
			S_\slab{e} = \int_{\lab{S}^D}\!\ud^D x\, \sqrt{g}\!\left[\tfrac{1}{2}(\partial \sigma)^2 + \tfrac{1}{2} m^2 \sigma^2\right],
		\end{equation}
		has a propagator $G(x, y)$ which obeys the Klein-Gordon equation
		\begin{equation}
			\big(\minus \nabla^2 + m^2\big)G(x, y) = \delta^{(D)}(x -y)/\sqrt{g}\,, \label{eq:kgEOM}
		\end{equation}
		where $\nabla^2 \equiv g^{\mu \nu} \nabla_\mu \nabla_\nu$ and $\sqrt{g}$ are the Laplacian and the square-root of the metric (\ref{eq:euclDSMet})'s determinant, respectively. Expanding in hyperspherical harmonics (\ref{eq:shCasimir}), this can be solved to find
		\begin{equation}
			G(x, y) = \sum_{\mb{J}} \frac{Y_{\mb{J}}(x) \SHc{\mb{J}}(y)}{J(J+2 \alpha) + m^2} = \frac{\Gamma(\alpha)}{2 \pi^{\alpha + 1}} \sum_{J = 0}^{\infty} \frac{J + \alpha}{J(J+2 \alpha) + m^2} C_{J}^{\alpha}(\emd)\,, \label{eq:propSH}
		\end{equation}
		which converges absolutely as long as $\xi$ lies in the Euclidean interval $\xi \in [\minus 1, 1]$.
		
		The propagator in momentum space is thus
		\begin{equation}
			[G]_J = \frac{1}{J(J+2\alpha)+m^2} = \frac{1}{(J + \Delta)(J+\bar{\Delta})}\,, \label{eq:propMom}
		\end{equation}
		the poles of which are the so-called scaling dimension of the field,
		\begin{equation}
				\Delta = \begin{dcases}
							\alpha + i \sqrt{m^2 - \alpha^2}\,, & m \geq \alpha \\
							\alpha - \sqrt{\alpha^2 - m^2}\,, & m < \alpha
						\end{dcases}
		\end{equation}
		and its conjugate or \emph{shadow} dimension $\bar{\Delta} \equiv 2\alpha - \Delta$. A ``light'' scalar field, with $m < \alpha$, is said to belong to the \emph{complementary series} with $\Delta \in (0, \alpha)$. Likewise, a ``heavy'' scalar field, with~$m \geq \alpha$, is said to belong to the \emph{principal series} with dimension $\Delta = \alpha + i \nu$, $\nu \in \mathbb{R}$.\footnote{There are other representations, but these are the only two which will concern us.}

		Harmonic expansions like (\ref{eq:propSH}) only converge for spacelike separations $\xi \in [\minus 1, 1]$ but, as we discussed in the previous section, we can analytically continue it to arbitrary Lorentzian $\xi \in \mathbb{R}$ via the Watson-Sommerfeld transformation (\ref{eq:watsonSom}) which, when applied to (\ref{eq:propMom}), yields
		\begin{equation}
				G(\emd) = \frac{\Gamma(\Delta) \Gamma(\bar{\Delta})}{(4 \pi)^{\alpha + \frac{1}{2}}\Gamma\big(\alpha + \frac{1}{2}\big)}\,  {}_2 F_1\big(\Delta, \bar{\Delta}; {\alpha + \tfrac{1}{2}}; \tfrac{1}{2}(1 + \emd)\big)\,. \label{eq:freeProp}
		\end{equation}
		This free propagator is analytic for all $\emd$ aside from a branch cut along $\emd \in [1, \infty)$ or, in terms of our $\imd$ variable (\ref{eq:imdDef}), $\imd \in [0, \infty)$. It will also be helpful to use (\ref{eq:hypConnectionFormula}) to rewrite the propagator as
		\begin{equation}
			G(\imd) = \mathcal{G}_{\Delta}(\imd) + \mathcal{G}_{\bar{\Delta}}(\imd) = \mathcal{A}(\Delta) (\minus 1/\imd)^{\sminus \Delta}  \tFo{\Delta}{\Delta - \alpha + \frac{1}{2}}{2 \Delta - 2 \alpha + 1}{\minus \imd } + (\Delta \to \bar{\Delta})\,, \label{eq:propAsymp}
		\end{equation}
		where we define the coefficient
		\begin{equation}
			\mathcal{A}(\Delta) = \frac{1}{(4 \pi)^{\alpha + \frac{1}{2}}} \frac{\Gamma(\Delta) \Gamma(2 \alpha - 2 \Delta)}{\Gamma\big(\alpha + \frac{1}{2} - \Delta\big)}\,. \label{eq:propCoeff}
		\end{equation}
		From this expression, we can see that the dimension $\Delta$ and its shadow $\bar{\Delta}$, and thus the field's mass $m$, controls the asymptotic behavior of the free propagator for very large separations $\imd \to 0$ or as $\emd \to \pm \infty$. At long distances, heavy fields oscillate at a ``rest mass'' frequency set by~$\nu = \sqrt{m^2 - \alpha^2}$ and decay at a rate determined entirely by the expansion of spacetime,~$\lab{Re}\, \Delta = \alpha$, independent of their mass. 
		In contrast, light fields decay at a rate set by their mass and do not oscillate.

		Finally, this free field propagator diverges in the short distance or coincident limit $\xi \to 1$ in physical spatial dimensions $\alpha > \frac{1}{2}$. However, since it will be useful for our loop calculations we may formally define this coincident limit via analytic continuation from $\alpha < \frac{1}{2}$, in which
		\begin{equation}
			\begin{aligned}
				G(1) &= \frac{1}{(4 \pi)^{\alpha + \frac{1}{2}}} \, \G{\frac{1}{2} - \alpha\,,\, \Delta\,,\, 2\alpha - \Delta}{\frac{1}{2} +\alpha - \Delta\,,\, \frac{1}{2} -\alpha + \Delta}\,.
			\end{aligned}
		\end{equation}
		As $\epsilon \to 0$ with $\alpha = \frac{1}{2}(3 - \epsilon)$, this behaves as
		\begin{equation}
			G(1) \sim -\frac{(\Delta - 1)(\bar{\Delta}-1)}{16 \pi^2}\left[\frac{2}{\epsilon}- \psi\big(\Delta-1\big) - \psi\big(\bar{\Delta} - 1\big) + \log 4 \pi \e^{\sminus \gamma_\slab{e}} + 1\right] - \frac{1}{8 \pi^2}\,, \label{eq:coincidentMassive}
		\end{equation}
		keeping the mass of the field held fixed.\footnote{It may be more natural to keep, say, the asymptotic behavior of the field fixed $\Delta$ held fixed as we send $\alpha \to \frac{3}{2}$, or its frequency $\nu = \lab{Im}\, \Delta$. In either case, this choice just changes the last term of (\ref{eq:coincidentMassive}) and reflects a choice of renormalization scheme.}

		Having reviewed the basics of both de Sitter space and its free scalar fields, we will now review the basic formulation of compact scalar fields at the level of the path integral. A compact scalar field $\varphi$ enjoys a \emph{gauge symmetry} in which $\varphi$ and $\varphi + 2 \pi$ are \emph{identical}---they are the same point in field space. Appropriately summing over field configurations that respect this identification requires that we modify the path integral measure and it will be convenient to first understand how this is done in flat space before moving onto de Sitter in Section~\ref{sec:compactDS}.

\section{Compact Scalar Fields in Flat Space} \label{sec:compactMink}

	The goal of this section is to review how compact scalar fields differ quantum mechanically from noncompact scalar fields in Minkowski space. We will focus on how the path integral over field configurations must be modified to correctly account for the field space's compactness and why it is difficult, without observing topological defects like axion strings, to distinguish between noncompact and compact scalar fields in Minkowski space. It will be clarifying to first study the simplest compact scalar field theory: the compact boson in $(0+1)$ spacetime dimensions, also known as the quantum mechanical particle on a circle. We will then extend this discussion to the compact boson in $(d+1)$ spacetime dimensions, in preparation for our discussion in Section~\ref{sec:compactDS} on the massless compact boson in de Sitter space.

	\subsection{The Particle on the Circle} \label{sec:poc}

		We begin by reviewing the simplest compact scalar field theory: the quantum mechanical particle on a circle or the rotor. This theory is defined by the Lagrangian and Hamiltonian
		\begin{equation}
			\mathcal{L} = \frac{1}{2} f^2 \dot{\varphi}^2 \qquad \text{and} \qquad \mathcal{H} = \frac{p_\varphi^2}{2 f^2}\,, \label{eq:pocLagHam}
		\end{equation} 
		where we take $\varphi \sim \varphi + 2 \pi$ to be an angle with $2\pi$-periodicity whose canonical momentum is~$p_\varphi = f^2 \dot{\varphi}$. The \emph{decay constant} $f$ controls how damped fluctuations in $\varphi$ are. If this system describes a particle of mass $m$ on a circle with radius $R$, then $f^2 = m R^2$ is its moment of inertia. We have assumed that this particle has no potential energy, though this is only for calculational simplicity and none of our conclusions rely on this assumption.  It will be helpful to review this simple system both from the canonical and path integral pictures to help calibrate our ruleset for quantization.

		Canonically, the Hilbert space of (\ref{eq:pocLagHam}) is given by the space of square-integrable functions on the circle~$L^2(\lab{S}^1)$. It is convenient to introduce a ``position'' operator $\hat{\varphi}$ with a complete set of eigenstates $\hat{\varphi} |\varphi\rangle = \varphi | \varphi\rangle$ that resolves the identity via
		\begin{equation}
			\mathbbm{1} = \int_{0}^{2 \pi}\!\ud \varphi\, |\varphi\rangle \langle \varphi|\,,
		\end{equation}
		where we integrate over a single ``fundamental domain'' of the circle, $\varphi \in [0, 2 \pi)$.\footnote{It is a gauge choice to take $\varphi \in [0, 2\pi)$ and we could have instead worked with different covering of the circle, for example $\varphi \in [\minus \pi, \pi)$ or $\varphi \in [2023 \pi, 2025 \pi)$. Of course, no physical results will depend on this choice.} 
		Crucially,~$\hat{\varphi}$ is not a well-defined operator~\cite{Barnett:2007qpo} on the Hilbert space since it does not map $L^2(\lab{S}^1)$ into itself, generally taking a single-valued function on $\lab{S}^1$ into a multi-valued one. Said differently, $\hat{\varphi}$ is not a gauge-invariant operator. As consequence, any physical state $|\Psi \rangle$ must be single-valued in this~$|\varphi \rangle$-basis and thus must obey $\langle \varphi + 2 \pi | \Psi \rangle = \langle \varphi |\Psi \rangle$. 

\begin{figure}
			\centering
			\includegraphics{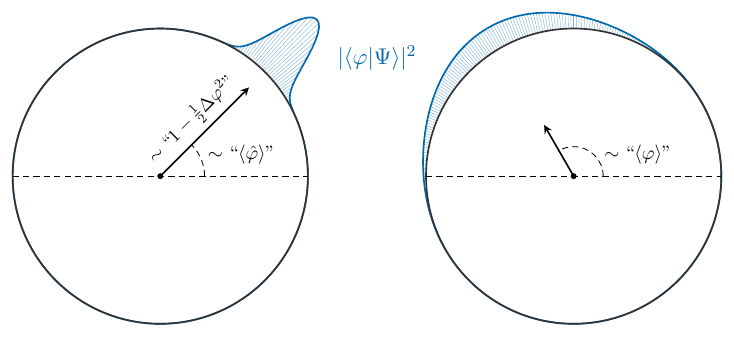}
			\caption{The expectation value of the vertex operator $\langle \e^{i \hat{\varphi}} \rangle$ measures both the particle's position and its spread. Roughly, the position of the particle on the circle is given by $\lab{arg}\, \langle \e^{i \hat{\varphi}}\rangle$ while the magnitude measures its variance, $|\langle \e^{i \hat{\varphi}} \rangle| \approx 1 - \frac{1}{2} \Delta \varphi^2 + \mathcal{O}\big(\Delta \varphi^4\big)$. As such, this magnitude is close to 1 when the wavefunction is well-localized [\textbf{left}], and close to 0 if it is not [\textbf{right}]. \label{fig:evs}}
		\end{figure}

		Similarly, expectation values like $\langle \Psi | \hat{\varphi} | \Psi\rangle$, which we would normally use to measure the position of the particle, are completely nonsensical and will lead to unphysical results. Instead, we must consider expectation values of the gauge-invariant ``vertex operators'' $\langle \Psi | \e^{i p \hat{\varphi}} | \Psi \rangle$, with~$p \in \mathbb{Z}$, to characterize the state $|\Psi\rangle$. As illustrated in Figure~\ref{fig:evs}, the phase and magnitude of $\langle \e^{i \hat{\varphi}} \rangle$ characterize where and how delocalized the particle is on the circle, respectively. If $\langle \e^{i \hat{\varphi}} \rangle = 0$, the particle is completely delocalized along the circle.

		The Hamiltonian (\ref{eq:pocLagHam}) has a discrete set of energy and momentum eigenstates,
		\begin{equation}
			\langle \varphi | n \rangle = \frac{1}{\sqrt{2 \pi}} \e^{i n \varphi}\,,
		\end{equation}
		with energy and momentum eigenvalues $E_n = n^2/(2 f^2)$ and $p_n = n$, respectively, for $n \in \mathbb{Z}$. These also form an orthonormal $\langle n | k \rangle = \delta_{nk}$ and complete basis $\mathbbm{1} = \sum_{n \in \mathbb{Z}} |n \rangle \langle n |$\,.
		Because it is simple to calculate using the Euclidean path integral, we will focus on the thermal trace
		\begin{equation}
			\mathcal{Z}(\beta) \equiv \lab{tr}\, \e^{\sminus \beta \mathcal{H}} = \sum_{n \in \mathbb{Z}} \e^{\sminus \beta E_n}\,. \label{eq:thermalTrace} 
		\end{equation}
		This can also be derived from the Euclidean propagator,
		\begin{equation}
			\langle \varphi | \e^{\sminus \beta \mathcal{H}} | \varphi'\rangle = \sum_{n \in \mathbb{Z}} \langle \varphi |n \rangle \langle n | \e^{\sminus \beta \mathcal{H}} | n \rangle \langle n | \varphi'\rangle = \frac{1}{2 \pi} \sum_{n \in \mathbb{Z}} \e^{\sminus \beta E_n + i n (\varphi - \varphi')} 
		\end{equation}
		in which case
		\begin{equation}
			\mathcal{Z}(\beta) = \int_{0}^{2\pi}\!\ud \varphi\, \langle \varphi | \e^{\sminus \beta \mathcal{H}} | \varphi \rangle = \sum_{n \in \mathbb{Z}} \e^{\sminus \beta E_n}\,. \label{eq:thermalTraceProp}
		\end{equation}
		We emphasize that, for (\ref{eq:thermalTrace}) and (\ref{eq:thermalTraceProp}) to agree,  the trace in ``position space'' (\ref{eq:thermalTraceProp}) must only integrate over a single fundamental domain of the circle,~$\varphi \in [0, 2\pi)$.

		As is well known, we can also calculate (\ref{eq:thermalTrace}) via the Euclidean path integral,
		\begin{equation}
			\mathcal{Z}(\beta) = \int\!\mathcal{D} \varphi_\lab{c} (\tau)\, \exp\!\left(-\int_{0}^{\beta}\!\ud \tau\, \tfrac{1}{2} f^2 \dot{\varphi}^2\right) \label{eq:thermalTracePI}
		\end{equation}
		on the thermal circle, $\tau \sim \tau + \beta$, where we impose periodic boundary conditions $\varphi(\tau) \sim \varphi(\tau + \beta)$. Here, we use $\mathcal{D} \varphi_\lab{c}$ to denote that we integrate over periodic paths $\varphi(\tau)$ from the Euclidean time circle $\tau \sim \tau + \beta$ to the \emph{field space circle $\varphi \sim \varphi + 2\pi$}. Since $\pi_1(\lab{S}^1) \in \mathbb{Z}$, these paths are characterized by a winding number $\ell$, which represents the number of times the path $\varphi(\tau)$ winds around field space as $\tau$ goes from $0$ to $\beta$. An arbitrary such path may be parameterized as
		\begin{equation}
			\varphi_\ell(\tau) = \varphi_0 + 2 \pi \ell \tau/\beta + \sum_{m = 1}^{\infty} \big[a_m \cos \omega_m \tau + b_m \sin \omega_m \tau\big]\,, \label{eq:pocPath}
		\end{equation}
		where we have introduced the Matsubara frequencies $\omega_m = 2\pi m/\beta$ and allowed $\varphi_\ell(\tau)$ to take values along the entire real line, $\varphi_\ell(\tau) \in \mathbb{R}$. We have also introduced the two sets of Fourier coefficients $a_m$, $b_m \in \mathbb{R}$, and the average value of the path along the fundamental domain,
		\begin{equation}
			\varphi_0 = \frac{1}{\beta} \int_{0}^{\beta}\!\ud\tau\,\left[\varphi_\ell(\tau) - 2 \pi \ell \tau/\beta\right]\,,
		\end{equation}
		which must take values in $\varphi_0 \in [0, 2\pi)$ in order to avoid overcounting paths. This is a common trope: we write a path integral over a compact degree of freedom $\varphi_\lab{c}(\tau)$ in terms of one over a noncompact degree of freedom $\varphi(\tau)$ by including a sum over topological sectors or windings, here labeled by the integer $\ell$. 

		Since the path integral (\ref{eq:thermalTracePI}) is Gaussian, we may evaluate it directly. Specifically, in terms of the path (\ref{eq:pocPath}), we have
		\begin{equation}
			\mathcal{Z}(\beta) = \sum_{\ell \in \mathbb{Z}} \int_{0}^{2\pi}\!\frac{\ud \varphi_0}{C_0} \left[\prod_{m = 1}^{\infty} \int_{\sminus \infty}^{\infty}\! \frac{\ud a_m \, \ud b_m}{C_m}\right] \exp\!\left(-\frac{(2 \pi \ell f)^2}{2 \beta} - \sum_{m = 1}^{\infty}\frac{1}{4} f^2\omega_m^2 (a_m^2 + b_m^2)\right)
		\end{equation}
		with the standard constants $C_m = 4 \pi/(\omega_m^2 f^2)$ and $C_0 = \sqrt{2 \pi \beta /f^2}$ that normalize the path integral measure~\cite{kleinert2009path}. Evaluating the Gaussian integrals, we arrive at
		\begin{equation}
			\mathcal{Z}(\beta) = \sqrt{\frac{2\pi f^2}{\beta}} \sum_{\ell \in \mathbb{Z}} \e^{\sminus \frac{1}{2} \ell^2 f^2/\beta}
 		\end{equation}
 		which equals the original expression (\ref{eq:thermalTrace}) upon Poisson resummation. As in (\ref{eq:thermalTraceProp}), that the integral over the \emph{zero mode} of the compact field is constrained to be in $\varphi_0 \in [0, 2\pi)$ is necessary to arrive at an answer that agrees with the canonical picture (\ref{eq:thermalTrace}). Furthermore, introducing a potential to (\ref{eq:pocLagHam}) will not change how the path integral is constructed, but will merely alter how the different paths are weighted.

 		To recap, we construct the thermal Euclidean path integral for a compact scalar field by integrating over all unique paths that come back to themselves. Since the field space is a circle,  these paths can wind $\ell$ times around it yet still arrive at the same point, and so we must also include a sum over windings or topological sectors in our definition of the path integral measure. Furthermore, the field space's compactness restricts the zero mode of these paths $\varphi_0$ to lie in the fundamental domain $\varphi_0 \in [0, 2\pi)$, lest we overcount. In Section~\ref{sec:compactDS}, we will find that this latter constraint fundamentally changes the behavior of compact scalars in de Sitter as compared to noncompact scalars. Before we move on, however, it will be useful to understand why this difference is unimportant for the compact scalar in $(d+1)$-dimensional Minkowski space.

	\newpage
	\subsection{Compact Scalar Field Theory}

		Let us now consider the higher-dimensional generalization of the particle on a circle, the compact scalar field in $(D = d+1)$-dimensional flat space, with Euclidean action
		\begin{equation}
				S_\slab{e} = \int\!\ud^{D} x\, \tfrac{1}{2} f^2 (\partial \varphi)^2 \label{eq:compMinkAction}
		\end{equation}
		and $\varphi(x) \sim \varphi(x) + 2\pi$. We will again study this theory, and its differences from the noncompact scalar with the same action, by computing the thermal trace. This would involve evaluating the Euclidean path integral on the cylinder $\lab{S}_\beta^1 \times \mathbb{R}^d$ with periodic Euclidean time, $x^0 = \tau \sim \tau + \beta$. However, to tame infrared divergences, it will be more convenient to also compactify the spatial directions $x^i \sim x^i + L$ and consider the theory on the $D$-dimensional torus $\lab{T}^D$ with spatial side lengths $L$. The goal of this section is to show that the dynamics that distinguish the compact scalar from the noncompact one freeze out in the limit of infinite volume, $L \to \infty$, and so it is impossible to distinguish between the two in Minkowski space from the dynamics of (\ref{eq:compMinkAction}) alone.\footnote{As mentioned in the introduction, this does not preclude distinguishing the two by observing finite energy topological defects like axion strings, whose existence relies on the compactness of the field space.}

		We are instructed to integrate over all single-valued field configurations on the torus $\lab{T}^D$. As in the previous section, the scalar field may shift by a multiple of $2 \pi$ as we wind around the thermal circle,
		\begin{equation}
			\varphi(\tau + \beta, \mb{x}) = \varphi(\tau, \mb{x}) + 2 \pi \ell\,,
		\end{equation}
		with $\ell \in \mathbb{Z}$.
		However, the field may also wind in the spatial directions,
		\begin{equation}
			\varphi(\tau, x^1, \ldots, x^k + L, \ldots, x^d) = \varphi(\tau, \mb{x}) + 2 \pi w_k\,,
		\end{equation}
		and so we must include $d$ additional winding numbers $w_i \in \mathbb{Z}$. So, any path will be characterized by a set of winding numbers $(\ell, \mb{w}) \in \mathbb{Z}^D$ which describe how many times the field $\varphi$ winds around its field space as we move around each of the $D$ nontrivial cycles in the $\lab{T}^D$. This is in contrast to the noncompact scalar field, with the same action (\ref{eq:compMinkAction}), which cannot wind and still respect the periodic boundary conditions.

		We may write the most general map from $\lab{T}^D \to \lab{S}^1$ as
		\begin{equation}
			\varphi(x) = \varphi_\ell(\tau) + 2 \pi w_i x^i/L + \tilde{\varphi}(x)\,, \label{eq:compactScalarFlat}
		\end{equation} 
		where we have split off the field's spatially isotropic \emph{zero mode}
		\begin{equation}
			\varphi_\ell(\tau) = \varphi_0 + 2 \pi \ell \tau/\beta + \sum_{m \neq 0} \big[a_{m} \e^{i \omega_m \tau} + \lab{c.c.}\big] \label{eq:compactZeroModeMink}
		\end{equation}
		and its non-zero mode
		\begin{equation}
			\tilde{\varphi}(x) = \frac{1}{L^{d/2}}\sum_{\omega_m, k_i \neq 0} \big[a_{\omega_m, k_i} \e^{i \omega_m \tau + i k_i x^i} + \lab{c.c.}\big]\,,
		\end{equation}
		where the sum is over all Matsubara frequencies $\omega_m = 2 \pi m/\beta$ and non-zero spatial momenta $k_i \in 2 \pi \mathbb{Z}/L$, and the Fourier coefficients are arbitrary complex numbers, $a_m \in \mathbb{C}$ and $a_{\omega_m, k_i} \in \mathbb{C}$. The zero mode (\ref{eq:compactZeroModeMink}) is equivalent to the particle on the circle (\ref{eq:pocPath}), and describes how the spatially constant mode on the torus $\lab{T}^d$ evolves in time $\tau$. As before, the average value of this mode $\varphi_0$ is constrained to $\varphi_0 \in [0, 2\pi)$, though our discussion here will not center on this.

		The Euclidean action (\ref{eq:compMinkAction}) evaluated on the configurations (\ref{eq:compactScalarFlat}) reduces to
		\begin{equation}
			S_\slab{e}[\varphi_0, \tilde{\varphi}] = \tfrac{1}{2} L^d  f^2  \int_0^\beta\!\ud \tau\,  \dot{\varphi}_\ell^2 + 4 \pi^2 \beta L^{d-2} \sum_{i =1}^d w^2_i + \int\!\ud^D x\, \tfrac{1}{2} f^2 (\partial \tilde{\varphi})^2\,. 
		\end{equation}
		In the limit $L \to \infty$, we see that for $d \geq 3$, the different winding sectors $w_i \neq 0$ acquire infinite action, and so the sum over windings localizes to the $\mb{w} = 0$ sector. Furthermore, fluctuations in the zero mode---including the $\ell \neq 0$ winding trajectories---completely freeze out. That is, in the limit $L \to \infty$, it is completely consistent to consider states in which the field's zero mode is localized at a \emph{definite} position in its field space. In the limit $L \to \infty$, the theory's dynamics will not delocalize the zero mode, and we can consider it to be a non-dynamical parameter of the theory. Indeed, this is often what we do for scalar fields or moduli in Minkowski space, where we quantize the field about a specific vacuum expectation value $\langle \varphi \rangle = \varphi_0$, integrating over fluctuations $\tilde{\varphi}$ which decay at spatial infinity $|\mb{x}| \to \infty$. The infinite weight of space ensures that states (or theories) with different expectation values $\langle \varphi\rangle$ are in superselection sectors.

		We conclude that, since the only modes that ``see'' the compactness of the scalar field space are completely frozen out in the infinite volume limit, we cannot distinguish between a noncompact and compact scalar field in Minkowski space in $D = 4$, without relying on physics beyond what is described by (\ref{eq:compMinkAction}). Said differently,  it is impossible to detect the topology of the field space in Minkowski space without observing finite energy objects that are charged under $\varphi \to \varphi + 2 \pi$. In $D = 3$ and $D =4$, these objects are vortices and axion strings, respectively. As we discuss in the next section, however, de Sitter space is different---compact and noncompact scalar fields can have completely different predictions, even if they are described by the same action.

\section{Massless Free Compact Scalars in de Sitter} \label{sec:compactDS}
	
	The sharpest difference between compact and noncompact scalars in de Sitter space is for massless fields. Said simply, the massless compact scalar exists while its noncompact counterpart does not. In this section, we will discuss the free massless compact scalar field in arbitrary $D$-dimensional de Sitter space. We first describe its Euclidean path integral and physical correlation functions. We then discuss their analytic continuation to Lorentzian signature in both position and momentum space. We will use these results in Section~\ref{sec:compact_int}, where we describe how a heavy field $\sigma$ propagates in the presence of this massless compact scalar and its effect on $\sigma$'s cosmological collider signal.

	\subsection{The Euclidean Path Integral}

	The theory of a massless compact scalar in de Sitter is defined by the Euclidean path integral
	\begin{equation}
		\mathcal{Z} = \int\!\mathcal{D} \varphi_\lab{c}\,  \exp\left(- \int_{\lab{S}^D}\!\!\ud^D x \, \sqrt{g}\,  \tfrac{1}{2} f^2 (\partial \varphi)^2 \right), \label{eq:compactDSPI}
	\end{equation}
	where $\varphi(x) \sim \varphi(x) + 2 \pi$. Here, the decay constant $f$ is measured in units of Hubble and is thus dimensionless.  As in Section~\ref{sec:compactMink}, we use $\mathcal{D}\varphi_c$ to denote the integration over compact field configurations. It will be convenient to expand $\varphi(x)$ into its hyperspherical harmonics,
	\begin{equation}
		\varphi(x)  = \varphi_0 + \tilde{\varphi}(x) =  \varphi_0 + \sum_{\mb{J} \neq 0} \varphi_{\mb{J}} Y_{\mb{J}}(x)\,, \label{eq:phiDecomp}
	\end{equation}
	explicitly separating out the Euclidean zero mode $\varphi_0$ and denoting the non-zero modes as~$\tilde{\varphi}(x)$. 

	Recall that to evaluate the path integral over compact field configurations, we could instead integrate over noncompact field configurations at the cost of including a sum over winding configurations and restricting the zero mode integral in $\mathcal{D}\varphi_\lab{c}$ to the field space's fundamental domain. Unlike the particle on the circle, here $\pi_D(\lab{S}^1) = 0$ is trivial for $D \geq 2$ and so there are no smooth topologically non-trivial maps from Euclidean de Sitter space $\lab{S}^D$ into $\varphi$'s field space $\lab{S}^1$. Any smooth field configuration $\varphi(x): \lab{S}^D \to \lab{S}^1$ may be generated by an appropriate choice of the coefficients $\varphi_{\mb{J}}$. Thus, the only effect the field space's compactness has on the path integral~(\ref{eq:compactDSPI}) is to restrict the integral over the zero mode $\varphi_0$ to the fundamental domain of the circle~\cite{Law:2020cpj},~$\varphi_0 \in [0, 2\pi)$, while the non-zero mode fluctuations $\tilde{\varphi}(x)$ are treated identically to those of the noncompact scalar field described in Section~\ref{sec:review}.\footnote{This is not to say that the sum over windings is trivial for all questions in de Sitter space. For example, to compute the path integral on the hemisphere to find the Bunch-Davies state, we need to include the sum over winding configurations. For equilibrium questions computed from (\ref{eq:compactDSPI}), however, this sum trivializes.}

	We work in Euclidean de Sitter space because there it is obvious how to define the path integral measure for a compact scalar field. That is not the case in the flat slicing. In Poincar\'{e} coordinates, naively one expects that the scalar field's zero mode $\varphi_0$ sees an infinite spatial volume and freezes out in a way analogous to how it behaves in flat space. However, it is unclear whether it is more appropriate to regulate this spatial volume using the Hubble volume as suggested by Starobinsky's stochastic picture of inflation~\cite{Starobinsky:1986fx,Starobinsky:1994bd,Gorbenko:2019rza}. Indeed, a Hubble-regulated spatial slice is more physically meaningful from the perspective of an observer living inside de Sitter space, as super-horizon modes are not observable. It is thus unclear exactly which modes in the flat slicing should ``see'' the compactness of $\varphi$'s field space. We sidestep this confusing conceptual question by instead working in Euclidean de Sitter space, where the answer is obvious. 

	Since the Euclidean action in (\ref{eq:compactDSPI}) only depends on the non-zero modes,
	\begin{equation}
		S_\slab{e}[\tilde{\varphi}] = \int_{\lab{S}^D}\!\!\ud^D x \, \tfrac{1}{2} f^2 (\partial \varphi)^2 =  \tfrac{1}{2} f^2 \sum_{\mb{J} \neq 0} J(J+2\alpha) |\varphi_{\mb{J}}|^2\,, \label{eq:masslessEuclAction}
	\end{equation}
	the Euclidean zero mode is completely unconstrained absent a potential for $\varphi$. In this massless limit, the Bunch-Davies state distributes the Euclidean zero mode uniformly along $\varphi$'s field space~\cite{Rajaraman:2010xd}. As illustrated in Figure~\ref{fig:varianceComparison}, this is disastrous for the noncompact scalar field. Writing the propagator (\ref{eq:propSH}) in the limit $m_\varphi \to 0$ as
	\begin{equation}
		\langle \varphi(x) \varphi(y) \rangle = \langle \varphi_0^2 \rangle + \frac{\Gamma(\alpha)}{2 \pi^{\alpha+1}} \frac{1}{f^2} \sum_{J = 1}^{\infty} \frac{J+\alpha}{J(J+2\alpha)} C_J^\alpha(\xi)\,, \label{eq:propSHMassless}
	\end{equation}
	the variance of the zero mode $\langle \varphi_0^2 \rangle$ for a noncompact scalar field, 
	\begin{equation}
		\langle \varphi_0^2 \rangle = \frac{\Gamma(\alpha+1)}{2 \pi^{\alpha+1}} \frac{1}{m_\varphi^{2} f^2}\, \label{eq:noncompactVar}
	\end{equation}
	diverges in the massless limit $m_\varphi \to 0$, rendering the propagator and this theory ill-defined. For the compact scalar field, however, $\varphi_0$ is constrained to lie in $[0, 2\pi)$, and so the variance of the zero mode cannot diverge but instead approaches a constant
	\begin{equation}
		\langle \varphi_0^2 \rangle = \frac{4 \pi^2}{3}\,,
	\end{equation}
	so that (\ref{eq:propSHMassless}) is well-defined. Physically, this makes sense. Absent a potential, $\varphi$ will approach its equilibrium state by random walking and will eventually explore its entire field space. For the noncompact field, this leads to a diverging variance since the field can walk arbitrarily far away from $\varphi = 0$ without penalty. The compact field cannot walk arbitrarily far because there is no ``arbitrarily far.'' The equilibrium state simply fills in the circle, as illustrated in Figure~\ref{fig:varianceComparison}, and the variance is finite.

	There have been attempts to render the massless scalar field in de Sitter consistent by either explicitly projecting out this divergent zero mode~\cite{Kirsten:1993ug,Tolley:2001gg} or by restricting to shift-symmetric observables like $\langle \big(\varphi(x) - \varphi(y)\big)^2\rangle$~\cite{Kirsten:1993ug,Page:2012fn} so that any dependence on the ill-behaved zero mode cancels out. We see that the compact scalar field, already so ubiquitous in beyond the Standard Model physics, automatically regulates this divergent zero mode.

	We will specifically be interested in the non-zero mode two-point function~\cite{Kirsten:1993ug,Tolley:2001gg},
	\begin{equation}
		\tilde{G}(\xi) = \langle \tilde{\varphi}(x) \tilde{\varphi}(y) \rangle = 2 \beta \sum_{J = 1}^{\infty} \frac{J + \alpha}{J(J+2\alpha)} C_J^\alpha(\xi)\,, \label{eq:nzmGeg}
	\end{equation}
	where, for convenience, we have defined the constant
	\begin{equation}
		\beta \equiv \frac{\Gamma(\alpha)}{4 \pi^{\alpha +1}} \frac{1}{f^2}\,. \label{eq:betaDef}
	\end{equation}
	For example, when $\alpha = \frac{3}{2}$, this is just half the square of the ratio between the Hubble scale and the circumference of the field space, $\beta = \frac{1}{2} (2\pi f)^{\sminus 2}$.
	We may analytically continue (\ref{eq:nzmGeg}) to arbitrary $\xi$ by taking the massless limit of the normal propagator (\ref{eq:freeProp}) without the zero mode, 
	\begin{equation}
		 \beta^{\sminus 1} \tilde{G}(\xi) = \lim_{\Delta \to 0}  \left[\frac{\Gamma(\Delta)\Gamma(\bar{\Delta})}{\Gamma(2 \alpha)}\, \tFo{\Delta}{\bar{\Delta}}{\alpha + \tfrac{1}{2}}{\frac{1 + \xi}{2}} -  \frac{2 \alpha}{\Delta \bar{\Delta}}\right].
	\end{equation}
	This limit can be explicitly evaluated to yield
	\begin{equation}
		 \langle \tilde{\varphi}(x) \tilde{\varphi}(y) \rangle = \tilde{G}(\xi) =  \frac{2 \alpha \beta}{\alpha + \frac{1}{2}}\left(\frac{1 + \xi}{2}\right) \pFq{3}{2}{1,\, 1,\, 2 \alpha + 1}{2,\, \alpha + \frac{3}{2}}{\frac{1+\xi}{2}} - \beta H_{2 \alpha}\,, \label{eq:nonZeroProp}
	\end{equation}
	where $H_{n}$ is the $n$'th harmonic number, defined in Appendix~\ref{app:formulary}. In three spatial dimensions,~$\alpha =\frac{3}{2}$, this reduces to 
	\begin{equation}
			\tilde{G}(\xi) = \frac{1}{16 \pi^2 f^2} \left[\frac{2}{1 - \xi} + 2 \log \left(\frac{2}{1 - \xi}\right) - \frac{14}{3}\right].
	\end{equation}
	Unfortunately, we see that even though the zero mode has been regulated by the compact field space, the massless scalar propagator \emph{diverges} logarithmically in the infrared. This is true in arbitrary dimensions as well, since in the coordinate (\ref{eq:imdDef}) the non-zero mode two-point function behaves as
	\begin{equation}
		\beta^{\sminus 1} \tilde{G}(\zeta) \sim - \log \big(\minus1/\zeta\big) + \frac{1}{2\sqrt{\pi}} \Gamma\big(\alpha + \tfrac{1}{2}\big) \Gamma(\minus \alpha) (\minus 2/\zeta)^{\sminus 2 \alpha } - \beta^{\sminus 1} \tilde{G}_0\label{eq:nZMPropAsymp}
	\end{equation}
	as $\zeta \to 0$ or $|\xi| \to \infty$, where we have defined the constant 	
	\begin{equation}
		\beta^{\sminus 1} \tilde{G}_0 = H_{2 \alpha} + H_{2 \alpha - 1}- H_{(2\alpha - 1)/2}\, 
	\end{equation}
	for later convenience. 

	That this propagator does not decay in the infrared is extremely problematic if we want to study interacting theories with a massless compact scalar, since it is not clear whether such interacting theories are even perturbatively stable in the infrared unless internal propagators decay at large Lorentzian separations, $\xi \to \minus \infty$~\cite{Marolf:2010nz}. Furthermore, even if the interacting theory is perturbatively stable, this poor infrared behavior can wreak havoc on the loop expansion as generally there are an infinite class of diagrams---beyond just the usual one-particle irreducible ones---that must be resummed to calculate a physically sensible answer in Lorentzian signature.

	Fortunately, the two-point function $\langle \varphi(x) \varphi(y) \rangle$ is neither gauge-invariant nor physical, and so it is not a problem if it does not decay as $|\xi| \to \infty$. Local operators $\varphi(x)$ are not gauge-invariant and so the usual $n$-point functions like $\langle \varphi(x) \varphi(y)\rangle$ and $\langle \varphi(x) \varphi(y) \varphi(z)\rangle$ are nonsensical for the compact scalar. We should instead compute correlation functions of gauge-invariant operators, and in the next two subsections we will focus on the properties of the so-called vertex operators in both position and momentum space. 

	\subsection{Vertex Propagator in Position Space} \label{sec:vertexPropPos}

	We will study the correlation functions of the vertex operators
	\begin{equation}
		\mathcal{V}_p(x) \propto \e^{i p \varphi(x)}\,,
	\end{equation}
	in the Bunch-Davies state prepared by the Euclidean path integral (\ref{eq:compactDSPI}). These operators are clearly invariant under the gauge symmetry $\varphi \to \varphi + 2\pi$ as long as the charge $p$ is an integer. We will suppress the subscript for the charge $p=1$ vertex operator, i.e. $\mathcal{V}(x) \equiv \mathcal{V}_1(x)$, and use the conjugate to negate the charge, $\mathcal{V}^\dagger_{p}(x) \equiv \mathcal{V}_{\sminus p}(x)$. Of course, these are composite operators and so they must be defined through some sort of normal-ordering scheme, but we will find that---unlike the correlation functions of $\varphi(x)$---the $n$-point functions of these vertex operators \emph{do} decay as $|\xi|\to \infty$ and thus they can be used to define a well-behaved perturbative expansion of an interacting field theory in de Sitter.

	Let us first consider an arbitrary product of vertex operators which, up to an overall normalization, we may compute via
	\begin{equation}
		\big\langle \e^{i \sum_{i} p_i \varphi(x_i) }\big\rangle = \frac{1}{\mathcal{Z}} \int\!\mathcal{D} \varphi_\lab{c} \, \e^{\sminus S_\slab{e}[\tilde{\varphi}]}\, \exp\!\Big[\es \es i \varphi_0 \big( {\textstyle\sum_{i}} p_i\big) + i \big( {\textstyle\sum_{i}} p_i \tilde{\varphi}(x_i)\big)\Big]\, .
	\end{equation}
	Since the action (\ref{eq:masslessEuclAction}) does not depend on the zero mode, the integration over $\varphi_0$ forces the correlator to vanish unless the correlator is ``charge neutral,''
	\begin{equation}
		\big\langle \e^{i \sum_i p_i \varphi(x_i)} \big\rangle \propto \int_{0}^{2 \pi}\!\ud \varphi_0\,  \e^{i \varphi_0 \sum_i p_i} \neq 0 \quad \text{iff}\quad \sum_i p_i = 0\,. \label{eq:chargeNeutral}
	\end{equation}
	If we add a potential for $\varphi(x)$, this condition will be relaxed and correlation functions like $\langle e^{i p \varphi(x)} \rangle$ develop non-vanishing $x$-independent expectation values whose phases roughly correspond to the minimum of the potential, cf. Figure~\ref{fig:evs}.

	In the massless theory, this charge neutrality condition forces the zero mode to drop out of correlation functions so that they only depend on shift-symmetric combinations like $\varphi(x) - \varphi(y)$, similar to~\cite{Kirsten:1993ug,Page:2012fn}. Since the path integral is Gaussian, we can then evaluate an arbitrary vertex correlation function in terms of the non-zero mode propagator (\ref{eq:nonZeroProp}),
	\begin{equation}
		\big\langle \e^{i \sum_i p_i \varphi(x_i)} \big\rangle = \exp\!\left[-\tfrac{1}{2} {\textstyle\sum_{i, j}} p_i p_j \big\langle \tilde{\varphi}(x_i) \tilde{\varphi}(x_j)\big\rangle \right].
	\end{equation}
	For simplicity, let us consider the gauge-invariant analog of the two-point function,
	\begin{equation}
		\langle \mathcal{V}(x) \mathcal{V}^\dagger(y) \rangle = \big\langle \e^{i (\varphi(x) - \varphi(y))}\big\rangle = \exp\!\Big[\minus \langle \tilde{\varphi}(x)\rangle^2 + \langle \tilde{\varphi}(x) \tilde{\varphi}(y) \rangle\Big]\,.
	\end{equation}
	Immediately, we can see that this correlator is ill-defined because it depends on the quantity
	\begin{equation}
		\langle \tilde{\varphi}^2(x) \rangle = \tilde{G}(1) \equiv \beta \big( \pi \tan \pi \alpha - H_{2 \alpha}\big) \sim \frac{1}{8 \pi^2 f^2}\left[\frac{2}{\epsilon} - \frac{11}{6}\right]\,,
	\end{equation}
	which diverges as $\epsilon \to 0$ with $\alpha = \tfrac{1}{2}(3 - \epsilon)$ and $\beta$ held fixed.\footnote{A priori, it is unclear whether to keep $\beta$ or $f$ held fixed as $\alpha \to \frac{3}{2}$. However, since we will find that it is $\beta$ that characterizes the physical behavior of the field over long distances and not $f$, it is more natural to approach~$\alpha \to \frac{3}{2}$ holding $\beta$ fixed.}
	This is a problem with all vertex correlators and we need to supply some notion of normal ordering to properly define the operator and its correlation functions. Thus, we will \emph{define} the normal-ordered vertex operator to remove this infinite constant~\cite{Coleman:1974bu},
	\begin{equation}
		\mathcal{V}_p(x) \equiv \nord{\e^{i p \varphi(x)}} \equiv \exp\!\left(\tfrac{1}{2} p^2\tilde{G}(1) + \tfrac{1}{2} \beta p^2 \tilde{G}_0\right) \e^{i p \varphi(x)}\,, \label{eq:vertexDef}
	\end{equation}
	plus an additional constant to ensure that $\langle \mathcal{V}(x) \mathcal{V}^\dagger(y) \rangle$ is ``canonically normalized'' as~$|\xi| \to \infty$. Arbitrary $n$-point functions of these vertex operators are then well-defined and given by
	\begin{equation}
		\left\langle {\textstyle \prod_{i}}\,  \nord{\e^{i p_i \varphi(x_i)}} \right\rangle = \exp\!\left[\tfrac{1}{2} \tilde{G}_0\, {\textstyle \sum_{i}} p_i^2  -{\textstyle \sum_{i > j}} p_i p_j \langle \tilde{\varphi}(x_i) \tilde{\varphi}(x_j)\rangle\right], \label{eq:vertexNPoint}
	\end{equation}
	as long as the correlator is ``neutral,'' $\sum_{i} p_i = 0$, and vanishes otherwise. Specifically, the vertex two-point function is simply
	\begin{equation}
			\mathcal{G}(\xi) \equiv \langle \mathcal{V}(x) \mathcal{V}^\dagger(y)\rangle = \mathcal{A} 
		\es \exp\!\left[\frac{2 \alpha \beta}{\alpha + \frac{1}{2}} \left(\frac{1+\xi}{2}\right)\pFq{3}{2}{1,\, 1,\, 2 \alpha + 1}{2,\, \alpha + \frac{3}{2}}{\frac{1+\xi}{2}}  \right]\,. \label{eq:vertexProp}
	\end{equation}
	with
	\begin{equation}
		\mathcal{A} \equiv \exp\!\left(\es \beta H_{2 \alpha-1} - \beta H_{(2\alpha - 1)/{2}}\right).
	\end{equation}
	The other two-point functions $\langle \mathcal{V}^{\phantom{\dagger}}_p(x) \mathcal{V}^\dagger_p(y)\rangle$ can be recovered by simply taking $\beta \to p^2 \beta$. Our goal for the remainder of this section is to study the two-point function (\ref{eq:vertexProp}) in both position and momentum space.

	Unlike the non-zero mode two-point function (\ref{eq:nZMPropAsymp}), this vertex two-point function \emph{does} decay at large separations,
	\begin{equation}
			\begin{aligned}
			&\mathcal{G}(\xi) \sim \left(\minus \frac{2}{\xi}\right)^{\beta} \left[1 - \frac{1}{16} \frac{\beta}{\alpha - 1} \left(\minus \frac{2}{\xi}\right)^{2} - \mathcal{O}\big(\xi^{\sminus 4}\big) \right] \\
			&\qquad\qquad - \frac{\beta \csc \pi \alpha \, \Gamma\big(\alpha + \frac{1}{2}\big)^2}{2 \Gamma(2 \alpha + 1)} \left(\minus \frac{2}{\xi}\right)^{2 \alpha + \beta} \Big[1  + \mathcal{O}\big(\xi^{\sminus 2}\big)\Big]
			\end{aligned} \label{eq:vertexPropAsymp}
	\end{equation}
	as $|\xi| \to \infty$, with the dominant decay rates set by the two real exponents $\beta$ and $2 \alpha + \beta$. We should contrast this with a light scalar field, whose correlations also decay as $|\xi| \to \infty$, $G(\xi) \sim \mathcal{C}_{\vphantom{\bar{\Delta}}\Delta} (\minus \xi)^{\sminus \Delta} + \mathcal{C}_{\bar{\Delta}} (\minus \xi)^{\sminus (2 \alpha - \Delta)}$ with $\Delta \in (0, \alpha]$. Not only can the dominant decay of the compact scalar be faster than a noncompact scalar, $\beta > \alpha$, but the subleading asymptotic scaling is also different. Note that this is not in conflict with the representation theory of de Sitter space~\cite{Penedones:2023uqc}, as the vertex operator does not transform in an irreducible representation of the de Sitter isometry group---it is a composite operator.

	As $\beta \to 0$, we find that the vertex correlation function decays more slowly. Intuitively, this makes sense as this limit also corresponds to sending $f \to \infty$ and penalizing fluctuations in $\varphi$ more and more, forcing the compact field to take more time or distance to relax back to its vacuum. Likewise, the propagator decays more quickly in the ultracompact limit $\beta \to \infty$, wherein $f \to 0$ so that $\varphi$ can fluctuate very easily and relax to its vacuum almost instantaneously.

	Furthermore, it is clear from (\ref{eq:vertexNPoint}) that the vertex operators do not behave like Gaussian free fields, in the sense that the $n$-point functions do not Wick factorize. Using (\ref{eq:vertexNPoint}), we may construct any charge neutral vertex $n$-point function by forming the product of one vertex propagator $\mathcal{G}(\xi_{ij})$ for each pair of oppositely-charged operators and an inverse propagator $\mathcal{G}^{\sminus 1}(\xi_{ij})$ for each pair of like-charged operators.  
	For instance, the four-point function is
	\begin{equation}
		\langle \mathcal{V}(x_1) \mathcal{V}^\dagger(x_2) \mathcal{V}(x_3) \mathcal{V}^\dagger(x_4) \rangle = \frac{\mathcal{G}(\xi_{12}) \mathcal{G}(\xi_{1 4}) \mathcal{G}(\xi_{2 3}) \mathcal{G}(\xi_{3 4})}{\mathcal{G}(\xi_{1 3}) \mathcal{G}(\xi_{2 4})} \neq \mathcal{G}(\xi_{12}) \mathcal{G}(\xi_{34}) + \mathcal{G}(\xi_{14})\mathcal{G}(\xi_{23})\,, \label{eq:vertexFour}
	\end{equation}
	which clearly does not Wick factorize. However, if we take any two charge neutral pairs, say $(x_1, x_2)$ and $(x_3, x_4)$, far apart from one another while keeping their constituents close together, e.g. $|\xi_{13}|, |\xi_{14}|, \ldots \gg |\xi_{12}|, |\xi_{34}|$ so that $\xi_{13} \approx \xi_{14}$, then using (\ref{eq:vertexPropAsymp}) this correlator \emph{does} factorize into a simple product
	\begin{equation}
		\langle \mathcal{V}(x_1) \mathcal{V}^\dagger(x_2) \mathcal{V}(x_3) \mathcal{V}^\dagger(x_4) \rangle \to \mathcal{G}(\xi_{12}) \mathcal{G}(\xi_{34})\,, \label{eq:longDistanceFact}
	\end{equation}
	since each member of a charge neutral pair sees both the oppositely- and like-charged member of the other, producing a factor like $\mathcal{G}(\xi_{14})/\mathcal{G}(\xi_{13})$ that approaches $1$ as we send the pairs apart~$|\xi_{14}| \sim |\xi_{13}| \to \infty$. So, if we integrate over the $x_i$ in Lorentzian de Sitter space, there will be some regions of the integral in which the vertex operators do effectively Wick factorize. We will rely on this approximate factorization in the next section to identify higher-order contributions that dominate when an internal particle can go ``on-shell.''

	Finally, the vertex two-point function (\ref{eq:vertexProp}) is obviously regular at $\xi = \minus 1$ but is extremely singular at coincident points,
	\begin{equation}
		\mathcal{G}(\xi) \sim \mathcal{A}\exp\!\left(\frac{\beta \, 2^{\frac{1}{2} - \alpha} \sqrt{\pi}\,  \Gamma\big(\alpha - \frac{1}{2}\big)}{\Gamma(\alpha) (1 - \xi)^{\alpha - \frac{1}{2}}} +  \pi\beta \tan \pi \alpha\right)\,,\mathrlap{\qquad \xi \to 1\,,}
	\end{equation}
	as we should expect from a propagator in the Bunch-Davies state. The vertex propagator is only regular as $\xi \to 1$ when $\alpha < \frac{1}{2}$, while for $\alpha > \frac{1}{2}$ this is an essential singularity. In one spatial dimension, $\alpha = \frac{1}{2}$, it is either a pole or branch point depending on the value of $\beta$.  Many of the manipulations we do, especially when we work with the momentum space representation of this propagator, are technically only well-defined when $\mathcal{G}(\xi)$ is not too singular along the Euclidean interval $\xi \in [\minus 1, 1]$. However, this usually does not matter in practice (even the free field propagator is strictly ``too singular'' when $\alpha = \frac{3}{2}$) but we will play it safe by keeping $\alpha$ arbitrary throughout and analytically continuing from $\alpha < \frac{1}{2}$ to $\alpha = \frac{3}{2}$~\cite{Chakraborty:2023qbp}. The singular nature of these functions then usually results in simple poles in $\epsilon$, with $\alpha = \frac{1}{2}(3 - \epsilon)$, which must be subtracted via appropriate counterterms. However, somewhat magically, we will find that even though the vertex two-point function is extremely singular as $\xi \to 1$, in loops it is extremely well-behaved in the ultraviolet.\footnote{We note that a similar mechanism, in which UV divergences can be softened by including all of the interactions from a periodic potential, has appeared in~\cite{Hook:2023pba}.}

	\subsection{Vertex Propagator in Momentum Space} \label{sec:vertexPropMom}

	It will also be useful, especially for our perturbative calculations, to compute the momentum space representation $[\mathcal{G}]_J$ of the vertex propagator. As discussed above, this is slightly awkward because expressions like
	\begin{equation}
		\mathcal{G}(\xi) \equiv \frac{\Gamma(\alpha)}{2\pi^{\alpha+1}}\sum_{J=0}^\infty (J+\alpha)[\mathcal{G}]_J \, C^\alpha_J(\xi)\,,
	\end{equation}
	and its inverse (\ref{eq:euclideanInv}) do not converge for $\mathcal{G}(\xi)$ in physical dimensions, $\alpha = \frac{3}{2}$, even for Euclidean separations $\xi \in [\minus 1, 1]$. We will ignore the Euclidean inversion formula and instead \emph{define}~$\mathcal{G}(\xi)$'s momentum space representation $[\mathcal{G}]_J$ for arbitrary $\alpha$ via the Lorentzian inversion formula~(\ref{eq:lInvForm}),
	\begin{equation}
		[\mathcal{G}]_{J} = - \frac{2 \pi^{\alpha+1}\Gamma(J+1)}{4^J\es \Gamma(J+\alpha+1)} \int_{0}^{\infty}\!\frac{\ud \imd}{2 \pi i} \, \imd^{J-1}\,  \tFo{J+\alpha + \frac{1}{2}}{J+1}{2 J + 2 \alpha + 1}{\minus \imd} \, \lab{disc}\, \mathcal{G}(\imd)\,. \label{eq:vertexLinv}
	\end{equation}
	We will then confirm that, when used with the Watson-Sommerfeld transformation (\ref{eq:watsonSom}), this momentum space representation indeed captures the correct asymptotic behavior~(\ref{eq:vertexPropAsymp}) of the position space propagator (\ref{eq:vertexProp}).

	To compute the vertex propagator's discontinuity along $\zeta \in [0, \infty)$, it will be extremely helpful to use the connection formula (\ref{eq:hypConnectionFormula}) to rewrite the propagator as
	\begin{equation}
		\begin{aligned}
			\beta^{\sminus 1} \log \mathcal{G}(\zeta) = \lim_{\Delta \to 0} \Bigg[ &\G{\Delta\,,\, \bar{\Delta} - \Delta\,,\, \alpha + \frac{1}{2}}{2 \alpha\,,\, \alpha + \frac{1}{2} - \Delta} \left(-\frac{1}{\zeta}\right)^{\sminus \Delta}\!\! \tFo{\Delta}{\Delta - \alpha + \frac{1}{2}}{\Delta - \bar{\Delta} + 1}{\minus \zeta} \\
			& \qquad + (\Delta \leftrightarrow \bar{\Delta})- \frac{2 \alpha}{\Delta \bar{\Delta}} + 2 H_{2 \alpha} - H_{(2 \alpha -1)/2}- \frac{1}{2 \alpha}\Bigg]\,.
		\end{aligned}
	\end{equation}
	On either side of the branch cut along $\zeta \in [0, \infty)$, we then find that $\mathcal{G}(\zeta)$ behaves as
	\begin{equation}
			\begin{aligned}
				\lim_{\epsilon \to 0^+} \beta^{\sminus 1} \log \mathcal{G}(\zeta \pm i \epsilon) &= \log \zeta -\frac{1}{2} \zeta\, \pFq{3}{2}{1,\, 1,\, \frac{3}{2} - \alpha}{2,\,2 - 2 \alpha}{\minus \zeta} \\
				&+ \frac{\e^{\mp 2 \pi i \alpha}}{2 \sqrt{\pi}} \big(\zeta/2\big)^{2 \alpha} \Gamma\big(\minus \alpha\big)\Gamma\big(\alpha + \tfrac{1}{2}\big) \tFo{2 \alpha}{\alpha+ \frac{1}{2}}{2 \alpha + 1}{\minus \zeta} \mp i \pi\,.
			\end{aligned} \label{eq:vertexAboveBelow}
	\end{equation}
	This expression will be especially helpful in the loop computations of Section~\ref{sec:compact_int}, but here we can use it to derive the (rather intimidating) discontinuity
	\begin{equation}
		\lab{disc}\, \mathcal{G}(\zeta) = \minus 2\es i \zeta^{\beta} e^{\sminus \beta \mathcal{F}(\zeta)} \sin \pi \beta \Phi(\zeta)\,, \label{eq:vertexPropDisc}
	\end{equation}
	where we have defined the functions 
	\begin{equation}
		\mathcal{F}(\zeta) = \frac{1}{2}\, \zeta \,\pFq{3}{2}{1,\, 1,\, \frac{3}{2} - \alpha}{2,\, 2 - 2 \alpha}{\minus \zeta} + \frac{\pi  \cot 2 \pi \alpha \, \Gamma\big(\alpha + \frac{1}{2}\big)\zeta^{2 \alpha}}{\Gamma\big(\frac{1}{2} - \alpha\big)\Gamma\big(2 \alpha + 1\big)}  \tFo{2 \alpha}{\alpha + \frac{1}{2}}{2 \alpha + 1}{\minus \zeta} \label{eq:fDef}
	\end{equation}
	and
	\begin{equation}
		\begin{aligned}
			\Phi(\zeta) &= 1 - \frac{ \Gamma\big(\alpha + \frac{1}{2}\big)\zeta^{2 \alpha}}{\Gamma\big(\frac{1}{2} - \alpha\big)\Gamma\big(2 \alpha + 1\big)}  \tFo{2 \alpha}{\alpha + \frac{1}{2}}{2 \alpha + 1}{\minus \zeta}.
		\end{aligned} \label{eq:phiDef}
	\end{equation}
	Even though this discontinuity is extremely complicated and we are not at present able to evaluate~(\ref{eq:vertexLinv}) exactly for arbitrary $\alpha$, we can still extract essential physics from it using the techniques presented in~\cite{Chakraborty:2023qbp} by appropriately expanding the integrand as a series in $\zeta$.

	In the deep ultraviolet, $\zeta \to \infty$, the discontinuity (\ref{eq:vertexPropDisc}) is shockingly well-behaved. There,
	\begin{equation}
		\mathcal{F}(\zeta) \sim \frac{\Gamma\big(\alpha + \frac{1}{2}\big) \Gamma\big(\alpha - \frac{1}{2}\big)  }{2 \sin \pi \alpha \, \Gamma(2 \alpha)}\,\zeta^{\alpha - \frac{1}{2}}\,,\mathrlap{\qquad \zeta \to \infty\,,}
	\end{equation}
	so that the discontinuity exponentially decays, specifically behaving as $\propto \e^{\sminus \beta \zeta/2}$ as $\alpha \to \frac{3}{2}$. 
	The ultraviolet part of the integral $\zeta \in [1, \infty)$ is thus completely regular and need not be regulated. We thus do not expect any $\epsilon$-poles as we analytically continue from $\alpha < \frac{1}{2}$ to $\alpha = \frac{3}{2}$. This is quite shocking when we consider the extremely poor behavior of $\mathcal{G}(\zeta)$ along any other direction in the complex plane. The point $\zeta \to \infty$ is an essential singularity for all $\alpha > \frac{1}{2}$, and by Picard's Great Theorem $\mathcal{G}(\zeta)$ takes on all possible complex values, with at most a single exception, infinitely many times in the punctured neighborhood around $\zeta = 0$~\cite{conway1978functions}. If we approach the point $\zeta \to \infty$ ($\xi \to 1$) from any direction other than along the positive real $\zeta$ line, then $\mathcal{G}(\zeta)$ and its discontinuity behave very poorly and are extraordinarily singular. Fortunately, as evidenced by the inversion formula (\ref{eq:vertexLinv}), physics only seems to care about the behavior of $\mathcal{G}(\zeta)$ along where it is most well-behaved. Somewhat amazingly, loop corrections involving this compact scalar will also be UV finite, aside from a multiplicative renormalization that results from the normal ordering (\ref{eq:vertexDef}). Of course, there may still be ultraviolet divergences from other sources.

	To confirm that this momentum space representation actually yields the propagator~(\ref{eq:vertexProp}) when used with the Watson-Sommerfeld transform (\ref{eq:watsonSom}), it will be helpful to rewrite it using the recurrence relation (\ref{eq:gegRecur}) as
	\begin{equation}
		\mathcal{G}(\xi) = \frac{\Gamma(\alpha+1)
			}{2 \pi^{\alpha}}  \! \int_{\gamma} \frac{\ud J}{2 \pi i} \, \big([\mathcal{G}]_{J} - [\mathcal{G}]_{J+2}\big) \csc \pi J \, C_{J}^{\alpha + 1}(\minus \xi)\,, \label{eq:watsonProp}
	\end{equation}
	where the contour $\gamma$ again runs parallel along, but slightly to the left of, the $\lab{Im}\, J$-axis. As described in~\cite{Marolf:2010zp}, since $C_{J}^{\alpha + 1}(\minus \xi) \sim \mathcal{C}_1 (2 \xi)^{J} + \mathcal{C}_2(2 \xi)^{\sminus(J + 2\alpha + 2)}$, we extract (\ref{eq:watsonProp})'s long-distance behavior by moving the contour $\gamma$ as far to the left as $\lab{Re}\, J = \minus(\alpha + 1)$, picking up any poles~$\{J_*\}$ that we encounter along the way while still suppressing the magnitude of the integral as $|\xi| \to \infty$,
	\begin{equation}
		\begin{aligned}
			\mathcal{G}(\xi) =& \minus \frac{\Gamma(\alpha + 1)}{2 \pi^{\alpha}}\sum_{\{J_*\}}\underset{{J = J_*}}{\lab{res}}\Big[\left([\mathcal{G}]_{J} - [\mathcal{G}]_{J+2}\right) \csc \pi J\,  C_{J}^{\alpha + 1}(\minus \xi)\Big]  \\
			&\quad + \frac{\Gamma(\alpha+1)
			}{2 \pi^{\alpha}}  \! \int_{\gamma'} \frac{\ud J}{2 \pi i} \, \big([\mathcal{G}]_{J} - [\mathcal{G}]_{J+2}\big) \csc \pi J \, C_{J}^{\alpha + 1}(\minus \xi)\,.
		\end{aligned} \label{eq:watsonPoles}
	\end{equation} 
	The poles closest to the $\lab{Im}\, J$-axis thus determine the asymptotic behavior of (\ref{eq:watsonProp}), and we can use their residues to confirm that the momentum space representation $[\mathcal{G}]_J$ encodes the same asymptotic behavior (\ref{eq:vertexPropAsymp}) as the position space representation (\ref{eq:vertexProp}).
	
	Following the techniques of~\cite{Chakraborty:2023qbp}, we can quickly extract the relevant singularities $\{J_*\}$ from the discontinuity (\ref{eq:vertexPropDisc}). Since the integrand in (\ref{eq:vertexLinv}) is extremely well-behaved as $\zeta \to \infty$, the only way a $J$-dependent singularity can develop in $[\mathcal{G}]_J$ is if the integrand diverges as $\zeta \to 0$. We can thus read off the singularity structure of $[\mathcal{G}]_J$ by studying the discontinuity of $\mathcal{G}(\zeta)$ as $\zeta \to 0$,
	\begin{equation}
			\begin{aligned}
				\lab{disc}\,\mathcal{G}(\zeta) &\sim -2 i \sin \pi \beta \, \zeta^{\beta} \left[1 - \tfrac{1}{2} \beta \zeta + \tfrac{1}{16} \beta\big(2(1 + \beta) + \tfrac{1}{1 - \alpha}\big) \zeta^2 + \cdots \right] \\
				&-\frac{i \beta}{{2^{2 \alpha} \sqrt{\pi}}} \sin (\pi \beta + 2 \pi \alpha)  \Gamma\big(\alpha + \tfrac{1}{2}\big) \Gamma(\minus \alpha) \zeta^{2 \alpha + \beta}\Big[ 1 - \tfrac{1}{2}(2 \alpha + \beta) \zeta + \cdots \Big] \\
				&-\frac{i \beta^2}{2^{4 \alpha + 2} \pi}\sin (\pi \beta + 4 \pi \alpha) \big[{\Gamma\big(\alpha + \tfrac{1}{2}\big) \Gamma(\minus \alpha)}\big]^2 \zeta^{4 \alpha + \beta} \Big[1 - \tfrac{1}{2}(4 \alpha + \beta) \zeta + \cdots \Big] \\
				&+ \mathcal{O}\big(\zeta^{\beta + 6 \alpha}\big) + \mathcal{O}\big(\zeta^{\beta + 8 \alpha}\big) + \cdots\,.
			\end{aligned} \label{eq:vertexDiscExp}
	\end{equation}
	As $\zeta \to 0$, the integrand of (\ref{eq:vertexLinv}) will have terms that go as $\zeta^{J + \beta -1}$, $\zeta^{J+ \beta}$, $\zeta^{J + \beta + 2 \alpha -1}$, and so on, which will generate poles in $[\mathcal{G}]_J$ upon integration at $J = \minus \beta$, $J = \minus(\beta + 1)$, $J = \minus(\beta + 2 \alpha)$, and so on, respectively. In total, $[\mathcal{G}]_J$ will potentially have poles at $J = \minus(\beta + 2\alpha n + k)$, for all positive integer $n$ and $k$, though many of their residues may (and will) vanish. As $\alpha \to \frac{3}{2}$ from below, the poles closest to the $\lab{Im}\, J$-axis are
	\begin{equation}
		\begin{aligned}
			[\mathcal{G}]_{J} &\sim \frac{\pi ^{\alpha }  \Gamma(1-\beta ) \sin \pi  \beta  }{2^{\sminus(2 \beta +1)}\Gamma (\alpha -\beta +1)}\frac{1}{J+\beta} + \frac{\pi ^{\alpha } \alpha   \beta ^2  \Gamma (-\beta -1) \sin \pi \beta }{2^{\sminus(2 \beta +1)}(\alpha -1) \Gamma (\alpha -\beta )}\frac{1}{J+ \beta +2} + \cdots \\
			& + \frac{\beta  \,  \sin (\pi \beta + 2 \pi \alpha) }{4^{\sminus(\alpha +\beta)} \pi^{\frac{1}{2}-\alpha}} \G{\alpha + \frac{1}{2}\,,\,\minus \alpha\,,\, 1 - \beta - 2 \alpha}{1 - \beta - \alpha}\frac{1}{J + \beta + 2 \alpha} + \cdots\,,
		\end{aligned} \label{eq:vertexPoles}
	\end{equation}
	and so (\ref{eq:watsonPoles}) reduces to
	\begin{equation}
		\begin{aligned}
			\mathcal{G}(\xi) =&4^\beta \es\es\G{\alpha + 1\,,\, 1 - \beta}{\alpha - \beta +1} C_{\sminus \beta }^{\alpha +1}(\minus \xi )+\frac{ \beta }{2^{1-2 \beta}} \, \G{2 \alpha + 1\,,\, \minus \alpha\,,\, 1-2\alpha - \beta}{1 -\alpha - \beta}C_{\sminus (\beta + 2 \alpha) }^{\alpha +1}(\minus \xi ) \\ & -\frac{ \beta  (1-\alpha +\beta) (\alpha  \beta -1) }{4^{\sminus \beta }(\alpha -1)} \, \G{\alpha+1\,,\, -\beta - 1}{\alpha - \beta + 1} C_{\sminus (\beta +2)}^{\alpha +1}(\minus \xi ) + \cdots\,,
		\end{aligned}
	\end{equation}
	where the $\cdots$ denote terms subleading as $|\xi| \to \infty$. Note that this no longer depends on the factor~$\sin \pi \beta$, and so this does not vanish as $\beta \to \mathbb{Z}_{\scriptscriptstyle > 0}$ even if (\ref{eq:vertexDiscExp}) does.\footnote{It would be interesting to understand what happens at such special values of $\beta$. Even at the free-field level, they seem special since the momentum space representation (\ref{eq:vertexPropDisc}) vanishes there, $[\mathcal{G}]_J \propto \sin \pi \beta$ when $\alpha = \frac{3}{2}$. We can understand this by noting that, for $\beta \in \mathbb{Z}_{\scriptscriptstyle > 0}$ and $\alpha = \frac{3}{2}$, the time- and anti-time-ordered two-point functions of the vertex operator are equivalent. This may be a flaw in our description---the position space propagator (\ref{eq:vertexProp}) certainly does not vanish, and again points to the need to keep $\alpha$ arbitrary until the very end of our calculations. On the other hand, there are other discrete values of $\beta$ that seem to be special for the self-energy of the vertex operators, presented in Appendix~\ref{app:vertexTwo}. Furthermore, the analogous model in flat two-dimensional space---the Sine-Gordon model---has an interesting structure as one changes the equivalent of~$\beta$. So, perhaps the limit $\beta \to \mathbb{Z}_{\scriptscriptstyle > 0}$, where the decay constant $f$ and Hubble constant have commensurate values, is physically interesting. However, here we will simply assume that $\beta$ avoids these special values and instead leave their study for future work.} This exactly recovers the large-distance expansion (\ref{eq:vertexPropAsymp}) of the position space propagator, and so we can have confidence that the inversion formula (\ref{eq:vertexLinv}) with the discontinuity (\ref{eq:vertexPropDisc}) defines the appropriate momentum space representation of the vertex two-point function $\mathcal{G}(\xi)$.

	This concludes our discussion of the free massless compact scalar field in de Sitter space. In the next section, we consider a theory with a light compact scalar field interacting with a heavy scalar $\sigma$ and compute how this interaction changes the long-distance behavior of $\sigma$'s two-point function, to leading order in perturbation theory.

\section{Light Interacting Compact Scalars in de Sitter}\label{sec:compact_int}
	
	In the previous section, we described the dynamics of free massless compact scalar fields in de Sitter space. There, we found that the two-point function of vertex operators was both well-defined and well-behaved, decaying at large Lorentzian separations $|\xi| \to \infty$. In this section, we will study how a massless (or very light) compact scalar $\varphi$ affects the propagation of a heavy scalar $\sigma$ over long distances and thus modifies $\sigma$'s cosmological collider signal. Specifically, we will compute the self-energies of both $\varphi$ and $\sigma$ in the presence of a simple interaction and potential.

	We will consider the theory defined by the Euclidean action
	\begin{equation}
		S_\slab{e} = \int_{\lab{S}^D}\!\ud^D z\, \sqrt{g}\left[\tfrac{1}{2}f^2 (\partial \varphi)^2 + \tfrac{1}{2}(\partial \sigma)^2 + \tfrac{1}{2} m_\sigma^2 \sigma^2 + g_\varphi \cos \varphi + \tfrac{1}{2} g \sigma^2 \cos \varphi\right]. \label{eq:interactingAction}
	\end{equation}
	One can interpret this as, for instance, a toy model of a light axion $\varphi$ interacting with a heavy saxion $\sigma$. Because the propagator $\langle \varphi(x) \varphi(y) \rangle$ does not decay at large distances, it is not useful to expand (\ref{eq:interactingAction}) in powers of $\varphi(x)$. It is instead much more helpful to keep the vertex operator $\mathcal{V}(x) = \nord{\e^{i \varphi(x)}}$ as a fundamental object, whose two-point function (\ref{eq:vertexProp}) does decay at large distances, and expand in terms of it. We may write the action (\ref{eq:interactingAction}) as the sum of a free part,
	\begin{equation}
		S_0 = \int_{\lab{S}^D}\!\ud^D z\, \sqrt{g}\left[\tfrac{1}{2}f^2 (\partial \varphi)^2 + \tfrac{1}{2}(\partial \sigma)^2 + \tfrac{1}{2} m_\sigma^2 \sigma^2\right], \label{eq:freeAction}
	\end{equation}
	and an interaction part
	\begin{equation}
		S_\lab{int} = \int_{\lab{S}^D}\!\ud^D z\, \sqrt{g}\left[g_\varphi \mathcal{V}  + \tfrac{1}{2} g \sigma^2 \mathcal{V} + \lab{c.c.} \right], \label{eq:intAction}
	\end{equation}
	where we have redefined the couplings $g$ and $g_\varphi$ to absorb the (divergent) normal-ordering constant (\ref{eq:vertexDef}) relating $\e^{i \varphi(x)}$ and $\mathcal{V}(x)$, and a factor of two. We also include a counterterm action,
	\begin{equation}
		S_\lab{ct} = \int_{\lab{S}^D}\!\ud^D z\, \sqrt{g}\left[\tfrac{1}{2} \delta_{Z_\sigma} (\partial \sigma)^2 + \tfrac{1}{2} \delta_{m_\sigma} \sigma^2 + \big(\delta_{g_\varphi} \mathcal{V} + \lab{c.c.}\big) + \mathcal{O}_{\mathcal{V}}\right], \label{eq:ctAction}
	\end{equation}
	where we omit the $\delta_{g} \sigma^2 \cos \varphi$ counterterm because it will not contribute at leading order in perturbation theory and we will define the operator $\mathcal{O}_\mathcal{V}$ below.
	We will allow the couplings and their associated counterterms to be complex, accounting for the two possible phases in the sinusoids. That is, we may shift the phase of $g_\varphi \cos \varphi \to g_\varphi \cos (\varphi + \delta)$ by taking $g_\varphi \to g_\varphi \e^{\sminus i \delta}$. 

	As we discussed in \S\ref{sec:vertexPropPos}, the vertex operators $\mathcal{V}(x)$ do not behave as Gaussian free fields and this unfortunately complicates any diagrammatic expansion we might perform especially since there is no longer a complete cancellation of ``disconnected'' diagrams. We will instead explicitly expand out the path integral to compute the interacting one-point function $\langle \mathcal{V}(x) \rangle$ and two-point functions $\langle \sigma(x) \sigma(y) \rangle$ and $\langle \mathcal{V}(x) \mathcal{V}^\dagger(y) \rangle$. However, it will still be useful (even if just from a psychological perspective) to represent these corrections diagrammatically, though we will not specify rules for their combinatorics. We will denote external $\sigma$, $\mathcal{V}$, $\mathcal{V}^\dagger$ operators using the black, blue, and red open circles, respectively,
	\begin{equation}
		\sigma(x) \to \begin{tikzpicture}[baseline=-3pt] \draw[fill=white, thick] (0, 0) circle (\extVert) node[below, shift={(0, -0.1)}] {$x$}; \end{tikzpicture}\quad , \qquad \mathcal{V}(x) \to \begin{tikzpicture}[baseline=-3pt] \draw[cornellBlue, fill=white, thick] (0, 0) circle (\extVert) node[below, shift={(0, -0.1)}, black] {$x$}; \end{tikzpicture}\quad , \qquad \mathcal{V}^\dagger(x) \to \begin{tikzpicture}[baseline=-3pt] \draw[cornellRed, fill=white, thick] (0, 0) circle (\extVert) node[below, shift={(0, -0.1)}, black] {$x$}; \end{tikzpicture}\,\, .
	\end{equation}
	These external vertices will always be associated with the coordinates $x$ or $y$. We also introduce a set of internal vertices for the two types of interactions in (\ref{eq:intAction}),
	\begin{equation}
		\begin{tikzpicture}[baseline=-3pt, thick]
			\draw[sigmaProp] (0, 0) --+(120:0.5);
			\draw[sigmaProp] (0, 0) --+(-120:0.5);
			\fill[cornellBlue, intVertSty] (0, 0) circle (\intVert);
		\end{tikzpicture} \to g \,\, , \quad\qquad \begin{tikzpicture}[baseline=-3pt, thick]
			\draw[sigmaProp] (0, 0) --+(120:0.5);
			\draw[sigmaProp] (0, 0) --+(-120:0.5);
			\fill[cornellRed, intVertSty] (0, 0) circle (\intVert);
		\end{tikzpicture} \to \bar{g} \,\, , \quad\qquad \begin{tikzpicture}[baseline=-3pt] 
			\draw[cornellBlue, fill=white, line width=0.18mm] (0, 0) circle (0.1);
			\fill[cornellBlue] (0, 0) circle (0.06); \end{tikzpicture} \to g_\varphi \,\, , \quad\qquad  \begin{tikzpicture}[baseline=-3pt] 
			\draw[cornellRed, fill=white, line width=0.18mm] (0, 0) circle (0.1);
			\fill[cornellRed] (0, 0) circle (0.06); \end{tikzpicture} \to \bar{g}_{\varphi}\,\,,
	\end{equation}
	each associated with an internal coordinate $z_i$ that we must integrate over the sphere. The charge neutrality condition (\ref{eq:chargeNeutral}) ensures that any diagram must have equal numbers of red and blue vertices. We introduce three types of internal lines,
	\begin{equation}
		\begin{aligned}
			\begin{tikzpicture}[thick, baseline=-3pt]
				\draw[sigmaProp] (-0.75, 0) node[below] {$x$} -- (0.75, 0)  node[below] {$y$} ;
				\draw[fill=white] (-0.75, 0) circle (\extVert);
				\draw[fill=white] (0.75, 0) circle (\extVert);
			\end{tikzpicture} = G^{\sigma}(x, y)\,,\quad
			\begin{tikzpicture}[thick, baseline=-3pt]
				\draw[cornellBlue, vertexProp] (-0.75, 0) node[below, black] {$x$} -- (0.75, 0) node[below, black] {$y$} ;
				\draw[cornellBlue, fill=white] (-0.75, 0) circle (\extVert);
				\draw[cornellRed, fill=white] (0.75, 0) circle (\extVert);
			\end{tikzpicture} = \mathcal{G}(x,y)\,, \quad
			\begin{tikzpicture}[thick, baseline=-3pt]
				\draw[cornellRed, vertexProp] (-0.75, 0) node[below, black] {$z_i$} -- (0.75, 0) node[below, black] {$z_j$} ;
				\fill[cornellBlue, intVertSty] (-0.75, 0) circle (\intVert);
				\fill[cornellBlue, intVertSty] (0.75, 0) circle (\intVert);
			\end{tikzpicture} = \mathcal{G}^{\sminus 1}(z_i,z_j)\,.
		\end{aligned}
	\end{equation}
	The black $\sigma$ propagator obeys standard Feynman rules---only one connects into the external vertex \begin{tikzpicture}[baseline=-3pt] \draw[fill=white, thick] (0, 0) circle (\extVert); \end{tikzpicture} while two must connect into each of the filled vertices \begin{tikzpicture}[baseline=-3pt, thick] \fill[cornellRed, intVertSty] (0, 0) circle (\intVert); \end{tikzpicture} and \begin{tikzpicture}[baseline=-3pt, thick] \fill[cornellBlue, intVertSty] (0, 0) circle (\intVert); \end{tikzpicture}\es \es . The vertex propagators are different however: we connect each pair of colored vertices with a blue line $\mathcal{G}(\xi)$ if they are different colors and a red line $\mathcal{G}^{\sminus 1}(\xi)$ if they are the same.

	We also introduce a set of counterterm vertices. First are the standard mass and wavefunction renormalization vertices for the heavy field $\sigma$,
	\begin{equation}
		\begin{tikzpicture}[baseline=-3pt, thick]
			\draw[sigmaProp] (-1, 0) -- (1, 0);
			\begin{scope}[rotate=45]
				\draw[fill=white, line width=0.35mm] (0, 0) circle (0.12);
				\draw[line width=0.2mm] (-0.12, 0)--(0.12, 0);
				\draw[line width=0.2mm] (0, -0.12) -- (0, 0.12);
			\end{scope}
		\end{tikzpicture} = -J(J+2\alpha) \delta_{Z_\sigma} - \delta_{m_\sigma} \label{eq:sigCT}
	\end{equation}
	whose contribution we give in momentum space. Furthermore, there are the ``tadpole'' counterterm vertices for both $\mathcal{V}$ and $\mathcal{V}^\dagger$,
	\begin{equation}
		\begin{tikzpicture}[baseline=-3pt] 
			\draw[cornellBlue, fill=white, line width=0.25mm] (0, 0) circle (0.12);
			\draw[cornellBlue, line width=0.25mm] (0, 0) circle (0.08); \begin{scope}[cornellBlue, rotate=45] \draw (-0.08, 0) -- (0.08, 0); \draw (0, -0.08) -- (0, 0.08); \end{scope} \end{tikzpicture} \to \delta_{g_\varphi} \,\, , \qquad \text{and}\qquad \begin{tikzpicture}[baseline=-3pt] 
			\draw[cornellRed, fill=white, line width=0.25mm] (0, 0) circle (0.12);
			\draw[cornellRed, line width=0.25mm] (0, 0) circle (0.08); \begin{scope}[cornellRed, rotate=45] \draw (-0.08, 0) -- (0.08, 0); \draw (0, -0.08) -- (0, 0.08); \end{scope} \end{tikzpicture} \to \delta_{\bar{g}_{\varphi}}\,\,,
	\end{equation}
	which behave identically to their associated interaction vertices. Finally, to obtain a finite quantum-corrected vertex two-point function, we introduce the counterterm vertex associated with the operator $\mathcal{O}_\mathcal{V}$ in 
	\begin{equation}
		\begin{tikzpicture}[baseline=-3pt, thick, cornellBlue]
			\draw[vertexProp] (-1, 0) -- (1, 0);
			\begin{scope}[rotate=45]
				\draw[fill=white, line width=0.35mm] (0, 0) circle (0.12);
				\draw[line width=0.2mm] (-0.12, 0)--(0.12, 0);
				\draw[line width=0.2mm] (0, -0.12) -- (0, 0.12);
			\end{scope}
		\end{tikzpicture} = -J(J+2 \alpha) \delta_{Z_\mathcal{V}} - \delta_{m_\mathcal{V}}\,.
	\end{equation}
	Unlike the other vertex interactions, we demand that $\mathcal{O}_\mathcal{V}$ is such that it only connects two vertex propagators. Unfortunately, we do not know how to write $\mathcal{O}_{\mathcal{V}}$ in terms of the vertex operators themselves, but this does not affect the predictivity of the model---we are simply defining this operator in terms of its matrix elements with the vertex operators.\footnote{Composite operators are notorious for needing their own counterterms~\cite{Collins:1984xc} which do not necessarily take a simple form in terms of the ``fundamental'' fields over which we path integrate. There are other computational schemes~\cite{Hollands:2009xf,Frob:2022jxi} that do not rely on such counterterms to renormalize the theory, and perhaps this perturbative expansion in vertex operators is more natural in those languages.} Indeed, we will mainly be interested in the corrected heavy field two-point function $\langle \sigma(x) \sigma(y) \rangle$ at leading order in perturbation theory which only knows about the tree-level behavior of the vertex two-point function $\langle \mathcal{V}(x) \mathcal{V}^\dagger(y) \rangle \to \mathcal{G}(x, y)$, and so our main results will be independent of precisely how we choose to renormalize the vertex two-point function. To that end, we will mainly relegate our discussion of the vertex two-point function to Appendix~\ref{app:vertexTwo}.

	Our goal now is to compute loop corrections to $\langle \sigma(x) \sigma(y) \rangle$ and the vertex propagator $\langle \mathcal{V}(x) \mathcal{V}^\dagger(y)\rangle$ at leading order in perturbation theory. That is, we want to compute the \emph{self-energies} $\Pi_\sigma(J)$ and $\Pi_\varphi(J)$, which we define in terms of the connected two-point functions
	\begin{equation}
		\langle \sigma(x) \sigma(y) \rangle_\lab{c} = \frac{\Gamma(\alpha)}{2 \pi^{\alpha}} \int_{\gamma} \frac{\ud J}{2 \pi i} \frac{(J+\alpha)}{[G^\sigma]^{\sminus 1}_J - \Pi_\sigma(J)} \frac{C_{J}^{\alpha}(\minus \xi)}{\sin \pi J}\, \label{eq:sigmaWatson}
	\end{equation}
	and
	\begin{equation}
		\langle \mathcal{V}(x) \mathcal{V}^\dagger(y) \rangle_\lab{c} = \frac{\Gamma(\alpha)}{2 \pi^{\alpha}} \int_{\gamma} \frac{\ud J}{2 \pi i} \frac{(J+\alpha)}{[\mathcal{G}]^{\sminus 1}_J - \Pi_\varphi(J)} \frac{C_{J}^{\alpha}(\minus \xi)}{\sin \pi J}\,. \label{eq:vertexWatson}
	\end{equation}
	As discussed in the Introduction, as long as $\sigma$ is not too heavy it will generate a decaying yet oscillatory cosmological collider signal in the primordial bispectrum, the frequency and decay rate of which are determined by the long-distance behavior of $\langle \sigma(x) \sigma(y) \rangle_\lab{c}$. Using the techniques of~\cite{Marolf:2010zp} and \S\ref{sec:vertexPropMom}, the long distance decay of $\langle \sigma(x) \sigma(y) \rangle_\lab{c} \propto \mathcal{C} (\minus \xi)^{J_*} + \bar{\mathcal{C}} (\minus \xi)^{\bar{J}_*}$ is determined by the poles $J_*$ and $\bar{J}_*$ closest to the $\lab{Im}\,  J$-axis. Assuming that the self-energy is a small correction, the pole originally at $J = \minus \Delta_\sigma$ is shifted to
	\begin{equation}
		J_* \approx -\!\left[\alpha + \frac{\lab{Im}\, \Pi_\sigma(\minus \Delta_\sigma)}{2 \sqrt{m_\sigma^2 - \alpha^2}} \right] - i \!\left[\sqrt{m_\sigma^2 - \alpha^2} + \frac{\lab{Re}\, \Pi_\sigma(\minus \Delta_\sigma)}{2 \sqrt{m_\sigma^2 - \alpha^2}}\right]\,,
	\end{equation}
	while its shadow at $J = \minus \bar{\Delta}_\sigma$ experiences the same (albeit complex conjugated) shift. The mass counterterm $\delta_{m_\sigma}$ can be used to ensure $\lab{Re}\, \Pi_\sigma(\minus \Delta_\sigma) = 0$ so that $\nu_\sigma = \sqrt{m_\sigma^2 - \alpha^2}$ corresponds to~$\sigma$'s measured ``rest mass frequency''~\cite{Lu:2021wxu}, but the change in $\sigma$'s decay produced by a non-trivial $\lab{Im}\, \Pi_\sigma(\minus \Delta_\sigma)\neq 0$ is real and physical. Our main goal, then, is to compute the $\lab{Im}\, \Pi_\sigma(\minus \Delta_\sigma)$ generated by $\sigma$'s interactions with a light compact scalar $\varphi$.

	As discussed~\cite{Marolf:2010zp,Lu:2021wxu,Chakraborty:2023qbp} interactions with a light \emph{noncompact} scalar field will always cause~$\sigma$ to dilute faster than any free field in de Sitter space, $\lab{Im}\, \Pi_\sigma(\varphi) > 0$, providing an inflationary detection channel for these light fields. Interestingly, we find that interactions with a light \emph{compact} scalar can be qualitatively different and depend sensitively on the circumference of the field space~$2 \pi f$. In the noncompact limit $f \to \infty$ or $\beta \ll 1$, we find that a compact scalar generates corrections very similar to those generated by a light noncompact scalar, enabling $\sigma$ to decay faster than any free field, $\lab{Im}\, \Pi_\sigma(\minus \Delta_\sigma) > 0$. However, as the circumference becomes smaller than Hubble $\beta \gg 1$, we find that $\lab{Im} \, \Pi_\sigma(\minus \Delta_\sigma)$ begins to oscillate, causing $\sigma$ to decay either \emph{faster} or \emph{slower} than any free field.

	To prevent our expressions from becoming overly long, we will introduce several abbreviations. We denote local fields evaluated at specific coordinates with a subscript, $\sigma_i \equiv \sigma(z_i)$, $\mathcal{V}_i \equiv \mathcal{V}(z_i)$, $G^\sigma_{i j} \equiv G^\sigma(z_i, z_j)$, and $\mathcal{G}_{ij} = \mathcal{G}(z_i, z_j)$. Here, the $z_i$ denote dummy variables over which we integrate, but we will also use $z_x = x$ and $z_y = y$ to denote the external coordinates. Furthermore, we use $\mathcal{Z} = \int\!\mathcal{D}\varphi_\lab{c} \, \mathcal{D}\sigma\, \e^{\sminus S_\slab{e}}$ and $\mathcal{Z}_0 = \int\!\mathcal{D}\varphi_\lab{c} \, \mathcal{D}\sigma \, \e^{\sminus S_0}$ to denote the interacting and free partition functions, respectively. We denote iterated integrals over the Euclidean sphere as $\int_{1,2,\cdots, n} \equiv \int_{\{n\}} \equiv \int_{\lab{S}^D} \!\ud^D z_1\, \sqrt{g(z_1)} \, \ud^D z_2\, \sqrt{g(z_2)} \cdots$. Finally, we will use $\langle \, \cdot \, \rangle_\lab{c}$ to denote the connected correlation functions in the \emph{interacting theory}, while we use $\langle \,\cdot\,\rangle$ to denote expectation values taken in the free theory. 

	We will first derive the leading order long-distance behavior of the propagators (\ref{eq:sigmaWatson}) and~(\ref{eq:vertexWatson}) in \S\ref{sec:leadingOrder}. We then analyze the self-energy $\Pi_\sigma(J)$ in the noncompact $\beta \to 0$ and ultracompact $\beta \to \infty$ limits in \S\ref{sec:noncompactLimit} and \S\ref{sec:ultracompactLimit}, respectively.

\subsection{Leading-Order Corrections} \label{sec:leadingOrder}
	
	We begin by considering the one-point functions $\langle \sigma(x) \rangle_\lab{c}$ and $\langle \mathcal{V}(x)\rangle_\lab{c}$. Clearly, $\langle \sigma(x)\rangle_\lab{c}$ always vanishes due to the $\sigma \to \minus \sigma$ symmetry of the action. However, there is no such symmetry that prevents a potential for $\varphi$, or equivalently an expectation value for $\mathcal{V}(x)$, from being quantum-mechanically generated. So, even absent an explicit classical potential for $\varphi$ with $g_\varphi = 0$ we should expect that the vertex operator will acquire a vacuum expectation value because, as depicted in Figure~\ref{fig:evs}, there will be a bias in where~$\varphi$ prefers to sit induced by fluctuations of $\sigma$.

	The interacting vertex one-point function
	\begin{equation}
		\langle \mathcal{V}(x) \rangle_\lab{c} = \frac{1}{\mathcal{Z}} \int\!\mathcal{D}\varphi_\lab{c}\, \mathcal{D}\sigma\, \mathcal{V}(x) \, \e^{\sminus S_\slab{e}}
	\end{equation}
	may be computed by expanding out the path integral in powers of the interaction (\ref{eq:interactingAction}) and counterterm actions (\ref{eq:ctAction}), though for presentation purposes we will ignore the counterterm action and instead add those contributions in explicitly by hand. Since $S_\lab{int}$ has no charge neutral interactions, expectation values involving any odd power of $S_\lab{int}$ alone vanishes, $\langle S_{\lab{int}}^{2k+1} \rangle = 0$ for~$k \in \mathbb{Z}_{\scriptscriptstyle \geq 0}$. The ``disconnected'' part of the path integral can then be written as
	\begin{equation}
		\begin{aligned}
			\frac{\mathcal{Z}_0}{\mathcal{Z}} &= \left[\frac{1}{\mathcal{Z}_0} \int\!\mathcal{D} \varphi_\lab{c} \, \mathcal{D} \sigma \, \e^{\sminus S_{0}} \left[1 - S_\lab{int} + \tfrac{1}{2} S^2_\lab{int} + \tfrac{1}{4!} S^4_\lab{int} \cdots \right]\right]^{\sminus 1} \\
			&= 1 - \tfrac{1}{2} \langle S_\lab{int}^2 \rangle + \tfrac{1}{4} \langle S_\lab{int}^2\rangle^2 - \tfrac{1}{4!} \langle S_\lab{int}^4 \rangle + \cdots\,. \label{eq:normalizingFactor}
		\end{aligned}
	\end{equation}
	To leading order in perturbation theory, we find that the vertex one-point function is
	\begin{equation}
		\begin{aligned}
			\langle \mathcal{V}(x) \rangle_\lab{c} &=  -\langle \mathcal{V}(x) S_\lab{int} \rangle = -\int_{1} \langle \mathcal{V}_x (\bar{g}_\varphi + \bar{\delta}_{g_\varphi} + \tfrac{1}{2}\bar{g} \sigma_1^2) \mathcal{V}_z \rangle \\
			&= \begin{tikzpicture}[thick, baseline=-3pt]
				\draw[cornellBlue, vertexProp] (0, 0) -- (1, 0);
				\draw[cornellRed, fill=white, line width=0.18mm] (1, 0) circle (0.1);
				\fill[cornellRed] (1, 0) circle (0.06);
				\draw[cornellBlue, fill=white, thick] (0, 0) circle (\extVert);
				\end{tikzpicture} +  \begin{tikzpicture}[thick, baseline=-3pt]
				\draw[cornellBlue, vertexProp] (0, 0) -- (1, 0);
				\begin{scope}[cornellRed, shift={(1, 0)}]
				\draw[cornellRed, fill=white, line width=0.25mm] (0, 0) circle (0.12);
				\draw[cornellRed, line width=0.25mm] (0, 0) circle (0.08);  
					\begin{scope}[rotate=45] \draw (-0.08, 0) -- (0.08, 0); \draw (0, -0.08) -- (0, 0.08);  \end{scope}
				\end{scope}
				\draw[cornellBlue, fill=white, thick] (0, 0) circle (\extVert);
				\end{tikzpicture} + 
				\begin{tikzpicture}[thick, baseline=-3pt]
				\draw[cornellBlue, vertexProp] (0, 0) -- (1, 0);
				\draw[sigmaProp] (1.4, 0) circle (0.4);
				\fill[cornellRed, intVertSty] (1, 0) circle (\intVert);
				\draw[cornellBlue, fill=white, thick] (0, 0) circle (\extVert);
				\end{tikzpicture} \\
				&= - \left(\bar{g}_\varphi + \bar{\delta}_{g_\varphi} + \tfrac{1}{2}\bar{g} \es G^{\sigma}(1)\right)\![\mathcal{G}]_{0}\,,
		\end{aligned} \label{eq:tadpole}
	\end{equation}
	where $[\mathcal{G}]_{0}$ is the $J = 0$ coefficient of the vertex propagator (\ref{eq:vertexLinv}), while a similar calculation yields $\langle \mathcal{V}^\dagger(x) \rangle_\lab{c} = -(g_\varphi + \delta_{g_\varphi} + \tfrac{1}{2} g G^{\sigma}(1))[\mathcal{G}]_0$. The counterterms $\delta_{g_\varphi}$ and $\bar{\delta}_{g_\varphi}$ can be tuned to absorb the UV divergence from the $\sigma$ loop, $\delta_{g_\varphi} = -\tfrac{1}{2} g \es G^{\sigma}(1)$. 

	We can follow a similar strategy for the heavy field two-point function $\langle \sigma(x) \sigma(y) \rangle_\lab{c}$. Expanding in the interaction $S_\lab{int}$ yields
	\begin{equation}
		\begin{aligned}
			&\langle \sigma(x) \sigma(y) \rangle_\lab{c} = \langle \sigma(x) \sigma(y) \rangle + \tfrac{1}{2}\! \left[\langle \sigma(x) \sigma(y) S_\lab{int}^2 \rangle - \langle \sigma(x) \sigma(y) \rangle \langle S_\lab{int}^2 \rangle \right] \\
			&\quad + \left[\tfrac{1}{4!} \langle \sigma(x) \sigma(y) S_\lab{int}^4 \rangle - \tfrac{1}{4} \langle \sigma(x) \sigma(y) S_\lab{int}^2 \rangle \langle S_\lab{int}^2 \rangle + \left(\tfrac{1}{4} \langle S_\lab{int}^2 \rangle^2 - \tfrac{1}{4!} \langle S_\lab{int}^4 \rangle \right)\langle \sigma(x) \sigma(y) \rangle\right] + \cdots\,,
		\end{aligned} \label{eq:sigExp}
	\end{equation}
	where we have also included the next-to-leading order corrections on the second line. The leading order corrections are given by
	\begin{equation}
		\begin{aligned}
			\langle \sigma(x) \sigma(y) \rangle_\lab{c} &\supset
			 \int_{1,2} \Big[\big\langle\!\left[\bar{g}_\varphi + \tfrac{1}{2} \bar{g} \sigma_1^2\right]\!\left[g_\varphi + \tfrac{1}{2} g \sigma_2^2 \right] \! \sigma_x \sigma_y \mathcal{V}^\dagger_1 \mathcal{V}_2 \big\rangle \\
			&\qquad\qquad\,\, - \langle \sigma_x \sigma_y \rangle \big\langle\!\left[\bar{g}_\varphi + \tfrac{1}{2} \bar{g} \sigma_1^2\right]\!\left[g_\varphi + \tfrac{1}{2} g \sigma_2^2 \right] \!\mathcal{V}^\dagger_1 \mathcal{V}_2 \big\rangle\Big] + \cdots
		\end{aligned} 
	\end{equation}
	where the $\cdots$ denote the contribution of $\sigma$'s mass and wavefunction renormalization counterterm~(\ref{eq:sigCT}). Clearly, the effect of the second term is to remove the ``disconnected'' contribution from the first line, and so diagrammatically this leading order correction can be represented as
	\begin{equation}
		\langle \sigma(x) \sigma(y) \rangle_\lab{c} \supset
		\begin{tikzpicture}[baseline=-3pt, thick]
			\draw[sigmaProp] (-1, 0)--(-0.5, 0);
			\draw[sigmaProp] (0.5, 0)--(1, 0);
			\draw[cornellBlue, vertexProp] (-0.5, 0) arc(180:0:0.5);
			\draw[sigmaProp] (-0.5, 0) arc(0:180:-0.5);
			\fill[cornellBlue, intVertSty] (-0.5, 0) circle (0.08);
			\fill[cornellRed, intVertSty] (0.5, 0) circle (0.08);
			\draw[fill=white] (-1, 0) circle (0.06);
			\draw[fill=white] (1, 0) circle (0.06);
		\end{tikzpicture}+  \begin{tikzpicture}[baseline=-3pt, thick]
			\draw[sigmaProp] (-0.75, 0) -- (0.75, 0);
			\begin{scope}[rotate=45]
				\draw[fill=white, line width=0.3mm] (0, 0) circle (0.12);
				\draw[line width=0.2mm] (-0.12, 0)--(0.12, 0);
				\draw[line width=0.2mm] (0, -0.12) -- (0, 0.12);
			\end{scope}
			\draw[fill=white] (-0.75, 0) circle (0.06);
			\draw[fill=white] (0.75, 0) circle (0.06);
		\end{tikzpicture} +
		\begin{tikzpicture}[baseline=-3pt, thick]
			\draw[sigmaProp] (-0.75, 0)--(0.75,0);
			\draw[fill=white] (-0.75, 0) circle (0.06);
			\draw[fill=white] (0.75, 0) circle (0.06);
			\draw[cornellBlue, vertexProp] (0, 0) -- (0, 0.5);
			\fill[cornellBlue, intVertSty] (0, 0) circle (0.08);
			\draw[cornellRed, fill=white, line width=0.18mm] (0, 0.5) circle (0.1);
			\fill[cornellRed] (0, 0.5) circle (0.06);
		\end{tikzpicture} + 
		\begin{tikzpicture}[baseline=-3pt, thick]
			\draw[sigmaProp] (-0.75, 0)--(0.75,0);
			\draw[fill=white] (-0.75, 0) circle (0.06);
			\draw[fill=white] (0.75, 0) circle (0.06);
			\draw[sigmaProp] (0, 0.75) circle (0.25);
			\draw[cornellBlue, vertexProp] (0, 0) -- (0, 0.5);
			\fill[cornellBlue, intVertSty] (0, 0) circle (0.08);
			\fill[cornellRed, intVertSty] (0, 0.5) circle (0.08);
		\end{tikzpicture} + \begin{tikzpicture}[baseline=-3pt, thick]
			\draw[sigmaProp] (-0.75, 0)--(0.75,0);
			\draw[fill=white] (-0.75, 0) circle (0.06);
			\draw[fill=white] (0.75, 0) circle (0.06);
			\draw[cornellBlue, vertexProp] (0, 0) -- (0, 0.5);
			\fill[cornellBlue, intVertSty] (0, 0) circle (0.08);
			\begin{scope}[cornellRed, shift={(0, 0.5)}]
				\draw[cornellRed, fill=white, line width=0.25mm] (0, 0) circle (0.12);
				\draw[cornellRed, line width=0.25mm] (0, 0) circle (0.08);  
					\begin{scope}[rotate=45] \draw (-0.08, 0) -- (0.08, 0); \draw (0, -0.08) -- (0, 0.08);  \end{scope}
			\end{scope}
		\end{tikzpicture} + \cdots \label{eq:sigLeadingOrderDiagrams}
	\end{equation}
	where the $\cdots$ denote diagrams with red and blue vertices interchanged. The last three diagrams represent $\sigma$ interacting with $\mathcal{V}$'s vacuum expectation value and can be completely absorbed by shifting the mass counterterm, leaving us with
	\begin{equation}
		\langle \sigma(x) \sigma(y) \rangle_c = G_{xy}^{\sigma} + 2 |g|^2 \!\int_{1,2}\! G_{x1}^{\sigma}\big(G^{\sigma} \mathcal{G}\big)_{12} G^{\sigma}_{2 y} - \int_{1,2} \! G_{x1}^{\sigma} \! \left[\minus \nabla_{z_2}^2 + \delta_{m_\sigma}\right] \! G_{2y}^{\sigma} + \cdots\,.
	\end{equation}
	where $\nabla_{z_2}^2$ is the Laplacian with respect to $z_2$. In momentum space, this takes the simple form
	\begin{equation}
		[\sigma^2]_{J} = [G^{\sigma}]_J + [G^{\sigma}]_J^2\left[2 |g|^2 [G^{\sigma} \mathcal{G}]_J - J(J+2\alpha) \delta_{Z_\sigma} - \delta_{m_\sigma} \right] + \cdots\,, \label{eq:sigLeadingOrder}
	\end{equation}
	and so the leading order correction to $\sigma$'s propagator is determined by the bubble diagram $[G^{\sigma} \mathcal{G}]_J$. If we were expanding in only Gaussian free fields, we would geometrically resum chains of this bubble diagram and identify the leading-order self-energy appearing in (\ref{eq:sigmaWatson}) as
	\begin{equation}
		\Pi_\sigma(J) = 2 |g|^2 [G^{\sigma} \mathcal{G}]_J - J(J+2\alpha) \delta_{Z_\sigma} - \delta_{m_\sigma}\,. \label{eq:sigmaSelfEnergy}
	\end{equation}
	However, the vertex operators do not Wick factorize, and we will see that this makes the standard Dyson resummation of one-particle-irreducible diagrams more complicated. We will argue, however, that this resummation indeed captures the corrected long-distance behavior of $\langle \sigma(x) \sigma(y) \rangle_\lab{c}$.

	To see the issue, let us consider the next-to-leading-order corrections generated by the quartic term $\frac{1}{4!}\langle \sigma(x) \sigma(y) S_\lab{int}^4\rangle$ on the second line of~(\ref{eq:sigExp}). This contains the term
	\begin{equation}
		\frac{1}{4!} \langle \sigma(x) \sigma(y) S_\lab{int}^4 \rangle \supset \frac{1}{4!} \frac{1}{2^4} |g|^4 \int_{\{4\}} \! \langle \sigma_x \sigma_1^2 \sigma_2^2 \sigma_3^2 \sigma_4^2 \sigma_y \rangle \! \left[\langle \mathcal{V}^{\vphantom{\dagger}}_1 \mathcal{V}_2^\dagger \mathcal{V}^{\vphantom{\dagger}}_3 \mathcal{V}_4^\dagger \rangle + \text{$\dagger$ perms}\right]\,,
	\end{equation}
	where ``$\dagger$ perms'' denote the permutations of which vertex operators are conjugated. If the vertex operators Wick factorized, there would be a contribution of the form
	\begin{equation}
		\frac{1}{4!} \langle \sigma(x) \sigma(y) S_\lab{int}^4 \rangle \supset \big(2 |g|^2\big)^2 \int_{\{4\}} \! G_{x1} \big(G^{\sigma} \mathcal{G}\big)_{12} G_{23}^\sigma \big(G^{\sigma} \mathcal{G}\big)_{34} G^{\sigma}_{4 y} = \begin{tikzpicture}[baseline=-3pt, thick]
			\def\circSize{0.4}
			\def\lSize{0.4}
			\begin{scope}
				\draw[sigmaProp] (-\circSize, 0) -- +(-\lSize, 0);
				\draw[sigmaProp] (\circSize, 0) -- +(\lSize, 0);
				\draw[fill=white] (-\circSize-\lSize, 0) circle (\extVert);
				\draw[cornellBlue, vertexProp] (-\circSize, 0) arc (180:0:\circSize);
				\draw[sigmaProp] (-\circSize, 0) arc(0:180:-\circSize);
				\fill[cornellBlue, intVertSty] (-\circSize, 0) circle (\intVert);
				\fill[cornellRed, intVertSty] (\circSize, 0) circle (\intVert);
			\end{scope}
			\begin{scope}[shift={(2*\lSize+\circSize, 0)}]
				\draw[sigmaProp] (\circSize, 0) -- +(\lSize, 0);
				\draw[fill=white] (\circSize+\lSize, 0) circle (\extVert);
				\draw[cornellBlue, vertexProp] (-\circSize, 0) arc (180:0:\circSize);
				\draw[sigmaProp] (-\circSize, 0) arc(0:180:-\circSize);
				\fill[cornellBlue, intVertSty] (-\circSize, 0) circle (\intVert);
				\fill[cornellRed, intVertSty] (\circSize, 0) circle (\intVert);
			\end{scope}
		\end{tikzpicture}\,, \label{eq:sigma1PI}
	\end{equation}
	which is the ``square'' of the leading order correction. Crucially, this higher-order correction is proportional to $|g|^4/[(J+\Delta_\sigma)(J+\bar{\Delta}_\sigma)]^{3}$ and so becomes arbitrarily important at the free-field poles $J = \minus \Delta_\sigma$ and $J = \minus \bar{\Delta}_\sigma$, regardless of how small $g$ is. To investigate the long-distance behavior of $\langle \sigma(x) \sigma(y) \rangle_\lab{c}$, we must then include all such corrections, equating the one-particle irreducible diagrams with the self-energy $\Pi_\sigma(J)$ and geometrically resumming them to find (\ref{eq:sigmaWatson}). The problem is, however, that the vertex operators do \emph{not} Wick factorize, and so the actual contribution is instead of the~form
	\begin{equation}
		\big(2 |g|^2\big)^2 \! \int_{\{4\}} \! G_{x1} G_{12}^{\sigma} G_{23}^{\sigma} G_{34}^{\sigma} G_{4y}^{\sigma}\,  \frac{\mathcal{G}_{12} \mathcal{G}_{14} \mathcal{G}_{23} \mathcal{G}_{34}}{\mathcal{G}_{13} \mathcal{G}_{24}} = \begin{tikzpicture}[baseline=-3pt, thick]
			\draw[sigmaProp] (-1, 0)--(-0.5, 0);
			\draw[sigmaProp] (0.5, 0)--(1, 0);
			\draw[cornellBlue, vertexProp] (-0.5, 0) arc(180:0:{0.8333/2} and 0.4);
			\draw[sigmaProp] (-0.5, 0) -- (0.5, 0);
			\draw[cornellBlue, vertexProp] (1.16666, 0) arc(180:0:{0.8333/2} and 0.4);
			\draw[cornellBlue, vertexProp] (0.33333, 0) arc(180:0:{0.8333/2} and 0.4);
			\draw[cornellBlue, vertexProp] (-0.5, 0) arc(180:0:1.25 and 0.75);
			\draw[cornellRed, vertexProp] (-0.5, 0) arc(0:-180:{-0.83333} and 0.5);
			\draw[cornellRed, vertexProp] (0.33333, 0) arc(0:-180:{-0.83333} and 0.5);
			\draw[sigmaProp] (1, 0) --(2, 0);
			\draw[sigmaProp] (2, 0)--(2.5, 0);
			\fill[cornellBlue, intVertSty] (-0.5, 0) circle (\intVert);
			\fill[cornellRed, intVertSty] (0.3333, 0) circle (\intVert);
			\fill[cornellBlue, intVertSty] (1.16666, 0) circle (\intVert);
			\fill[cornellRed, intVertSty] (2, 0) circle (\intVert);
			\draw[fill=white] (-1, 0) circle (\extVert);
			\draw[fill=white] (2.5, 0) circle (\extVert);
		\end{tikzpicture}\,, \label{eq:sigmaNPI}
	\end{equation}
	and so it is unclear whether such a Dyson resummation makes physical sense or whether it is legitimately capturing the physics of these higher order corrections.

	Physically, these divergent corrections arise because an internal $\sigma$ particle can go on on-shell and propagate over very large distances, accumulating probability. In Lorentzian signature, the chain of one-particle-irreducible bubbles in (\ref{eq:sigma1PI}) corresponds to a process in which a $\sigma$ particle propagates over a large distance, spontaneously splits into virtual $\sigma$ and $\mathcal{V}$ ``particles,'' which then subsequently decay into another long-lived $\sigma$ particle, which then propagates over a long distance until it repeats this process again and again. Of course, an arbitrarily long chain of these events can occur, necessitating the Dyson resummation. Our goal now is to argue using the long-distance factorization of vertex correlators (\ref{eq:longDistanceFact}) that the actual higher-order correction~(\ref{eq:sigmaNPI}) can be well-approximated by (\ref{eq:sigma1PI}) for $J$ near the free-field poles, $J = \minus \Delta_\sigma$ and $J = \minus \bar{\Delta}_\sigma$. Thus, we may indeed approximate the leading-order self-energy $\Pi_\sigma(J)$ in (\ref{eq:sigmaWatson}) by the bubble diagram~(\ref{eq:sigLeadingOrderDiagrams}) and use it to infer the long-distance behavior of $\langle \sigma(x) \sigma(y) \rangle_\lab{c}$.

	For completeness, we will identify these dangerous corrections at arbitrary order in perturbation theory. At $\mathcal{O}(|g|^{2n})$, $\sigma$'s two-point function receives a contribution from
	\begin{equation}
		\frac{1}{(2 n)!} \langle \sigma_x \sigma_y S_\lab{int}^{2n}\rangle \supset \frac{|g|^{2n}}{2^{2n} (2n)!}  \int_{\{2n\}} \! \langle \sigma_x \sigma_1^2 \sigma_2^2 \cdots \sigma_{2n}^2 \sigma_y \rangle \left[\langle \mathcal{V}^{\vphantom{\dagger}}_1 \mathcal{V}_2^\dagger \cdots \mathcal{V}_{2n}^\dagger \rangle + \text{$\dagger$ perms}\right].
	\end{equation}
	Corrections that diverge as $J \to \minus \Delta_\sigma$ only result from a long chain of $\sigma$ propagators. There are~$2^{2n} (2n)!$ ways contracting each $\sigma$ with another and so, after relabeling the vertices, we have
	\begin{equation}
		\frac{1}{(2 n)!} \langle \sigma_x \sigma_y S_\lab{int}^{2n}\rangle \supset |g|^{2n}  \int_{\{2n\}} \! G_{x1}^{\sigma} G_{12}^{\sigma} G_{23}^{\sigma} \cdots G_{2n, y}^{\sigma} \left[\langle \mathcal{V}^{\vphantom{\dagger}}_1 \mathcal{V}_2^\dagger \cdots \mathcal{V}_{2n}^\dagger \rangle + \text{$\dagger$ perms}\right].
	\end{equation}
	The $\sigma$ propagators now set a particular ordering of the points and group the internal vertices into pairs $(z_1, z_2)$, $(z_3, z_4)$, \dots, $(z_{2n-1}, z_{2n})$. Furthermore, ``$\dagger$ perms'' contain $2^n$ terms which have identical correlation functions (i.e. interchanging $\mathcal{V}^{\vphantom{\dagger}}_1 \mathcal{V}_2^\dagger \to \mathcal{V}^\dagger_1 \mathcal{V}^{\vphantom{\dagger}}_{2}$ does not change the integrand) and so grouping these yields
	\begin{align}
			&\frac{1}{(2 n)!} \langle \sigma_x \sigma_y S_\lab{int}^{2n}\rangle \supset 2^{n}|g|^{2n}  \int_{\{2n\}} \! G_{x1}^{\sigma} G_{12}^{\sigma} G_{34}^{\sigma} \cdots G_{2n, y}^{\sigma} \langle \mathcal{V}^{\vphantom{\dagger}}_1 \mathcal{V}_2^\dagger \mathcal{V}^{\vphantom{\dagger}}_3 \mathcal{V}_4^\dagger  \cdots \mathcal{V}_{2n}^\dagger \rangle  \label{eq:2nCorr}
	\end{align}
	where we have dropped all other $\dagger$ permutations that are not of this form, as they will be subleading in the long-distance limit.

	Since we are interested in Lorentzian signature physics, $J \to \minus \Delta_\sigma$, it will be helpful to analytically continue this diagram to the Poincar\'{e} patch, where the metric takes the form
	\begin{equation}
		\ud s^2 = \frac{1}{\lambda^2} \left(-\ud \lambda^2 + \ud \tilde{\mb{z}}^2\right),
	\end{equation}
	with $\lambda \in [0, \infty)$ and $\tilde{\mb{z}} \in \mathbb{R}^d$. In these coordinates, the embedding distance between $z_i = (\lambda_i, \tilde{\mb{z}}_i)$ and $z_j = (\lambda_j, \tilde{\mb{z}}_j)$~is
	\begin{equation}
		\xi_{ij} = 1 - \frac{|\tilde{\mb{z}}_i - \tilde{\mb{z}}_j|^2 - (\lambda_i - \lambda_j)^2}{2 \lambda_i \lambda_j}\,.
	\end{equation}
	Usefully, there is a diagram-by-diagram equivalence~\cite{Higuchi:2010xt} between perturbation theory in Euclidean de Sitter space and in-in perturbation theory on the Poincar\'{e} patch.  Up to an appropriate choice of time-ordering the different propagators---which we will not specify since it will not affect our conclusions---the Euclidean contribution (\ref{eq:2nCorr}) maps to in-in one with the same integrand but which is now integrated over the entire Poincar\`{e} patch,
	\begin{equation}
		\int_{\lab{S}^D} \ud^D z_i \, \sqrt{g(z_i)} \to \int_{\mathcal{C}} \frac{\ud \lambda_i}{\lambda_i^D} \int_{\mathbb{R}^d} \!\ud^d \tilde{z}_i 
	\end{equation}
	with $\mathcal{C}$ the closed time contour necessitated by the in-in formalism. Crucially, the loop integrals in (\ref{eq:2nCorr}) now span all separations and the embedding distances $|\xi_{ij}|$ between the internal vertices are no longer bounded by 1.

	We can now identify a particularly dangerous region of the integral (\ref{eq:2nCorr}) that produces a singularity as $J \to \minus \Delta_\sigma$. Let us consider the region in which there is a hierarchy in the embedding distances, $|\xi_{2i -1, 2i}| \ll |\xi_{2i -1, j}|$ for all $i = 1, \cdots, n$ and $j \neq 2i$.  This is the situation in which there are $n$ pairs of points close to one another, but each pair is far away from all others. Here, the vertex propagator drastically simplifies. Note that the charge-neutral $2n$-point function of vertex operators in the free theory may be constructed by taking the product of $\mathcal{G}_{ij}$ for each pair of oppositely charged vertex operators (that is, each pair of $\mathcal{V}_i$ and $\mathcal{V}^\dagger_j$) with the product of $\mathcal{G}^{\sminus 1}_{ij}$ for each pair of like charged operators,
	\begin{equation}
		\langle \mathcal{V}^{\vphantom{\dagger}}_1 \mathcal{V}_2^\dagger \mathcal{V}^{\vphantom{\dagger}}_3 \mathcal{V}_4^\dagger \mathcal{V}^{\vphantom{\dagger}}_5 \mathcal{V}^{\dagger}_6 \cdots\rangle = \mathcal{G}_{12}\! \left[\frac{\mathcal{G}_{14}}{\mathcal{G}_{13}} \frac{\mathcal{G}_{16}}{\mathcal{G}_{15}} \frac{\mathcal{G}_{23}}{\mathcal{G}_{24}} \frac{\mathcal{G}_{25}}{\mathcal{G}_{26}} \cdots \right] \! \mathcal{G}_{34} \! \left[\frac{\mathcal{G}_{36}}{\mathcal{G}_{35}} \frac{\mathcal{G}_{45}}{\mathcal{G}_{46}}\cdots \right] \! \mathcal{G}_{56}  \cdots\,. \label{eq:vertexExample}
	\end{equation}
	For example, we see that the operator $\mathcal{V}_1$ contributes a factor of $\mathcal{G}_{1, 2i}/\mathcal{G}_{1, 2i-1}$ for each of the other charge neutral pairs of operators, $i = 2, \cdots, n$, since it sees both the $\mathcal{V}$ and $\mathcal{V}^\dagger$ of that pair. As we move those pairs further and further away from $z_1$, the correlation between $\mathcal{V}_1$ and that pair vanishes, $\mathcal{G}_{1, 2i}/\mathcal{G}_{1, 2i-1} \to 1$. The same is true for every other operator in~(\ref{eq:vertexExample}). 

	Thus, in the region of integration where each pair of points is close to one another but far away from all other pairs, the vertex $2n$-point function factorizes and reduces to
	\begin{equation}
		\langle \mathcal{V}^{\vphantom{\dagger}}_1 \mathcal{V}_2^\dagger \mathcal{V}^{\vphantom{\dagger}}_3 \mathcal{V}_4^\dagger \mathcal{V}^{\vphantom{\dagger}}_5 \mathcal{V}^{\dagger}_6 \cdots\rangle = \mathcal{G}_{12}\mathcal{G}_{34} \mathcal{G}_{56} \cdots\,.
	\end{equation}
	In this region, (\ref{eq:2nCorr}) can be approximated as
	\begin{equation}
		\begin{aligned}
			\langle \sigma(x) \sigma(y) \rangle_\lab{c} &\supset |g|^{2n} \int_{\{2n\}}\!\! G_{x1}^{\sigma} \left(G^{\sigma} \mathcal{G}\right)_{12} G^{\sigma}_{23} \left(G^{\sigma} \mathcal{G}\right)_{34} G^{\sigma}_{45} \cdots \left(G^{\sigma} \mathcal{G}\right)_{2n-1, 2n} G^{\sigma}_{2n, y} + \cdots \\
			&\supset \def\lLength{0.45}
						\def\circSize{0.45}
		 \begin{tikzpicture}[thick,baseline=-3pt]
						\draw[sigmaProp] (-\lLength-\circSize, 0) -- (-\circSize, 0);
						
						\draw[sigmaProp] (\circSize, 0) -- (\circSize+\lLength, 0) ;
						\draw[fill=white] (-\lLength-\circSize, 0) circle (\extVert);
						\draw[cornellBlue, vertexProp] (0, 0)++(-\circSize,0) arc(180:0:\circSize);
						\draw[sigmaProp] (0, 0)++(\circSize,0) arc(0:-180:\circSize);
						\fill[cornellBlue, intVertSty] (-\circSize, 0) circle (\intVert);
						\fill[cornellRed, intVertSty] (\circSize, 0) circle (\intVert);
						\begin{scope}[shift={(2*\lLength+\circSize, 0)}]
							\draw[sigmaProp] (\circSize, 0) -- (\circSize+\lLength, 0);
							\draw[cornellBlue, vertexProp] (0, 0)++(-\circSize,0) arc(180:0:\circSize);
							\draw[sigmaProp] (0, 0)++(\circSize,0) arc(0:-180:\circSize);
							\fill[cornellBlue, intVertSty] (-\circSize, 0) circle (\intVert);
							\fill[cornellRed, intVertSty] (\circSize, 0) circle (\intVert);
						\end{scope}
					\end{tikzpicture}\cdots\begin{tikzpicture}[thick,baseline=-3pt]
						\draw[sigmaProp] (-\lLength-\circSize, 0) -- (-\circSize, 0);
						
						\draw[sigmaProp] (\circSize, 0) -- (\circSize+\lLength, 0);
						\draw[cornellBlue, vertexProp] (0, 0)++(-\circSize,0) arc(180:0:\circSize);
						\draw (0, 0)++(\circSize,0) arc(0:-180:\circSize);
							\fill[cornellBlue, intVertSty] (-\circSize, 0) circle (\intVert);
							\fill[cornellRed, intVertSty] (\circSize, 0) circle (\intVert);
						\begin{scope}[shift={(2*\lLength+\circSize, 0)}]
							\draw[sigmaProp] (\circSize, 0) -- (\circSize+\lLength, 0);
							\draw[cornellBlue, vertexProp] (0, 0)++(-\circSize,0) arc(180:0:\circSize);
							\draw[sigmaProp] (0, 0)++(\circSize,0) arc(0:-180:\circSize);
							\fill[cornellBlue, intVertSty] (-\circSize, 0) circle (\intVert);
							\fill[cornellRed, intVertSty] (\circSize, 0) circle (\intVert);
							\draw[fill=white] (\lLength+\circSize, 0) circle (\extVert);
						\end{scope}
					\end{tikzpicture} + \cdots\,,
		\end{aligned}
	\end{equation}
	or in momentum space as 
	\begin{equation}
		[\sigma^2]_J \supset [G^{\sigma}]_J^{n+1} \left[2|g|^{2} [G^{\sigma} \mathcal{G}]_{J} - J(J+2\alpha) \delta_{Z_\sigma} - \delta_{m_\sigma}\right]^{\!\es\es n} + \cdots\,.
	\end{equation}
	This contribution diverges strongly as $J \to \minus \Delta_\sigma$ and so we expect that the contributions from the other regions of integration, denoted here by the $\cdots$, are subleading in this limit. We find that, indeed, even if the vertex operators do not factorize and there is not a strict notion of the set one-particle-irreducible diagrams that can be chained together, it still makes sense to Dyson resum the leading order correction to the propagator and identify the self-energy as (\ref{eq:sigmaSelfEnergy}). 

	A similar argument applies to the vertex two-point function. Since it will distract from the flow of this section and will have no bearing on our main results, we will reserve that analysis for Appendix~\ref{app:vertexTwo}. There we show that $\sigma$'s effect on the leading long-distance behavior of $\langle \mathcal{V}(x) \mathcal{V}^\dagger(y) \rangle$ can be completely absorbed by a counterterm and is thus unphysical, similar to what happens for a light noncompact field~\cite{Lu:2021wxu,Chakraborty:2023qbp}. That said, the self-energy is still nontrivial, and so $\sigma$ still affects the subleading asymptotic behavior. We will not dwell on this here, and instead work to evaluate $\sigma$'s self-energy~(\ref{eq:sigmaSelfEnergy}).  

	So, at leading order in perturbation theory and near the free field poles $J = \minus \Delta_\sigma$ and $J = \minus \bar{\Delta}_\sigma$, the self-energy $\Pi_\sigma(J)$ is determined by the bubble diagram with one internal $\sigma$ and one vertex propagator, $[G^\sigma \mathcal{G}]_J$. This loop diagram can be easily computed~\cite{Chakraborty:2023qbp} by applying the Lorentzian inversion formula (\ref{eq:lInvForm}) to the product $G^\sigma(\zeta) \mathcal{G}(\zeta)$,
	\begin{equation}
		[G^\sigma \mathcal{G}]_{J} = - \frac{2 \pi^{\alpha+1}\Gamma(J+1)}{4^J\es \Gamma(J+\alpha+1)} \int_{0}^{\infty}\!\frac{\ud \imd}{2 \pi i} \, \imd^{J-1}\,  \tFo{J+\alpha + \frac{1}{2}}{J+1}{2 J + 2 \alpha + 1}{\minus \imd} \lab{disc}\, G^{\sigma} \mathcal{G}(\imd)\,. \label{eq:bubbleLinv}
	\end{equation}
	Using both (\ref{eq:propAsymp}) and (\ref{eq:vertexAboveBelow}), the discontinuity can be written as a sum of two components,
	\begin{equation}
		\lab{disc}\, G^\sigma \mathcal{G}(\zeta) = \minus 2 i\es\es  \zeta^{\beta + \Delta_\sigma}\es\es \mathcal{F}_{\Delta_\sigma}(\zeta) \sin \Phi_{\Delta_\sigma}(\zeta) + (\Delta_\sigma \to \bar{\Delta}_\sigma) \label{eq:vertexBubDisc}
	\end{equation}
	where we define the amplitude
	\begin{equation}
		\mathcal{F}_{\Delta_\sigma}(\zeta) = \mathcal{A}(\Delta_\sigma) \, \tFo{\Delta}{\Delta - \alpha + \frac{1}{2}}{2 \Delta - 2 \alpha + 1}{\minus \zeta} \e^{\sminus \mathcal{F}(\zeta)}
	\end{equation}
	and phase
	\begin{equation}
		\Phi_{\Delta_\sigma}(\zeta) = \pi \Delta_\sigma + \pi \beta \Phi(\zeta)
	\end{equation}
	in terms of $\mathcal{F}(\zeta)$, $\Phi(\zeta)$ and $\mathcal{A}(\Delta)$ given by (\ref{eq:fDef}), (\ref{eq:phiDef}), and (\ref{eq:propCoeff}), respectively. Since at worst the hypergeometric function ${}_2 F_1(a, b, c;\minus \zeta)$ grows polynomially as $\zeta \to \infty$ and  $\mathcal{F}_{\Delta_\sigma}(\zeta)$ is exponentially suppressed as $\zeta \to \infty$ for any finite $\beta > 0$, (\ref{eq:bubbleLinv}) is finite in the ultraviolet~$\zeta \to \infty$. Again, even though the vertex propagator is extraordinarily singular in the coincident limit~$\xi \to 1^\subm$ (or equivalently $\zeta \to \minus \infty$), loop corrections involving it are ultraviolet finite since it is extremely well-behaved as we approach the coincident limit along time-like separations $\xi \to 1^\subp$. Furthermore, if we series expand the inversion formula's (\ref{eq:bubbleLinv}) integrand around $\zeta = 0$, using~(\ref{eq:vertexDiscExp}) we find that $[G^\sigma \mathcal{G}]_{J}$ can have poles at $J = -(\beta + 2 n \alpha + \Delta_\sigma + k)$ and $J = \minus(\beta + 2 n \alpha + \bar{\Delta}_\sigma + k)$ for non-negative integers $n$ and~$k$, though many will be spurious.

	\begin{figure}
		\centering
		\includegraphics{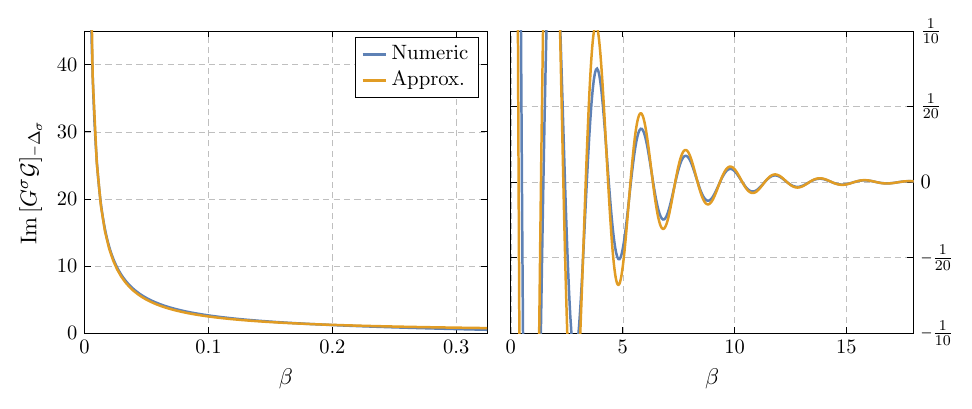}
		\caption{The imaginary part the bubble $\lab{Im}\, [G^\sigma \mathcal{G}]_{J}$ evaluated at $J = \minus\Delta_\sigma$, $\alpha = \frac{3}{2}(1 - 10^{\sminus 4})$, and $\nu_\sigma = 2$, as a function of $\beta$. For small $\beta$ [\textbf{left}], we compare (\ref{eq:bubbleLinv}) integrated numerically [{\color{Mathematica1}\textbf{blue}}] to the approximation found by truncating (\ref{eq:smallBetaApprox}) to its first term [{\color{Mathematica2}\textbf{yellow}}]. For large $\beta$ [\textbf{right}], we compare the numeric result [{\color{Mathematica1}\textbf{blue}}] to the saddle-point approximation (\ref{eq:largeBetaApprox}) [{\color{Mathematica2}\textbf{yellow}}]. In both cases, we find excellent agreement in the ranges of $\beta$ where our approximations should apply. Qualitatively, as $\nu_\sigma \to \infty$ the amplitude of $\lab{Im}\, [G^{\sigma} \mathcal{G}]_{\sminus \Delta_\sigma}$ shrinks and it shifts slightly to~the~left. \label{fig:compare}}
	\end{figure}

	At this order in perturbation theory, a non-zero potential for $\varphi$, or equivalently a non-zero vacuum expectation value for $\langle\mathcal{V}(x)\rangle_{\lab{c}} \propto g_\varphi$, does not impact $\sigma$'s self-energy. We will thus assume, for simplicity, that $\varphi$ remains massless in the presence of interactions $\langle \mathcal{V}(x) \rangle = 0$. Since it is the only free parameter that affects predictions, we thus want to quantify how $\lab{Im}\, \Pi_\sigma(\minus \Delta_\sigma)$ changes as a function of $\beta$ or equivalently $f$. Of course, the integral (\ref{eq:vertexBubDisc}) converges for $J = \minus\Delta_\sigma$ and so~$\Pi_\sigma(\minus \Delta_\sigma)$ can be easily numerically evaluated for arbitrary $\beta$, as we show in Figure~\ref{fig:compare}. We find two qualitatively different behaviors, depending on whether the field space is much larger ($\beta \ll 1$) or much smaller ($\beta \gg 1$) than the Hubble scale. We analyze both in the following sections.

\subsection{The Noncompact Limit} \label{sec:noncompactLimit}

	In the noncompact limit $\beta \to 0$, where the physical circumference of $\varphi$'s field space~$2 \pi f$ becomes much larger than the Hubble scale, the fluctuations of the massless compact scalar field~$\varphi$ become highly suppressed and so long-distance correlations $\mathcal{G}(\xi) \sim (\minus 2/\xi)^{\beta}$ between vertex operators decay very slowly. This agrees with our intuition that, in the limit $f \to \infty$, a massless compact scalar should qualitatively behave like a light noncompact scalar whose long-distance correlations also decay very slowly $\propto (\minus 2/\xi)^{\Delta_\varphi}$, with $\Delta_\varphi = \alpha - \sqrt{\alpha^2 - m_{\smash{\varphi}}^2}$. Furthermore, in this limit, we will show that a single pole at $J = \minus (\beta + \Delta_\sigma)$ encroaches upon $J = \minus \Delta_\sigma$ and dominates $\sigma$'s self-energy. This is also directly analogous to what happens when a light \emph{noncompact} scalar interacts with $\sigma$~\cite{Chakraborty:2023qbp} in the limit that its mass goes to zero. We can thus import the techniques used to study that case to approximate $\lab{Im}\, \Pi_\sigma(\minus \Delta_\sigma)$ as $\beta \to 0$.

	We will use the same strategy~\cite{Chakraborty:2023qbp} used in \S\ref{sec:vertexPropMom} to extract the singularities of $[\mathcal{G}]_J$. We first divide the inversion integral (\ref{eq:bubbleLinv}) into an integral over $\zeta \in [0, 1]$ and one over $\zeta \in [1, \infty)$. Since the latter integral is regular in $J$, we may extract $[G^\sigma \mathcal{G}]_J$'s singularities by series expanding the integrand of the former and integrating term-by-term, 
	\begin{equation}
		\begin{aligned}
			[G^\sigma \mathcal{G}]_J &\approx \frac{1}{2(\alpha - \Delta_\sigma)} \frac{1}{J + \Delta_\sigma + \beta} + \frac{\alpha  \beta  (2 \alpha -\Delta_\sigma -2)}{2 (\alpha -1) (\Delta_\sigma +1) (\alpha -\Delta_\sigma -1)} \frac{1}{J + \Delta_\sigma + \beta + 2} \\
			&+\frac{4^{\alpha -1} \beta  }{\sqrt{\pi }} \G{\Delta_\sigma\,,\, \alpha - \Delta_\sigma\,,\, \alpha + \frac{1}{2} \,,\, \minus \alpha}{1 - \alpha - \Delta_\sigma\,,\, 2\alpha + \Delta_\sigma}\frac{1}{J + \Delta_\sigma + \beta + 2 \alpha} + \cdots\,, 
		\end{aligned} \label{eq:smallBetaApprox}
	\end{equation}
	where we have approximated each of the residues in the limit $\beta \to 0$. We find, near $J= \minus \Delta_\sigma$, the first pole dominates while the contributions from the others are suppressed by a factor of $\beta^2$. 

	Near the free-field pole $J = \minus \Delta_\sigma$ and as $\beta \to 0$, the self-energy is thus well-approximated by the closest singularity 
	\begin{equation}
		\Pi_\sigma(J) \approx \frac{i |g|^2}{\nu_\sigma} \frac{1}{J+\Delta_\sigma + \beta}\,. \label{eq:noncompactCompactSelfEnergy}
	\end{equation}
	As shown in the left panel of Figure~\ref{fig:compare}, we find excellent agreement between this approximation and the direct numerical evaluation of (\ref{eq:bubbleLinv}) in the limit $\beta \to 0$. The free-field pole is then shifted to
	\begin{equation}
		J_* \approx -\!\left[\alpha + \frac{|g|^2}{2 \beta (m_\sigma^2 - \alpha^2)}\right] - i \sqrt{m_\sigma^2 - \alpha^2}\,, 
	\end{equation}
	having used the mass counterterm $\delta_{m_\sigma}$ to set $\lab{Re}\, \Pi_\sigma(\minus \Delta_\sigma) = 0$.  This correction is always positive and so~$\sigma$'s interaction with $\varphi$ will cause it to decay \emph{faster} than if it were just a free field. Furthermore, we find that we must make the fields more and more weakly coupled $|g|^2 \to 0$ as we send $\beta \to 0$ if we want to retain control over our perturbative expansion. This is to be expected since light noncompact scalars are infamously plagued with these types of divergences and so we should not be able to send $\beta \to 0$ without something drastic happening.

	However, the scalar's compact field space still improves the calculability of this model, even in this noncompact limit. We can contrast this result with that of a light noncompact scalar field with mass $m_\varphi$, which interacts with the heavy scalar $\sigma$ via the quartic interaction~$\frac{1}{2} g \sigma^2 \varphi^2$~\cite{Lu:2021wxu,Chakraborty:2023qbp}. In the limit $m_\varphi \to 0$ or equivalently $\Delta_\varphi \to 0$, it was found that the self-energy of $\sigma$ near the free-field pole $J = \minus \Delta_\sigma$ is  well-approximated by
	\begin{eqnarray}
		\Pi_{\sigma}(J) &\propto&\frac{g^2}{\Delta_\varphi^2 \nu_\sigma}\frac{1}{J+\Delta_\sigma + 2\Delta_\varphi}\,.
	\end{eqnarray}
	As discussed in \cite{Chakraborty:2023qbp}, there are two sources of infrared enhancement as $\Delta_\varphi \to 0$. The first is the prefactor $\propto \Delta_\varphi^{-2}$, which can be attributed to the Euclidean zero mode of the noncompact $\varphi$, whose variance diverges in the massless limit (\ref{eq:noncompactVar}). The other is due to the Lorentzian long-distance behavior of $\varphi$. Since $\varphi$'s propagator at long distances behaves as $G^\varphi(\xi) \sim (\minus \xi)^{\sminus \Delta_\varphi}+\cdots$, infrared fluctuations are enhanced as $\Delta_\varphi \to 0$, and this is captured by the appearance of an additional~$\Delta_\varphi$ enhancement of the self-energy at $J= \minus \Delta_\sigma$. As emphasized in \cite{Chakraborty:2023qbp}, the diverging variance of the Euclidean zero mode causes a loss in the perturbativity of the model, as contributions which are nominally suppressed by the coupling $g$ can be greatly enhanced by fluctuations of the zero mode, making the model strongly-coupled. This led to the requirement that the theory had to become much more weakly coupled as $m_\varphi \to 0$, $g \ll \frac{128}{27} \pi^4 m_\varphi^4$, to maintain perturbative control.

	The compact scalar is different because its zero mode always has finite variance. We find that the self-energy in the noncompact limit (\ref{eq:noncompactCompactSelfEnergy}) only sees an enhancement around the free field pole $J = \minus \Delta_\sigma$ from the long-distance behavior of the vertex propagator, $\mathcal{G}(\xi) \propto \mathcal{C} (\minus \xi)^{\sminus \beta}$. Higher order contributions will not be enhanced by wild fluctuations of the zero mode and so the compact scalar provides a \emph{large} and \emph{calculable} correction to the long-distance behavior of $\sigma$---we only require that $g \ll 1$.\footnote{If we work with the canonically normalized field $\theta = f \varphi$, interactions in $\theta$ are always dressed by inverse powers of $f$, i.e. $\tfrac{1}{2} g \sigma^2 \cos (\theta/f) \supset -\frac{1}{4} (g/f^2) \sigma^2 \theta^2 + \cdots$ and so we can also interpret $f \to \infty$ with $g$ held fixed as a weak coupling limit. }

\subsection{The Ultracompact Limit} \label{sec:ultracompactLimit}

	From Figure~\ref{fig:compare}, we can see that the ultracompact limit $\beta \to \infty$, in which the physical circumference of the field space becomes much smaller than Hubble, leads to qualitatively different behavior than the noncompact limit. As evident from that figure, the approximation we derive in the~$\beta \to \infty$ limits is qualitatively accurate even for $\beta \sim \mathcal{O}(1)$. We may write (\ref{eq:bubbleLinv}) in a form that can be readily approximated via the saddle-point method as
	\begin{equation}
		\begin{aligned}
			[G^{\sigma} \mathcal{G}]_{J} &= \mathcal{N}_{J, \Delta_\sigma} \int_{0}^{\infty}\!\ud \zeta\, \mathcal{H}_{J, \Delta_\sigma}(\zeta) \left[\e^{i \pi \Delta_\sigma} \e^{\sminus \beta\tilde{\mathcal{F}}_+(\zeta)} - \e^{\sminus i \pi \Delta_\sigma} \e^{\sminus\beta\tilde{\mathcal{F}}_-(\zeta)}\right] + (\Delta_\sigma \to \bar{\Delta}_\sigma)
		\end{aligned} \label{eq:saddleBubble}
	\end{equation}
	where we define the exponent, coefficient, and smoothly-varying amplitude
	\begin{equation}
		\begin{aligned}
			\tilde{\mathcal{F}}_\pm(\zeta) &=   \mathcal{F}(\zeta) \pm i \pi \Phi(\zeta) -\log \zeta \\ 
			\mathcal{N}_{J, \Delta} &= \minus \frac{i}{ \pi 4^{J+\Delta+1}} \G{J+1\,,\, \Delta \,,\,\alpha - \Delta}{J+\alpha +1}\\
			\mathcal{H}_{J, \Delta}(\zeta) &= \zeta^{J+\Delta - 1} \tFo{J+1}{J+\alpha +\frac{1}{2}}{2J+2\alpha+1}{\minus \zeta} \tFo{\Delta}{\Delta-\alpha + \frac{1}{2}}{2 \Delta - 2 \alpha +1}{\minus \zeta}\,,
		\end{aligned}
	\end{equation}
	respectively. 
	In the limit $\beta \to \infty$ with $J$ fixed, each of these integrals is dominated by a saddle point at $\zeta = \zeta_{\pm}$, defined by the equations $\tilde{\mathcal{F}}'(\zeta_\pm) = 0$ or, explicitly,
	\begin{equation}
		-\frac{2}{\zeta_\pm} + \frac{ 2^{2 - 2\alpha} \pi^{3/2} \zeta_\pm^{2\alpha -1} (\mp i +  \cot 2 \pi \alpha)}{(1 + \zeta_\pm)^{\alpha + \frac{1}{2}} \Gamma\big(\frac{1}{2} - \alpha\big)\Gamma(\alpha)} + \tFo{1}{\frac{3}{2} - \alpha}{2 - 2 \alpha}{\minus \zeta_\pm} = 0\,.
	\end{equation}
	We are not able to solve this equation exactly for arbitrary $\alpha$, but in the limit $\alpha \to \frac{3}{2}$ both roots converge to $\zeta_\pm \to \zeta_* = 2$ and
	\begin{equation}
		\tilde{\mathcal{F}}_\pm(\zeta_\pm) = 1 - \log 2 \pm i \pi \quad \text{with} \quad \mathcal{F}_\pm''(\zeta_\pm) = \frac{1}{4}\,.
	\end{equation}
	Defining the constant $\tilde{\mathcal{F}}_0 \equiv \lab{Re}\, \tilde{\mathcal{F}}_\pm(\zeta_\pm) = 1 - \log 2 \approx 0.3069$, (\ref{eq:saddleBubble}) is well-approximated by
	\begin{equation}
		[G^\sigma \mathcal{G}]_J \sim \sqrt{ \frac{8 \pi}{\beta}} \mathcal{N}_{J, \Delta_\sigma} H_{J, \Delta_\sigma}(\zeta_*) \,\e^{\sminus \beta \tilde{\mathcal{F}}_0} \sin \pi(\beta + \Delta_\sigma) + (\Delta_\sigma \to \bar{\Delta}_\sigma)\,, \label{eq:largeBetaApprox}
	\end{equation}
	which, as shown in Figure~\ref{fig:compare}, shows excellent agreement with the numerical evaluation of (\ref{eq:bubbleLinv}) for large $\beta$. We thus find that, as $\beta \to \infty$ and for fixed $J$, the imaginary part of the self-energy is well-approximated by an exponentially decaying sinusoid with decay rate $\tilde{\mathcal{F}}_0$ and frequency $\pi$, whose amplitude and phase are set by $J$ and $\Delta_\sigma$. Intuitively, as $\beta \to \infty$ or $f \to 0$, $\varphi$ can relax very quickly back to its vacuum and so vertex correlations die very quickly---the vertex operator is thus very ``heavy,'' and the self-energy it induces becomes very small. Similarly, we find that these corrections also die off as $\sigma$ becomes heavier, $\nu_\sigma \to \infty$. 

	In this limit, $\sigma$'s anomalous dimension
	\begin{equation}
		\gamma_\sigma \equiv \frac{1}{2 \nu_\sigma} \lab{Im}\, \Pi_\sigma(\minus \Delta_\sigma) \approx  \frac{|g|^2}{\nu_\sigma}\es\es \lab{Im}\, [G^\sigma \mathcal{G}]_{\sminus \Delta_\sigma}\,, \label{eq:anomDim}
	\end{equation}
	is no longer necessarily positive, but oscillates between positive and negative values as we increase~$\beta$. In contrast to its behavior in the presence of a light noncompact field or a light compact field with a large field space, the heavy field $\sigma$ may decay \emph{faster}, \emph{slower}, or at the same rate when interacting with a very compact massless scalar field $\varphi$. Given that this depends on the ratio of the circumference of the field space to the Hubble scale, we interpret this as a sort of resonance or interference effect: for some ranges of $H/(2 \pi f)$, the ``thermally'' fluctuating compact scalar can \emph{add} energy to $\sigma$, while for other ranges it removes it. It would be interesting to have a better physical explanation of this phenomenon.

	\begin{figure}
		\centering
		\includegraphics{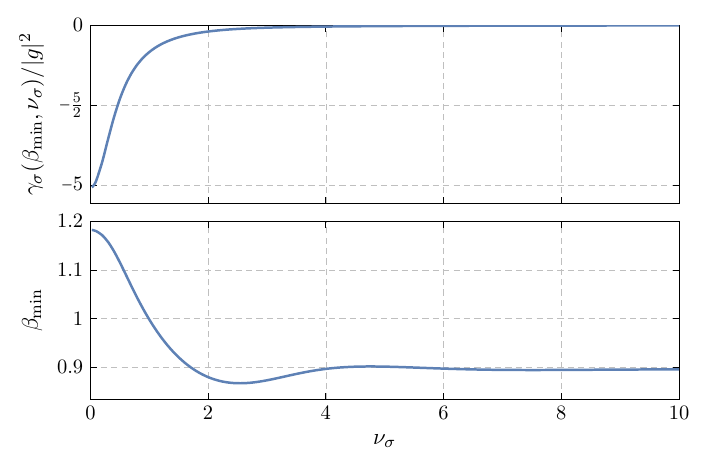}
		\caption{The minimum value [\textbf{top}] of the anomalous dimension (\ref{eq:anomDim}) in ``units'' of $|g|^2$, and the corresponding $\beta_\lab{min}$ [\textbf{bottom}], as functions of $\nu_\sigma$. This minimum anomalous dimension is bounded from below and quickly approaches zero as $\nu_\sigma \to \infty$.   \label{fig:minImag}}
	\end{figure}

	It is interesting to understand how negative this anomalous dimension can be. For each $\nu_\sigma$, the anomalous dimension $\gamma_\sigma$ is minimized at a different value of $\beta$, $\lab{min}_{\beta} \gamma_\sigma(\beta, \nu_\sigma) = \gamma_\sigma(\beta_\lab{min}, \nu_\sigma)$. In Figure~\ref{fig:minImag}, we plot both this minimum value $\gamma_\sigma(\beta_\lab{min}, \nu_\sigma)$ and the corresponding $\beta_\lab{min}$ as functions of~$\nu_\sigma$. As we should expect, we find that the anomalous dimension can be most negative for light~$\sigma$, approaching $\gamma_\sigma(\beta_\lab{min}, \nu_\sigma) \approx -5.07 |g|^2$ as $\nu_\sigma = \sqrt{m_\sigma^2 - \alpha^2} \to 0$. As we make $\sigma$ heavier,~$\nu_\sigma \to \infty$, the self-energy is suppressed for all $\beta$ and correspondingly this minimum anomalous dimension quickly approaches zero from below.\footnote{It would be interesting to understand precisely how this negative anomalous dimension fits with the positivity bounds of~\cite{Green:2023ids}, which argued from positivity of classical statistics that anomalous dimensions of \emph{heavy} fields must always be positive. Strictly speaking, this bound on $\gamma_\sigma$ only holds in the limit $\nu_\sigma \to \infty$, and any violation of unitarity may be avoided via a non-zero contribution to the co-moving curvature perturbation's trispectrum. As such, their bounds still allow for $\gamma_\sigma < 0$ for $\nu_\sigma \sim \mathcal{O}(1)$.  Since the minimum anomalous dimension induced by a compact scalar (\ref{eq:anomDim}) vanishes as $\nu_\sigma \to \infty$, it seems consistent with their bounds but we would need to compute its contribution to the trispectrum to be sure. We save this for future work. }

\section{Conclusions} \label{sec:conclusions}

	The compact scalar field is a fundamentally different object from the noncompact scalar field. Often, these differences can be attributed to how we treat the field's spatially isotropic zero mode~$\varphi_0$. For the compact scalar, the gauge symmetry $\varphi(x) \sim \varphi(x) +2\pi$ forces the zero mode to live in the finite interval $\varphi_0 \in [0, 2\pi)$, while the noncompact field sees no such restriction and its zero mode is allowed to fluctuated over the entire real line $\varphi_0 \in \mathbb{R}$. While the distinction between the two is usually unimportant in Minkowski space, since there this zero mode is non-dynamical, we have argued that the situation is dramatically different in de Sitter space---the zero mode is dynamically active and so there are observables that are sensitive to the topology of field space.

	We argued that the largest differences between compact and noncompact scalars arise when they are light, as then they can more easily explore their field spaces. We argued that these differences are most extreme in the massless limit: the massless compact scalar field exists in de Sitter space while the noncompact scalar field does not. By careful analytic continuation from Euclidean de Sitter space to Lorentzian signature, we computed the gauge-invariant $n$-point functions of vertex operators of the free massless compact scalar field and showed that they are both de-Sitter invariant and well-behaved, decaying appropriately at infinity. Furthermore, we computed the vertex two-point function's momentum space representation and showed how it accurately encodes the propagator's long-distance behavior.

	To demonstrate that differences in field space topology can be observable, we determined how a light compact scalar $\varphi$ changes the long-distance behavior of a heavy scalar $\sigma$. During inflation, a spontaneously produced $\sigma$ can decay into scalar gravitational fluctuations and impart a unique oscillatory signature onto the squeezed limit of the primordial bispectrum, commonly called the cosmological collider signal. This signal primarily depends on how the heavy particle $\sigma$ propagates over long distances. We showed that there can be qualitative differences between how compact and noncompact light scalars affect $\sigma$'s long-distance behavior, and so $\sigma$'s cosmological collider signal can serve as a concrete detection channel for the topology of $\varphi$'s field~space.

	Specifically, we showed that a light compact field's impact on a heavy field $\sigma$'s long-distance behavior primarily depends on how large its field space is compared to the Hubble scale. In the noncompact limit $2 \pi f \gg H$, $\varphi$ behaves much like a noncompact field does and always enables $\sigma$ to decay faster than any free field. When the field space becomes smaller than Hubble, however, the compact $\varphi$ can behave in a qualitatively different way compared to a noncompact field, as there are some ranges of $2 \pi f/H$ in which a weakly coupled compact field can cause $\sigma$ to decay \emph{slower} than any free field. Such a thing is impossible for a weakly coupled noncompact field.

	The goal of this work was to initiate the study of compact scalar fields in de Sitter space, and to establish that compact and noncompact fields can have qualitatively distinct physics in inflationary backgrounds. Said simply, the topology of field space matters in de Sitter space and so it must be accounted for. We find that these differences are the most stark for light fields. Since the existence of such light compact fields is theoretically well-motivated, there are many future research directions worth pursuing. For instance,
	\begin{itemize}
		\item Are there any other inflationary detection channels that are sensitive to field space topology? We focused only on interacting two-point functions---how do higher-point functions behave? And how do our predictions change for more realistic axion models? 

		\item Is it possible to recover our results by working in another cosmological patch? How do we correctly impose the gauge symmetry $\varphi(x) \sim \varphi(x) + 2 \pi$ in the Poincar\'{e} patch? 

		\item Our analysis of the interacting theory relies on an approximate summation of higher-order contributions that we expect to only be valid in the long-distance limit. It would be useful to develop a more systematic framework to compute the self-energies of $\sigma$ and $\varphi$.

		\item We found that the compact scalar field exhibited strange behavior at discrete values of the axion decay constant, $({H}/{2 \pi f})^2 \in 2\mathbb{Z}_{\scriptscriptstyle > 0}$. For example, there correlation functions of time- and anti-time-ordered vertex operators were equivalent. What is special about these values of $\beta$? Are they associated with other interesting physics?

	\end{itemize}
	We hope to pursue some of these questions in future work.

\vspace{14pt}
\newpage
\noindent \textbf{Acknowledgements}

\noindent We would like to thank Arindam Bhattacharya, Dan Green, Cody Long, Manuel Loparco, Qianshu Lu, Austin Joyce, Rashmish Mishra, Matt Reece, and Zhong-Zhi Xianyu for very helpful discussions. We would especially like to thank Matt Reece for comments on a draft. The work of both PC and JS is supported by the DOE grant \texttt{DE-SC0013607}, while the work of JS is also supported NASA Grant \texttt{80NSSC20K0506}. This work was performed in part at Aspen Center for Physics, which is supported by National Science Foundation grant \texttt{PHY-2210452}.

\appendix

\section{Formulary} \label{app:formulary}	

	In this Appendix, we present the definitions of the various special functions and useful formulae commonly used throughout the main text.

	Some of our expressions will contain large ratios and products of gamma functions so it will be convenient to adopt the common shorthand
	\begin{equation}
		\G{a, b, c, \cdots}{d, e, f, \cdots} = \frac{\Gamma(a)\Gamma(b)\Gamma(c)\cdots}{\Gamma(d)\Gamma(e)\Gamma(f)\cdots} \quad \text{and} \quad \Gamma[a, b, c, \cdots] = \Gamma(a)\Gamma(b)\Gamma(c)\cdots\,.
	\end{equation}
	The $n$'th harmonic number $H_{n} \equiv \sum_{k = 1}^{n} 1/k$ for $n \in \mathbb{Z}_{\scriptscriptstyle > 0}$ can be analytically continued for arbitrary arguments by the digamma function $\psi(z)$,
	\begin{equation}
		H_{z} = \psi(z - 1) - \gamma_\slab{e}\,, \label{eq:harmonicNumber}
	\end{equation} 
	where $\psi(z) = \frac{\ud}{\ud z} \log \Gamma(z)$ and $\gamma_\slab{e} \equiv \minus \psi(1) \approx = \minus 0.577216\ldots$ is the Euler-Mascheroni constant.

	The Gaussian or ordinary hypergeometric function is defined in the disk $|z| < 1$ by the series
	\begin{equation}
		{}_2 F_1(a, b; c; z) \equiv \tFo{a}{b}{c}{z} = \frac{\Gamma(c)}{\Gamma(a)\Gamma(b)} \sum_{n = 0}^{\infty} \frac{\Gamma(a+n)\Gamma(b+n)}{\Gamma(c+n)} \frac{z^n}{n!}\,. \label{eq:hypSum}
	\end{equation}
	It obeys the following connection formula when $a - b \notin \mathbb{Z}$,
	\begin{equation}
		\begin{aligned}
			\tFo{a}{b}{c}{1 + \frac{1}{\zeta}} &= \G{c, b-a}{b, c-a} \left(\minus \frac{1}{\zeta}\right)^{\sminus a}\!\! \tFo{a}{c-b}{1+a-b}{\minus \zeta} + (a \leftrightarrow b)\,,
		\end{aligned} \label{eq:hypConnectionFormula}
	\end{equation}
	which can be used to compute the discontinuity of functions involving (\ref{eq:hypSum}) across its branch cut along $z \in [1,\infty)$. Similarly, the generalized hypergeometric function ${}_3F_2(a,b,c;d,e;z)$ can be defined inside the unit disk $|z| < 1$,
	\begin{equation}
		\pFq{3}{2}{a\,,\,b\,,\,c}{d\,,\,e}{z} \equiv \G{d\,,\,e}{a\,,\,b,\,\,c}\sum_{n = 0}^{\infty} \G{a+n\,,\,b+n\,,\,c+n}{d+n\,,\,e+n} \frac{z^n}{n!}\,. \label{eq:hyp3F2}
	\end{equation}
	This generalized hypergeometric function can be related to derivatives of ${}_2 F_1(a, b;c, z)$ with respect to its parameters. For instance, the identity
	\begin{equation}
		{}_2 F_1(a, b;c;z) = 1 + \frac{a b}{c}\, z \,  \pFq{3}{2}{1,\, 1,\, b + 1}{2,\, c+1}{z} + \mathcal{O}\big(a^2\big) \label{eq:hypDer}
	\end{equation}
	will be particularly useful.

	The Gegenbauer-$C$ function~\cite{Durand:1976efa} is defined as the solution to the differential equation
	\begin{equation}
		\left[(\emd^2 - 1) \frac{\ud^2}{\ud \emd^2} + (2 J + 1) \, \emd \, \frac{\ud}{\ud \emd} - J(J+2 \alpha)\right]\! C_{J}^{\alpha}(\emd) = 0\,\label{eq:gegDeq}
	\end{equation}
	that is regular around $\xi = 1$. It can be expressed in terms of hypergeometric function
	\begin{equation}
		C_{J}^{\alpha}(\emd) \equiv \frac{\Gamma(J+2\alpha)}{\Gamma(J+1)\Gamma(2\alpha)}\, \tFo{\minus J}{J+2 \alpha}{\alpha + \tfrac{1}{2}}{ \frac{1 - \emd}{2}}\,.\label{eq:gegC}
	\end{equation}
	These functions obey the useful recursion relation
	\begin{equation}
		(J + \alpha) C_{J}^{\alpha}(\minus \xi) = \alpha \! \left[C_{J}^{\alpha+1}(\minus \xi) -C_{J - 2}^{\alpha+1} (\minus \xi)\right], \label{eq:gegRecur}
	\end{equation}
	as do any appropriately normalized solution to (\ref{eq:gegDeq}). They also reduce to the Gegenbauer polynomials for non-negative integer $J$. As $|\emd|\to \infty$, they behave as
	\begin{equation}
		C_{J}^{\alpha}(\emd) \sim \G{J+\alpha}{\alpha\,,\,J+1} (2\es  \emd)^{J} + \G{J+2 \alpha\,,\, -(J+\alpha)}{\alpha\,,\,\minus J\,,\, J+1} (2 \emd)^{-(J+2\alpha)}\,, \label{eq:gegAsymp}
	\end{equation}
	and thus contain both of (\ref{eq:gegDeq})'s asymptotic behaviors when $J \notin \mathbb{Z}_{\scriptscriptstyle > 0}$. 
	At the endpoints of the interval $\xi \in [\minus 1, 1]$, they simplify to
	\begin{equation}
		C_{J}^{\alpha}(1) = \frac{\Gamma(J+2 \alpha)}{\Gamma(J+1) \Gamma(2 \alpha)} \qquad \text{and} \qquad C_{J}^{\alpha}(\minus 1) = \frac{\cos \pi(J+\alpha)}{\cos \pi \alpha} \frac{\Gamma(J+2 \alpha)}{\Gamma(J+1) \Gamma(2 \alpha)}\,, \label{eq:gegEndpoints}
	\end{equation}
	which both grow as $J^{2\alpha - 1}$ for positive real $J$, though $C_{J}^{\alpha}(\minus 1)$ grows exponentially as $\lab{Im}\, J \to \pm \infty$.

	The other linearly independent solution to (\ref{eq:gegDeq}) is the Gegenbauer-$Q$ function, given by
	\begin{equation}
		Q_J^{\alpha}(\emd) \equiv \frac{2^{1 - J - 2 \alpha} \pi \Gamma(J+ 2 \alpha)}{\Gamma(\alpha) \Gamma(J+\alpha + 1)} (\emd - 1)^{-J - 2 \alpha} \tFo{J+\alpha + \frac{1}{2}}{J+2 \alpha}{2 J + 2 \alpha + 1}{\frac{2}{1 - \emd}}\,. \label{eq:gegQ}
	\end{equation}
	Crucially, this solution decays like $\xi^{-J-2\alpha}$ as $|\xi| \to \infty$, a fact that allows the Lorentzian inversion formulae~(\ref{eq:lorentzianInversion}) and (\ref{eq:lInvForm}) to work.

\newpage
\section{Interacting Vertex Two-Point Function} \label{app:vertexTwo}
	
	In Section~\ref{sec:compact_int}, we computed the long-distance behavior of $\sigma$'s two-point function~$\langle \sigma(x) \sigma(y) \rangle_\lab{c}$ by first explicitly expanding the path integral out order-by-order in interactions and then identifying the infinite class of contributions relevant to this limit. There, it was not possible to identify a class of one-particle-irreducible diagrams that could be chained together and readily summed, as our interactions were mediated by vertex operators that do not Wick factorize like free Gaussian fields. However, we argued that at long distances certain higher-order corrections could be well-approximated as an effective chain of diagrams, and so that it was possible to Dyson resum the leading-order correction to $\sigma$'s two-point function to determine its corrected long-distance behavior. In this Appendix, we do the same for the connected vertex two-point function $\langle \mathcal{V}(x) \mathcal{V}^\dagger (y)\rangle_\lab{c}$ and determine how interactions with the heavy scalar $\sigma$ affect its long-distance behavior.

	For the sake of brevity, we will ignore the counterterm interactions in intermediate steps and instead only include them in final expressions. To leading order, the vertex two-point function is
	\begin{equation}
		\langle \mathcal{V}(x)\mathcal{V}^\dagger(y) \rangle_{\rm c} = \langle \mathcal{V}(x)\mathcal{V}^\dagger(y) \rangle+\tfrac{1}{2}\big[ \langle \mathcal{V}(x)\mathcal{V}^\dagger(y)S_{\rm int}^2 \rangle-\langle \mathcal{V}(x)\mathcal{V}^\dagger(y) \rangle \langle S_{\rm int}^2 \rangle \big]-\big|\langle \mathcal{V}(x) S_\lab{int}\rangle\big|^2+\cdots\,,
	\end{equation}
	or, by using (\ref{eq:vertexFour}),
	\begin{equation}
		\begin{aligned}
			&\langle \mathcal{V}(x)\mathcal{V}^\dagger(y) \rangle_{\rm c}
			=\mathcal{G}(x,y) + \tfrac{1}{2}|g|^2 \int_{\{2\}}\! (G^\sigma_{12})^2\! \left[\frac{\mathcal{G}_{xy}\mathcal{G}_{x1} \mathcal{G}_{12} \mathcal{G}_{2y}}{\mathcal{G}_{x2}\mathcal{G}_{1y}}-\mathcal{G}_{xy}\mathcal{G}_{12}\right] \label{eq:vert_leading_ord} \\
			&\qquad +\big|g_\varphi+\tfrac{1}{2}g \es G^\sigma(1)\big|^2 \int_{\{2\}} \!\left[\frac{\mathcal{G}_{xy}\mathcal{G}_{x1} \mathcal{G}_{12} \mathcal{G}_{2y}}{\mathcal{G}_{x2}\mathcal{G}_{1y}}-\mathcal{G}_{xy}\mathcal{G}_{12}\right] -|g_\varphi+\tfrac{1}{2}g\es G^\sigma(1)|^2 [\mathcal{G}]_0^2 + \cdots \,,
		\end{aligned}
	\end{equation}
	where we have isolated terms related to the one-point function (\ref{eq:tadpole}) on the second line. Even at leading order in perturbation theory, we can see the trouble caused by the fact that free vertex correlators do not Wick factorize since the ``disconnected'' contribution in $\langle \mathcal{V}(x) \mathcal{V}^\dagger(y) S_\lab{int}^2 \rangle$ is no longer canceled by $\langle \mathcal{V}(x) \mathcal{V}^\dagger(y)\rangle \langle S_\lab{int}^2 \rangle$. However, if we are interested in the behavior of this correction at long distances, or analogously around $J \approx \minus \beta$, we can follow the discussion of Section~\ref{sec:compact_int} and analytically continue (\ref{eq:vert_leading_ord}) into an in-in contribution in which the integration variables $z_1$ and $z_2$ span the entire Poincar\'{e} patch. 

	In the long-distance limit $|\xi_{xy}| \to \infty$, there are two relevant regions of integration. The first is when $z_1$ and $z_2$ are close to one another yet far from $x$ and $y$, such that $|\xi_{12}| \ll |\xi_{x1}|, |\xi_{2y}|, \ldots, |\xi_{xy}|$. In this region, the vertex four-point function factorizes into two disconnected pieces,
	\begin{equation}
		 \frac{\mathcal{G}_{xy}\mathcal{G}_{x1} \mathcal{G}_{12} \mathcal{G}_{2y}}{\mathcal{G}_{x2}\mathcal{G}_{1y}} = \begin{tikzpicture}[thick,baseline=-3pt]
				\draw[cornellBlue, vertexProp] (-1., 0) -- (-0.3333, 0);
				\draw[cornellBlue, vertexProp] (0.33333, 0) -- (1., 0) ;
				
				\draw[cornellBlue, vertexProp] (-1, 0) arc(180:0:1 and 0.8);
				\draw[cornellRed, vertexProp] (-1, 0) arc(180:360:0.6666 and 0.5);
				\draw[cornellRed, vertexProp] (-0.3333, 0) arc(180:360:0.6666 and 0.5);
				\draw[cornellBlue, vertexProp] (-0.333333, 0) -- (0.33333, 0);
				\draw[cornellBlue, fill=white] (-1., 0) circle (\extVert);
				\draw[cornellRed, fill=white] (1., 0) circle (\extVert);
				\fill[cornellRed, intVertSty] (-0.3333,0) circle (\intVert);
				\fill[cornellBlue, intVertSty] (0.3333,0) circle (\intVert);
			\end{tikzpicture} \, \approx \, \begin{tikzpicture}[baseline=-3pt, thick]
				\draw[cornellBlue, vertexProp] (-0.75, -0.25)--(0.75, -0.25);
				\draw[cornellBlue, vertexProp] (-0.75, 0.25)--(0.75, 0.25);
				\fill[cornellRed, intVertSty] (0.75, 0.25) circle (\intVert);
				\fill[cornellBlue, intVertSty] (-0.75 ,0.25) circle (\intVert);
				\draw[cornellBlue, fill=white] (-0.75, -0.25) circle (\extVert);
				\draw[cornellRed, fill=white] (0.75, -0.25) circle (\extVert);
			\end{tikzpicture}
			\,=\,  \mathcal{G}_{xy}\mathcal{G}_{12}\,,
	\end{equation}
	but we see from (\ref{eq:vert_leading_ord}) that the contribution from this region is canceled by the disconnected contributions. However, there is also the region of integration in which $\xi_{x1}$ and $\xi_{2y}$ are smaller than all other separations so that the vertex four-point function factorizes as
	\begin{equation}
		\frac{\mathcal{G}_{xy}\mathcal{G}_{x1} \mathcal{G}_{12} \mathcal{G}_{2y}}{\mathcal{G}_{x2}\mathcal{G}_{1y}} \, =\,  \begin{tikzpicture}[thick,baseline=-3pt]
				\draw[cornellBlue, vertexProp] (-1., 0) -- (-0.3333, 0);
				\draw[cornellBlue, vertexProp] (0.33333, 0) -- (1., 0) ;
				
				\draw[cornellBlue, vertexProp] (-1, 0) arc(180:0:1 and 0.8);
				\draw[cornellRed, vertexProp] (-1, 0) arc(180:360:0.6666 and 0.5);
				\draw[cornellRed, vertexProp] (-0.3333, 0) arc(180:360:0.6666 and 0.5);
				\draw[cornellBlue, vertexProp] (-0.33333, 0) -- (0.33333, 0);
				\draw[cornellBlue, fill=white] (-1., 0) circle (\extVert);
				\draw[cornellRed, fill=white] (1., 0) circle (\extVert);
				\fill[cornellRed, intVertSty] (-0.3333,0) circle (\intVert);
				\fill[cornellBlue, intVertSty] (0.3333,0) circle (\intVert);
			\end{tikzpicture} \, \approx \, \begin{tikzpicture}[thick, baseline=-3pt]
				\draw[cornellBlue, vertexProp] (-1., 0) -- (-0.3333, 0);
				\draw[cornellBlue, vertexProp] (0.3333, 0) -- (1., 0) ;
				\draw[cornellBlue, fill=white] (-1., 0) circle (\extVert);
				\draw[cornellRed, fill=white] (1., 0) circle (\extVert);
				\fill[cornellRed, intVertSty] (-0.3333,0) circle (0.08);
				\fill[cornellBlue, intVertSty] (0.3333,0) circle (0.08);
			\end{tikzpicture} \, =\,   \mathcal{G}_{x1}\mathcal{G}_{2y}\,,
	\end{equation}
	We find then that~(\ref{eq:vert_leading_ord}) can be approximated diagrammatically as
	\begin{equation}
			\langle \mathcal{V}(x) \mathcal{V}^\dagger(y) \rangle_{\rm c} = \begin{tikzpicture}[thick, baseline=-3pt]
			\draw[cornellBlue, vertexProp] (-1., 0) -- (1., 0)  ;
			\draw[cornellBlue, fill=white] (-1., 0) circle (\extVert);
			\draw[cornellRed, fill=white] (1., 0) circle (\extVert);
		\end{tikzpicture} \, + \, \begin{tikzpicture}[thick,baseline=-3pt]
				\draw[cornellBlue, vertexProp] (-1., 0) -- (-0.5, 0);
				\draw[cornellBlue, vertexProp] (0.5, 0) -- (1., 0) ;
				\draw[cornellBlue, fill=white] (-1., 0) circle (\extVert);
				\draw[cornellRed, fill=white] (1., 0) circle (\extVert);
				\draw[sigmaProp] (0, 0)++(0.5,0) arc(0:180:0.5);
				\draw[sigmaProp] (0, 0)++(0.5,0) arc(0:-180:0.5);
				\fill[cornellRed, intVertSty] (-0.5,0) circle (\intVert);
				\fill[cornellBlue, intVertSty] (0.5,0) circle (\intVert);
			\end{tikzpicture} + \begin{tikzpicture}[baseline=-3pt, thick]
			\def\cSize{0.1}
			\draw[cornellBlue, vertexProp] (-1, 0) -- (1, 0);
			\draw[cornellBlue, fill=white] (-1., 0) circle (\extVert);
			\draw[cornellRed, fill=white] (1., 0) circle (\extVert);
			\begin{scope}[shift={(0, 0)}, rotate=45]
			\draw[cornellBlue, fill=white] (0, 0) circle (\cSize);
			\draw[cornellBlue] (-\cSize, 0)--(\cSize, 0);
			\draw[cornellBlue] (0, -\cSize)--(0, \cSize);
			\end{scope}
		\end{tikzpicture}\, + \cdots\,, \label{eq:vertDiag}
	\end{equation}
	where we have restored the counterterm contributions and used the $\cdots$ to denote contributions from other regions of integration. In momentum space, (\ref{eq:vertDiag}) is 
	\begin{equation}
		[\mathcal{V}\mathcal{V}^\dagger]_J = [\mathcal{G}]_J + [\mathcal{G}]_J^2\left[\tfrac{1}{2}|g|^2[G^\sigma G^\sigma]_J -J(J+2\alpha)\delta_{Z_\varphi}-\delta_{m_\varphi}\right] +\cdots\,,
	\end{equation}
	and so the region of integration we have identified dominates in the long-distance limit $J \to \minus \beta$. 

	We can repeat this argument for arbitrary orders in perturbation theory. For instance, at~$\mathcal{O}(|g|^{2n})$ the vertex correlator receives a contribution of the form
	\begin{equation}
		\frac{1}{(2n)!} \langle \mathcal{V}(x) \mathcal{V}^\dagger(y) S_\lab{int}^{2n}\rangle \supset  \frac{|g|^{2n} }{(2n)! \, 2^{2n}} \! \int_{\{2n\}} \langle \sigma_1^2 \sigma_2^2 \cdots \sigma_{2n}^{2} \rangle\,  \big\langle \mathcal{V}_x  \Big[{\textstyle \prod_{i}}\big(\mathcal{V}_i + \mathcal{V}_i^\dagger\big)\Big] \mathcal{V}^\dagger_y\big\rangle\,.
	\end{equation}
	We want to argue that there is a region of integration that yields a long chain of the bubble diagrams in (\ref{eq:vertDiag}) connected by vertex propagators. In momentum space, this contribution would diverge as $J \to \minus \beta$ and so it must be included to determine the vertex two-point function's asymptotic behavior. We thus focus on the class of contributions in which all $\sigma$'s are contracted into propagator pairs, e.g. $\langle \sigma_1^2 \sigma_2^2 \cdots \sigma_{2n}^{2} \rangle \supset (G^\sigma_{12})^2 \cdots (G^\sigma_{2n-1, 2n})^2$, of which there are $(2n)!$ ways of doing so. Using our freedom to relabel the integration variables, we can choose a canonical ordering of the $z_i$ and write
	\begin{equation}
		\frac{1}{(2n)!} \langle \mathcal{V}(x) \mathcal{V}^\dagger(y) S_\lab{int}^{2n}\rangle \supset   \frac{|g|^{2n}}{2^{2n}}  \! \int_{\{2n\}}\! \!\big(G_{12}^{\sigma}\big)^2 \cdots \big(G_{2n-1,2n}^{\sigma}\big)^2  \left[\big\langle \mathcal{V}^{\phantom{\dagger}}_x \mathcal{V}_1^\dagger \mathcal{V}_{2}^{\phantom{\dagger}} \cdots \mathcal{V}^\dagger_y\big\rangle + \text{$\dagger$ perms}\right],
	\end{equation}
	where ``$\dagger$ perms'' again denotes all permutations of which vertex operators are conjugated. These permutations will interchange, say, $\mathcal{V}^{\dagger}_1 \mathcal{V}_{2}^{\phantom{\dagger}} \leftrightarrow \mathcal{V}^{\phantom{\dagger}}_1 \mathcal{V}_2^{\dagger}$ but this can be undone by taking $z_1 \leftrightarrow z_2$ without affecting the form of the $\sigma$ propagators. There are $2^n$ such permutations, and so we may write
	\begin{equation}
		\frac{1}{(2n)!} \langle \mathcal{V}(x) \mathcal{V}^\dagger(y) S_\lab{int}^{2n}\rangle \supset   \frac{|g|^{2n}}{2^{n}}  \! \int_{\{2n\}}\! \!\big(G_{12}^{\sigma}\big)^2 \cdots \big(G_{2n-1,2n}^{\sigma}\big)^2  \big\langle \mathcal{V}^{\phantom{\dagger}}_x \mathcal{V}_1^\dagger \mathcal{V}_{2}^{\phantom{\dagger}}\mathcal{V}_3^\dagger \cdots \mathcal{V}^\dagger_y\big\rangle\,. \label{eq:vertexCorrIntermediate}
	\end{equation}
	In the limit the separations $\xi_{x1}$, $\xi_{23}$, \dots, $\xi_{2n,y}$ are small while all others are large,\footnote{Note that this region of integration is different than the one which we identified for $\langle \sigma(x) \sigma(y) \rangle_\lab{c}$ in \S\ref{sec:leadingOrder}, in which pairs of internal vertices were kept close to one another. As illustrated with the leading order correction~(\ref{eq:vert_leading_ord}), this region of integration cancels once we include the disconnected components like $\langle \mathcal{V}(x) \mathcal{V}(y) \rangle \langle S_\lab{int}^{2n} \rangle$.} the vertex correlator again factorizes as
	\begin{equation}
		\langle \mathcal{V}^{\vphantom{\dagger}}_x \mathcal{V}_1^\dagger \mathcal{V}^{\vphantom{\dagger}}_2 \mathcal{V}_3^\dagger \cdots \mathcal{V}_y^\dagger \rangle = \mathcal{G}_{x1}\! \left[\frac{\mathcal{G}_{x3}}{\mathcal{G}_{x2}} \frac{\mathcal{G}_{12}}{\mathcal{G}_{13}} \cdots \right] \! \mathcal{G}_{23} \! \left[\frac{\mathcal{G}_{25}}{\mathcal{G}_{24}} \cdots \right] \! \cdots \mathcal{G}_{2n,y} \approx \mathcal{G}_{x1} \mathcal{G}_{23} \cdots \mathcal{G}_{2n,y}
	\end{equation}
	and the contribution from this region may be approximated as 
	\begin{equation}
		\frac{1}{(2n)!} \langle \mathcal{V}(x) \mathcal{V}^\dagger(y) S_\lab{int}^{2n}\rangle \supset   \frac{|g|^{2n}}{2^{n}}  \! \int_{\{2n\}} \! \mathcal{G}_{x1} \big(G_{12}^{\sigma}\big)^2\mathcal{G}_{23} \big(G_{34}^{\sigma}\big)^2 \cdots \big(G_{2n-1,2n}^{\sigma}\big)^2 \mathcal{G}_{2n,y} + \cdots\,, 
	\end{equation}
	or diagrammatically as
	\begin{equation}
		\begin{aligned}
			\langle \mathcal{V}(x) \mathcal{V}^\dagger(y) \rangle_\lab{c} &\supset \begin{tikzpicture}[baseline=-3pt, thick]
			\def\arcSize{0.45}
			\def\lSize{0.7}

			\draw[cornellBlue, vertexProp] (-\lSize/2, 0)--(\lSize/2, 0);
			\begin{scope}[shift={({-\arcSize-\lSize/2}, 0)}]
				\draw[cornellBlue, vertexProp] (-\arcSize-\lSize, 0)--(-\arcSize, 0);
				\draw[cornellBlue, fill=white] (-\arcSize-\lSize, 0) circle (\extVert);
				\draw[sigmaProp] (0, 0)++(-\arcSize, 0) arc (180:0:\arcSize);
				\draw[sigmaProp] (0, 0)++(-\arcSize, 0) arc (0:180:{-\arcSize});
				\fill[cornellRed, intVertSty] (-\arcSize, 0) circle (\intVert);
				\fill[cornellBlue, intVertSty] (\arcSize, 0) circle (\intVert);
			\end{scope}
			\begin{scope}[shift={({\arcSize+\lSize/2}, 0)}]
				\draw[cornellBlue, vertexProp] (\arcSize+\lSize/1.5, 0)--(\arcSize, 0);
				\draw[sigmaProp] (0, 0)++(-\arcSize, 0) arc (180:0:\arcSize);
				\draw[sigmaProp] (0, 0)++(-\arcSize, 0) arc (0:180:{-\arcSize});
				\fill[cornellRed, intVertSty] (-\arcSize, 0) circle (\intVert);
				\fill[cornellBlue, intVertSty] (\arcSize, 0) circle (\intVert);
			\end{scope}
			\begin{scope}[shift={({4*\arcSize+3*\lSize/2+0.35}, 0)}]
				\draw[cornellBlue, vertexProp] (-\arcSize-\lSize/1.5, 0) node[left, shift={(0.15, -0.015)}] {$\cdots$}--(-\arcSize, 0);
				\draw[cornellBlue, vertexProp] (\arcSize+\lSize, 0)--(\arcSize, 0);
				\draw[cornellRed, fill=white] (\arcSize+\lSize, 0) circle (\extVert);
				\draw[sigmaProp] (0, 0)++(-\arcSize, 0) arc (180:0:\arcSize);
				\draw[sigmaProp] (0, 0)++(-\arcSize, 0) arc (0:180:{-\arcSize});
				\fill[cornellRed, intVertSty] (-\arcSize, 0) circle (\intVert);
				\fill[cornellBlue, intVertSty] (\arcSize, 0) circle (\intVert);
			\end{scope}
		\end{tikzpicture} + \cdots\,,
		\end{aligned}
	\end{equation}
	where the $\cdots$ denote contributions from other regions of integration. In momentum space, this contribution strongly diverges as we send $J \to \minus \beta$,
	\begin{equation}
		[\mathcal{V}\mathcal{V}^\dagger] \supset [\mathcal{G}]_J^{n+1}\left[\tfrac{1}{2}|g|^2[G^\sigma G^\sigma]_J -J(J+2\alpha)\delta_{Z_\varphi}-\delta_{m_\varphi}\right]^{\!\es\es n},
	\end{equation}
	where we expect it to dominate over the regions of integration we have ignored.

	To summarize, we have identified contributions at each order in $g$ that become equally important and diverge as $J \to \minus \beta$. Diagrammatically, they form a geometric series of bubble diagrams~$[G^{\sigma} G^{\sigma}]_J$ connected by vertex propagators that can be straightforwardly summed. We can therefore approximate the self-energy $\Pi_\varphi(J)$ in (\ref{eq:vertexWatson}) at leading order in perturbation theory around $J \approx \minus \beta$ as
	\begin{equation}
		\Pi_\varphi(J) = \tfrac{1}{2} |g|^2 [G^\sigma G^{\sigma}]_J - J(J+2 \alpha) \delta_{Z_\varphi} - \delta_{m_\varphi} +  \cdots
	\end{equation}
	where the bubble diagram $[G^\sigma G^\sigma]_J$ can be computed by applying the inversion formula (\ref{eq:lInvForm}) to the product $G^\sigma(\zeta) G^\sigma(\zeta)$. This diagram has been studied in detail in \cite{Marolf:2010zp,Chakraborty:2023qbp}. It diverges as~$[G^\sigma G^\sigma]_{J} \sim (8\pi^2 \epsilon)^{\sminus 1}$ as $\alpha = \frac{1}{2}(3 - \epsilon) \to \frac{3}{2}$ and has poles at $J = -2\Delta_\sigma - 2k$, $J = -\Delta_\sigma - \bar{\Delta}_\sigma - 2k = -2\alpha - 2k$, and $J = -2 \bar{\Delta}_\sigma - 2k$ for all non-negative integers $k \in \mathbb{Z}_{\scriptscriptstyle \geq 0}$. It is thus convenient to define the regular part of this diagram as $[G^{\sigma} G^{\sigma}]_J = \rho_\sigma(J) + 1/(8 \pi^2 \epsilon)$ and write
	\begin{equation}
		\Pi_\varphi(J)  = \frac{1}{2} |g|^2 \rho_\sigma(J) + \frac{|g|^2}{16 \pi^2 \epsilon} - J(J+2 \alpha) \delta_{Z_\varphi} - \delta_{m_\varphi} \!\es\es+ \cdots\,.
	\end{equation}
	Unless $\beta$ is tuned to be an odd integer, the bubble is completely regular at the ``free-field pole'' $J = \minus \beta$ and, since $\Delta_\sigma$ and $\bar{\Delta}_\sigma$ are complex conjugates, $\rho_\sigma(J)$ is also real along the~$\lab{Re}\, J$ axis.

	As discussed in Section~\ref{sec:compact_int}, the asymptotic behavior of the corrected two-point function is determined by the zero $J_*$ of
	\begin{equation}
		[\mathcal{G}]_{J_*}^{\sminus 1} - \frac{1}{2}|g|^2 \rho_\sigma(J_*) - \frac{|g|^2}{16 \pi^2 \epsilon} + \delta_{m_\varphi} = 0
	\end{equation}
	closest to the $\lab{Im}\, J$-axis. We will choose our counterterms such that this zero is at $J_* = \minus \beta$ and $\Pi_\varphi'(\minus \beta) = 0$,
	\begin{equation}
		\begin{aligned}
			\delta_{m_\varphi} &= \tfrac{1}{2} |g|^2 [G^\sigma G^\sigma]_{\sminus \beta} + \beta(2 \alpha - \beta)\delta_{Z_\varphi}\,, \\
			\delta_{Z_\varphi} &= \tfrac{1}{4}|g|^2 \rho'_\sigma(\minus \beta)/(\alpha - \beta)
		\end{aligned}\,.
	\end{equation}
	so that $\beta$ is \emph{defined} to measure the connected vertex two-point function asymptotic decay such that $\langle \mathcal{V}(x) \mathcal{V}^\dagger(y) \rangle_\lab{c} \sim (\minus 2/\xi)^{\beta} + \cdots$ as $|\xi| \to \infty$. The self-energy is nontrivial
	\begin{equation}
		\Pi_\varphi(J) = \frac{1}{2} |g|^2 \left[ \rho_\sigma(J) - \rho_\sigma(\minus \beta) - \frac{(J+\beta)(J+2 \alpha - \beta)}{2(\alpha - \beta)} \rho'_{\sigma}(\minus \beta)\right]\,,
	\end{equation}
	and affects the locations and residues of the other poles of $[\mathcal{G}]_J$, and thus changes the subleading behavior of the vertex propagator, but unlike $\langle \sigma(x) \sigma(y) \rangle_\lab{c}$ the dominant asymptotic behavior is unaffected by the interactions, as their effect can be completely absorbed by our counterterms.

\linespread{1.05}
\phantomsection
\addcontentsline{toc}{section}{References}
\bibliographystyle{utphys}
\bibliography{compact.bib}

\end{document}